\pdfoutput=1
\documentclass[11pt]{article}
\usepackage[
  letterpaper,
  top = 1in,
  bottom = 1in,
  left = 1in,
  right = 1in
]{geometry}
\usepackage[T1]{fontenc}
\usepackage{mathpazo}
\usepackage{amsmath, amssymb, amsthm, mathtools}
\usepackage{aliascnt}
\usepackage{bbm}
\usepackage{microtype}
\usepackage{enumitem}
\usepackage[dvipsnames]{xcolor}
\usepackage{booktabs}
\usepackage{graphicx}
\usepackage{pgfplots}
\pgfplotsset{compat=1.18}
\usetikzlibrary{arrows.meta}
\usepackage{tikz-3dplot}
\usepackage{adjustbox}
\usepackage{float}
\usepackage{algorithm}
\usepackage{array}
\usepackage{longtable}
\usepackage{fancyhdr}
\usepackage[title, titletoc]{appendix}
\usepackage{url}
\usepackage[pagebackref]{hyperref}
\usepackage[nameinlink, capitalize]{cleveref}

\hypersetup{
  colorlinks = true,
  urlcolor = blue,
  linkcolor = RoyalBlue,
  citecolor = OliveGreen,
  linktocpage = true,
  pdftitle = {State Evolution and Universality of Iterated Linear Programs for the Binary Perceptron},
  pdfauthor = {Jianan Zhang and Shuangping Li},
  pdfsubject = {Iterated linear programming, binary perceptron, state evolution, and universality}
}

\makeatletter
\renewcommand\paragraph{\@startsection{paragraph}{4}{\z@}{1.1ex \@plus .4ex \@minus .2ex}{-1em}{\normalfont\normalsize\bfseries}}
\makeatother
\setlist{topsep = 0.3em, itemsep = 0.15em, parsep = 0.05em, leftmargin = 2.0em}
\allowdisplaybreaks

\newtheorem{theorem}{Theorem}[section]

\newaliascnt{proposition}{theorem}
\newtheorem{proposition}[proposition]{Proposition}
\aliascntresetthe{proposition}

\newaliascnt{lemma}{theorem}
\newtheorem{lemma}[lemma]{Lemma}
\aliascntresetthe{lemma}

\newaliascnt{corollary}{theorem}
\newtheorem{corollary}[corollary]{Corollary}
\aliascntresetthe{corollary}

\newaliascnt{conjecture}{theorem}

\aliascntresetthe{conjecture}

\theoremstyle{definition}
\newaliascnt{definition}{theorem}
\newtheorem{definition}[definition]{Definition}
\aliascntresetthe{definition}

\newaliascnt{assumption}{theorem}

\aliascntresetthe{assumption}

\theoremstyle{remark}
\newaliascnt{remark}{theorem}

\aliascntresetthe{remark}

\numberwithin{equation}{section}

\newcommand{\R}{\mathbb{R}}
\newcommand{\E}{\mathbb{E}}
\newcommand{\Pp}{\mathbb{P}}
\newcommand{\1}{\mathbbm{1}}
\newcommand{\law}{\mathsf{Law}}
\newcommand{\op}{\mathrm{op}}

\newcommand{\Wtwo}{W_2}
\newcommand{\Ent}{\mathrm{Ent}}
\newcommand{\clip}{\operatorname{clip}}
\newcommand{\dist}{\operatorname{dist}}

\newcommand{\diag}{\operatorname{diag}}
\newcommand{\argmax}{\operatorname*{arg\, max}}
\newcommand{\argmin}{\operatorname*{arg\, min}}
\newcommand{\cC}{\mathcal{C}}
\newcommand{\cA}{\mathcal{A}}

\newcommand{\eps}{\varepsilon}

\newcommand{\norm}[1]{\left\lVert #1\right\rVert}
\newcommand{\abs}[1]{\left\lvert #1\right\rvert}
\newcommand{\spanop}{\operatorname{span}}

\newcommand{\IterLP}{\textnormal{\textsc{Iterated-LP}}}
\newcommand{\coderepo}{\url{https://github.com/fifalsp/iterated-lp-perceptron}}

\title{State Evolution and Universality of Iterated Linear Programs for the Binary Perceptron}
\author{Jianan Zhang\thanks{School of the Gifted Young, University of Science and Technology of China. \texttt{jnzhang@mail.ustc.edu.cn}.}
\and
Shuangping Li\thanks{Department of Statistics and Data Science, Yale University. \texttt{shuangping.li@yale.edu}.}}
\date{November 2026}

\begin{document}
\hypersetup{pageanchor = false}
\maketitle

\begin{abstract}
The binary perceptron is a random constraint satisfaction problem that asks for a sign vector satisfying a system of random linear inequalities. Its solution space has a strikingly extreme geometry: at every positive constraint density, all but a vanishing fraction of the solutions are isolated. Yet polynomial-time algorithms find solutions at small enough constraint density, which raises the question of how efficient algorithms navigate such a landscape. Linear programming is a natural algorithm to study here, but its analysis has so far been limited to a single round. Li, Schramm, and Zhou (STOC 2025)~\cite{LSZ2025} analyzed one linear program followed by iterative rounding, and their analysis does not extend to further rounds.

In this work, we overcome this limitation and give a precise analysis of iterated linear programming for the binary perceptron. Our contributions are threefold. (1)~For Gaussian matrices and any fixed number of rounds, we characterize the iterated linear programs exactly through a deterministic state-evolution recursion. (2)~Adding a final rounding step, we obtain a randomized polynomial-time algorithm for the binary perceptron. Numerical evaluations of the recursion provide evidence that it succeeds at constraint density $\alpha=0.60$ with zero margin, well beyond the range $\alpha\le0.1$ guaranteed by the single-LP analysis and $\alpha\approx0.327$~\cite{JoLee2026}, the best guarantee so far. (3)~We establish universality, showing that the same characterization holds for matrices with i.i.d.\ entries drawn from any centered, unit-variance subgaussian law.

The key idea of our proof is that, although conditioning on earlier rounds makes the matrix non-Gaussian, its conditional law can be computed explicitly and remains close to a Gaussian law, so the convex Gaussian min--max theorem applies in every round. For universality, we compare Gaussian and subgaussian matrices through the Lindeberg principle, using a regularized version of each linear program.
\end{abstract}

\thispagestyle{empty}
\newpage
\setcounter{tocdepth}{1}
\tableofcontents
\thispagestyle{empty}
\newpage
\hypersetup{pageanchor = true}
\setcounter{page}{1}

\section{Introduction}\label{sec:introduction}

\subsection{The binary perceptron}\label{subsec:binary-perceptron-algorithmic-question}
The binary perceptron is a random constraint satisfaction problem that asks for a sign vector satisfying a system of random linear inequalities. The model originated in the study of the storage capacity of perceptrons~\cite{Rosenblatt1958,Cover1965,Gardner1988}, in particular with binary weights~\cite{GardnerDerrida1988,KrauthMezard1989}. It has since been studied extensively in probability~\cite{Talagrand1999,Xu2021,PerkinsXu2021,AbbeLiSlyContiguity2022,BolthausenEtAl2022,NakajimaSun2023,Altschuler2023,SahSawhney2023,Huang2024,DingSun2025,GongLi2026}, statistical physics~\cite{EngelVanDenBroeck2001,BraunsteinZecchina2006,BaldassiEtAl2007,HuangWongKabashima2013,HuangKabashima2014,BaldassiEtAl2015,AubinEtAl2019}, and theoretical computer science~\cite{KimRoche1998,AbbeLiSly2022,GamarnikEtAl2022,LSZ2025,VafaVaikuntanathan2025,ElAlaouiGamarnik2025,GongEtAl2026,JoLee2026}, and it is closely connected to discrepancy theory~\cite{Spencer1985,Bansal2010,LovettMeka2015,BansalSpencer2020} and random integer programming~\cite{ChandrasekaranVempala2014,BorstEtAl2023,AltschulerEtAl2026}.

The binary perceptron is particularly intriguing because its solution space has a strikingly unusual geometry. However small the ratio of constraints to variables (the \emph{constraint density}), the solution space shatters in an extreme form: all but a vanishing fraction of the solutions are isolated, in the sense that each of them differs from every other solution in a constant fraction of the coordinates. This picture was conjectured in~\cite{KrauthMezard1989,HuangWongKabashima2013,HuangKabashima2014,AubinEtAl2019}, proved first for a symmetric variant of the model~\cite{PerkinsXu2021,AbbeLiSlyContiguity2022}, and recently proved for the binary perceptron itself~\cite{GongLi2026}. Shattering also occurs in other random constraint satisfaction problems, such as random $k$-SAT and random graph $q$-coloring, but only once the constraint density is large enough~\cite{MezardParisiZecchina2002,KrzakalaEtAl2007,ZdeborovaKrzakala2007,AchlioptasCojaOghlan2008}. In those problems, the onset of shattering and freezing is commonly associated with the failure of known polynomial-time algorithms~\cite{AchlioptasCojaOghlan2008,AchlioptasRicciTersenghi2009,Molloy2018}. The binary perceptron defies this intuition, as polynomial-time algorithms provably find solutions at small enough constraint density~\cite{KimRoche1998,AbbeLiSly2022,LSZ2025,JoLee2026}. This tension leaves several fundamental questions open, including which kinds of algorithms can find solutions, how they navigate such a landscape, and up to what constraint density they succeed.

We now define the model. Let $A\in\R^{M\times N}$ be a random matrix with i.i.d.\ entries of mean zero and variance $1/N$. For a margin $\kappa\in\R$, we define the solution set
\begin{equation}\label{eq:intro-solution-set}
  \mathcal S_{N,M}(A,\kappa)
  =\left\{\sigma\in\{\pm1\}^N:A\sigma\ge\kappa\1\right\}.
\end{equation}
We work in the \emph{proportional regime}, in which $M,N\to\infty$ with $M/N\to\alpha\in(0,\infty)$, and we call $\alpha$ the \emph{constraint density}. Geometrically, we seek a vertex of the hypercube that lies in the intersection of $M$ random halfspaces. The margin $\kappa$ prescribes how much slack each inequality must have.

\subsection{Algorithmic barriers and linear programming}\label{subsec:one-lp-to-many}

The tension between shattering and efficient algorithms raises a central question about the binary perceptron: how do efficient algorithms find solutions in a landscape where all but a vanishing fraction of the solutions are isolated? Answering this question requires more than proving that an algorithm succeeds. It calls for a precise description of what the algorithm does along the way, such as how many coordinates it fixes and how the slack of the constraints evolves.

What is known about the limits of efficient algorithms concerns restricted classes of algorithms. The best rigorous guarantee so far comes from online algorithms, which choose each sign irrevocably before seeing later columns. Their exact threshold was recently determined by Jo and Lee~\cite{JoLee2026}, and at zero margin it lies between $0.327$ and $0.367$. For Lipschitz algorithms, such as gradient descent, Langevin dynamics, and approximate message passing, the threshold for optimizing perceptron Hamiltonians is characterized by a stochastic control problem~\cite{HuangSellkeSun2026} (its exact value for the binary perceptron at zero margin requires numerical evaluation). Yet solutions exist up to density about $0.833$~\cite{KrauthMezard1989,DingSun2025,Huang2024,Shmalo2026}, and message-passing heuristics find them empirically up to density about $0.74$~\cite{BraunsteinZecchina2006,BaldassiEtAl2007}. Going beyond these results therefore requires algorithms outside these classes, together with tools that describe them precisely.

In this paper, we focus on linear programming, for three reasons. First, it lies outside these classes. It reads the whole matrix before fixing any sign, so it is not online. Its optimizer is a vertex of a polytope determined by the matrix, and this vertex can jump under small perturbations of the matrix, so it is not Lipschitz. Second, it is the classical approach to integer feasibility problems~\cite{Schrijver1986,RaghavanThompson1987}. For the binary perceptron, we relax the sign constraint $\sigma\in\{\pm1\}^N$ to the cube constraint $x\in[-1,1]^N$ and maximize a linear objective over the polytope $P=\{x\in[-1,1]^N:Ax\ge\kappa\1\}$. The optimizer typically has many coordinates equal to $\pm1$, and the strategy of fixing such coordinates and recursing on the rest underlies iterative rounding and much of algorithmic discrepancy theory, where it has been remarkably successful~\cite{BeckFiala1981,LenstraShmoysTardos1990,LauRaviSingh2011,Bansal2010,LovettMeka2015,Rothvoss2017,EldanSingh2018,BansalEtAl2019,JainSahSawhney2023}. The connection is most direct for the symmetric binary perceptron~\cite{AubinEtAl2019}, a symmetric variant of the model that is an average-case discrepancy problem. Third, linear programs are among the few algorithms whose behavior on random instances can be analyzed exactly, through Gaussian comparison and related techniques, in settings ranging from sparse recovery and phase retrieval to robust regression, max-margin classification, and the negative spherical perceptron~\cite{Gordon1985,ThrampoulidisOymakHassibi2015,DonohoTanner2009,AmelunxenEtAl2014,DhifallahEtAl2017,ElKarouiEtAl2013,ThrampoulidisEtAl2018,LiangSur2022,MontanariEtAl2024}.

Li, Schramm, and Zhou (STOC 2025)~\cite{LSZ2025} were the first to bring these ideas to the binary perceptron. They analyzed a single linear program with a random objective and showed that it fixes a positive fraction of the signs while keeping the constraint margins under control. Their analysis, however, applies only to this first round. Once the optimizer is revealed, the remaining constraints have been selected by the same matrix, and conditioning on this selection makes the matrix non-Gaussian, so Gaussian comparison~\cite{Gordon1985,ThrampoulidisOymakHassibi2015} no longer applies directly. To complete the signing, their algorithm therefore switches to a different procedure, the Lovett--Meka random walk~\cite{LovettMeka2015}, and it provably succeeds at zero margin for $\alpha\le0.1$. This leaves open how linear programming itself behaves when it is iterated.

In this paper, we consider \IterLP, an algorithm that iterates the linear program of~\cite{LSZ2025} on the coordinates that remain free, in the spirit of~\cite{EldanSingh2018,JainSahSawhney2023}, and we overcome this conditioning barrier. Our contributions are threefold.

\paragraph{1.~Exact analysis.}
For Gaussian matrices and any fixed number of rounds, we characterize \IterLP\ exactly through a deterministic state-evolution recursion (Theorem~\ref{thm:intro-state-evolution}). The recursion determines, in every round, the fraction of frozen coordinates and the limiting distributions of the solution and of the constraint margins. To our knowledge, this is the first exact analysis of linear programs iterated on the same random matrix.

\paragraph{2.~An algorithm and numerical evidence beyond the current best guarantee.}
We obtain a randomized polynomial-time algorithm for the binary perceptron (Theorem~\ref{thm:intro-algorithm}). It runs \IterLP\ and then rounds the remaining free coordinates with the Lovett--Meka stage of~\cite{LSZ2025}, and its success reduces to a condition on the deterministic recursion. Numerical evaluation of this condition indicates that the algorithm succeeds at zero margin and density $\alpha=0.60$. This density is well above the current best guarantee $0.327$ and above $0.367$, beyond which no online algorithm succeeds~\cite{JoLee2026}, and it is six times the density $0.1$ guaranteed by the single-LP analysis.

\paragraph{3.~Universality.}
The same limits hold for matrices with i.i.d.\ centered, unit-variance subgaussian entries (Theorem~\ref{thm:intro-universality}). Since neither \IterLP\ nor the rounding stage uses the entry law, the algorithm and its guarantee carry over unchanged.

We state the precise results in Section~\ref{subsec:intro-main-results} and outline the proofs in Section~\ref{sec:technical-overview}.

\subsection{The \IterLP\ algorithm}\label{subsec:iterlp-algorithm}

We now describe the algorithm \IterLP\ that we analyze. It runs for a fixed number $K$ of rounds and solves one linear program in each round. In round $t$, it draws a fresh Gaussian direction $v_t$ and finds the feasible point that lies farthest in this direction. That is, it maximizes $v_t^\top x$ over all $x\in[-1,1]^N$ with $Ax\ge\kappa_t\1$, keeping the coordinates that reached $\pm1$ in earlier rounds fixed. Every coordinate of the new optimizer $x^{(t)}$ that equals $\pm1$ is then frozen for the rest of the algorithm. We write $B_t$ for the set of frozen coordinates after round $t$, which lie on the \emph{boundary} of the cube, and $F_t=[N]\setminus B_t$ for the remaining \emph{free} coordinates. In this way, each round turns some free coordinates into signs, and the next round continues with those that remain free.

\begin{algorithm}[H]
\caption{\IterLP}
\label{alg:intro-iterlp}
\small
\textbf{Input:} a matrix $A$, a number $K$ of rounds, and margins $\kappa_1>\cdots>\kappa_K$. Set $B_0=\varnothing$.\\
\textbf{For $t=1,\ldots,K$:}
\begin{enumerate}[label=\arabic*.,leftmargin=1.8em,itemsep=.1em,topsep=.2em]
  \item Draw $v_t\sim N(0,I_N)$, independent of $A$ and of $v_1,\ldots,v_{t-1}$.
  \item Solve the linear program
  \[
    \max_{x\in[-1,1]^N}\ v_t^\top x
    \quad\text{subject to}\quad
    Ax\ge\kappa_t\1,\quad x_j=x_j^{(t-1)}\ \text{for } j\in B_{t-1},
  \]
  and let $x^{(t)}$ be an optimal solution.
  \item Freeze the new signs: $B_t=\{j:|x_j^{(t)}|=1\}$ and $F_t=[N]\setminus B_t$.
\end{enumerate}
\textbf{Output:} the final optimizer $x^{(K)}$ and the set $B_K$ of frozen coordinates.
\end{algorithm}

The margins decrease, $\kappa_1>\cdots>\kappa_K$, so that each new round has room to make progress. Since $Ax^{(t-1)}\ge\kappa_{t-1}\1>\kappa_t\1$, every row constraint has positive slack at the previous optimizer, including the rows that were tight in round $t-1$. Hence $x^{(t-1)}$ is strictly feasible for the row constraints of round $t$, and the free coordinates can move in every direction. Algorithm~\ref{alg:intro-iterlp} summarizes the procedure, and Figure~\ref{fig:intro-iterated-lp} illustrates its first three rounds.

\begingroup
\setlength{\intextsep}{6pt}
\setlength{\abovecaptionskip}{4pt}
\setlength{\belowcaptionskip}{0pt}
\begin{figure}[!t]
\centering
\input{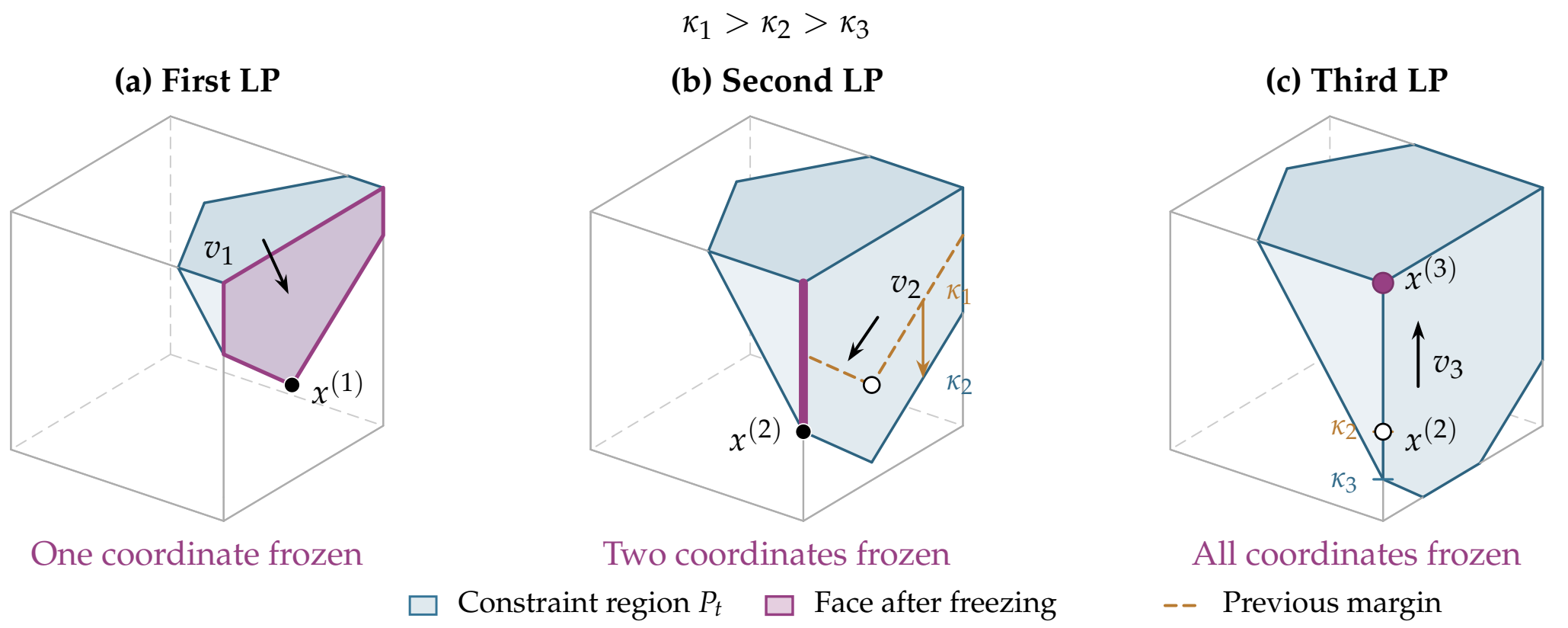}
\caption{Three rounds of \IterLP\ in dimension three, with margins $\kappa_1>\kappa_2>\kappa_3$. In panel $t$, blue shows the constraint region $P_t=\{x\in[-1,1]^3:Ax\ge\kappa_t\1\}$, the black arrow shows the objective $v_t$ along the coordinates that are still free, the filled point is the new optimizer $x^{(t)}$, and the open point is the previous optimizer. In the first round, one coordinate of the optimizer reaches $\pm1$ and freezes, so the second LP moves only within the purple face. The second round freezes another coordinate, so the third LP moves only along the purple edge and ends at a vertex of the cube, which is a sign vector. Dashed orange lines show the previous margin, which is lowered so that the previous optimizer has slack in every row.}
\label{fig:intro-iterated-lp}
\end{figure}
\endgroup

\subsection{Main results}\label{subsec:intro-main-results}

Our main results describe precisely what \IterLP\ does in every round. For any fixed number of rounds, we determine how many coordinates have frozen to signs, how the remaining coordinates are distributed, and how much slack the constraints retain. Writing $1-q_t$ for the limiting fraction of frozen coordinates after round $t$ and $U_t$ for the limiting law of a typical row margin, we compute these quantities, together with the limiting law of the coordinates, from a deterministic state-evolution recursion. Moreover, these limits are universal, in the sense that they are the same for every centered, unit-variance subgaussian entry law.

We now state these conclusions precisely. Fix $M=\lfloor\alpha N\rfloor$, a number $K$ of rounds independent of $N$, and internal margins $\kappa_1>\cdots>\kappa_K>\kappa$, where $\kappa$ is the target margin of the binary perceptron. All limits are taken as $N\to\infty$, and an event holds \emph{with high probability} if its probability tends to one. Write $u^{(t)}=Ax^{(t)}$ for the row margins in round $t$, and let
\[
  \widehat\mu_{N,t}^{\rm col}=\frac1N\sum_{j=1}^N\delta_{x_j^{(1:t)}},
  \qquad
  \widehat\mu_{N,t}^{\rm row}=\frac1M\sum_{i=1}^M\delta_{u_i^{(1:t)}}
\]
be the empirical laws of the coordinate trajectories, one for each column of $A$, and of the margin trajectories, one for each row of $A$, up to round $t$.

The first LP must be feasible. Following~\cite[Eq.~(37)]{LSZ2025}, we assume the first-round feasibility condition
\begin{equation}\label{eq:stage1-feasibility}
  \alpha<\sup_{0<\rho\le1}\frac{\varphi_1(\rho)^2}{\E(\kappa_1-\rho G)_+^2},
  \qquad
  \varphi_1(\rho)=\sup\bigl\{\E[GY]:|Y|\le1,\ \E Y^2=\rho^2\bigr\},
\end{equation}
where $G\sim N(0,1)$ and the supremum defining the Gaussian width $\varphi_1$ allows independent randomization. This condition ensures that the first LP is feasible with high probability.

Our first theorem identifies their limits for Gaussian matrices.

\begin{theorem}[Gaussian state evolution]\label{thm:intro-state-evolution}
  Let $A_{ij}\stackrel{\rm iid}{\sim}N(0,1/N)$, and fix a number $K$ of rounds. Suppose that $(\alpha,\kappa_1)$ satisfies the first-round feasibility condition~\eqref{eq:stage1-feasibility}, and let $q_t$, $\mu_t^{\rm col}$, and $\mu_t^{\rm row}$ be given by the state-evolution recursion (see Algorithm~\ref{alg:state-evolution} in Section~\ref{subsec:state-evolution} for details). Then, jointly for $1\le t\le K$,
  \[
    \Wtwo\bigl(\widehat\mu_{N,t}^{\rm col},\mu_t^{\rm col}\bigr)\xrightarrow{\Pp}0,
    \qquad
    \Wtwo\bigl(\widehat\mu_{N,t}^{\rm row},\mu_t^{\rm row}\bigr)\xrightarrow{\Pp}0,
    \qquad
    \frac{|B_t|}{N}\xrightarrow{\Pp}1-q_t.
  \]
\end{theorem}

\begin{figure}[!ht]
\centering
\includegraphics[width=0.85\linewidth]{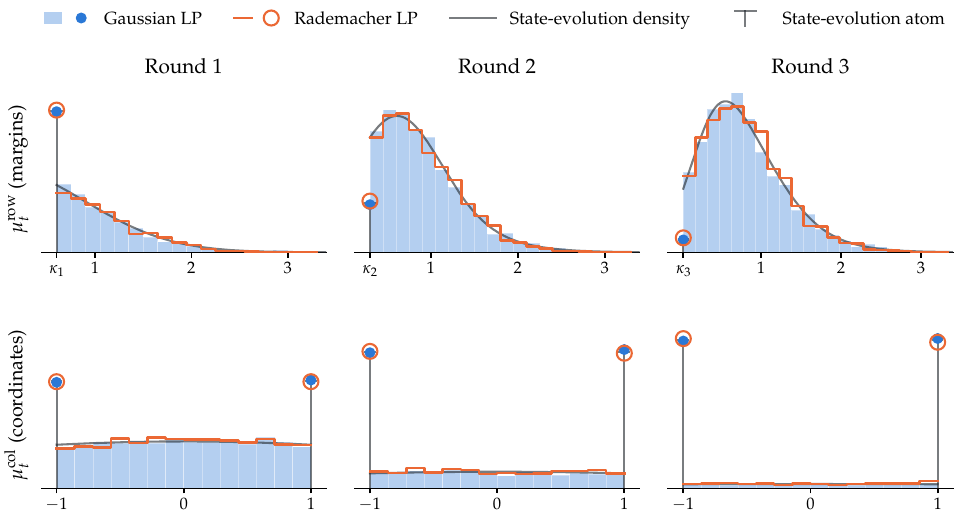}
\caption{The laws of Theorems~\ref{thm:intro-state-evolution} and~\ref{thm:intro-universality} in round $t$ at $\alpha=0.40$, with $(\kappa_1,\kappa_2,\kappa_3)=(0.60,0.30,0.03)$, for the row margins $u_i^{(t)}$ (top, from $\mu_t^{\rm row}$) and the coordinates $x_j^{(t)}$ (bottom, from $\mu_t^{\rm col}$). Gray curves and stems show the state-evolution limits, and histograms and markers show the empirical laws for Gaussian and Rademacher matrices with $N=2048$. Stem heights are the masses of the atoms formed by tight rows and frozen coordinates.}
\label{fig:intro-state-evolution}
\end{figure}

Figure~\ref{fig:intro-state-evolution} illustrates Theorem~\ref{thm:intro-state-evolution} at density $\alpha=0.40$.

Our second theorem shows that the same limits hold for subgaussian matrices (see Figure~\ref{fig:intro-state-evolution}). A random variable $\xi$ is \emph{subgaussian} if its tails decay at least as fast as Gaussian tails, that is, if its subgaussian norm $\|\xi\|_{\psi_2}=\inf\{s>0:\E e^{\xi^2/s^2}\le2\}$ is finite. For example, Gaussian and Rademacher random variables are subgaussian.

\begin{theorem}[Subgaussian universality]\label{thm:intro-universality}
  Fix a density $\alpha$ and an LP margin schedule $(\kappa_t)_{t=1}^K$ satisfying the hypotheses of Theorem~\ref{thm:intro-state-evolution}. Let $A_{ij}=\xi_{ij}/\sqrt N$, where the $\xi_{ij}$ are i.i.d.\ copies of a random variable $\xi$ whose law does not depend on $N$, with $\E\xi=0$, $\E\xi^2=1$, and $\|\xi\|_{\psi_2}<\infty$. Then \IterLP\ is well defined with high probability, and, for the iterates $x^{(t)}$, row margins $u^{(t)}=Ax^{(t)}$, and frozen sets $B_t$ of \IterLP, jointly for $1\le t\le K$,
  \[
    \Wtwo\bigl(\widehat\mu_{N,t}^{\rm col},\mu_t^{\rm col}\bigr)\xrightarrow{\Pp}0,
    \qquad
    \Wtwo\bigl(\widehat\mu_{N,t}^{\rm row},\mu_t^{\rm row}\bigr)\xrightarrow{\Pp}0,
    \qquad
    \frac{|B_t|}{N}\xrightarrow{\Pp}1-q_t,
  \]
  with exactly the same limits $\mu_t^{\rm col}$, $\mu_t^{\rm row}$, and $q_t$ as in Theorem~\ref{thm:intro-state-evolution}.
\end{theorem}

We now briefly describe the recursion in Theorems~\ref{thm:intro-state-evolution} and~\ref{thm:intro-universality}, and we defer its full statement to Algorithm~\ref{alg:state-evolution} in Section~\ref{subsec:state-evolution}. Besides the primal values $x^{(t)}$ and $u^{(t)}$, the recursion tracks the dual solution of each LP, namely the optimal multipliers $\lambda^{(t)}\ge0$ of the row constraints and $r^{(t)}=A^\top\lambda^{(t)}$. Entering round $t$, its state consists of the free fraction $q_{t-1}$ and random vectors $(X,R)$ and $(U,\Lambda)$ in $\R^{t-1}\times\R^{t-1}$, whose laws are the limiting joint laws of the past values $(x_j^{(s)},r_j^{(s)})_{s<t}$ over the free coordinates and of $(u_i^{(s)},\lambda_i^{(s)})_{s<t}$ over the rows. Round $t$ first solves a deterministic optimization problem in $2t-1$ variables, whose value $V_t$ is the limiting normalized value of the round-$t$ LP and whose optimizer determines a few scalar coefficients. It then updates a free coordinate and a row by
\begin{equation}\label{eq:intro-recursion}
  Y_t=\max\Bigl\{-1,\,\min\Bigl\{1,\,\frac{S_t+\eta_{x,t}^\top X+\eta_{r,t}^\top R}{\tau_t}\Bigr\}\Bigr\},
  \qquad
  U_t=\max\{\kappa_t,\,Z_t-\eta_{\lambda,t}^\top\Lambda\},
\end{equation}
where $S_t\sim N(0,1)$ is independent of the state, $Z_t$ is an independent Gaussian plus a linear function of $(U,\Lambda)$, and $\tau_t>0$ and $\eta_{x,t},\eta_{r,t},\eta_{\lambda,t}\in\R^{t-1}$ are the coefficients from the first step. Thus a free coordinate freezes when $|Y_t|=1$, and a row is tight when $U_t=\kappa_t$. These two events produce the atoms in Figure~\ref{fig:intro-state-evolution}.

We now turn \IterLP\ into an algorithm for the binary perceptron. After $K$ rounds, a fraction $q_K$ of the coordinates is still free, and every row margin is at least $\kappa_K$. To round the free coordinates to signs, we use the rounding stage of~\cite{LSZ2025}, which repeatedly applies the partial coloring algorithm of Lovett and Meka~\cite{LovettMeka2015}. We run it with every number of partial-coloring steps up to $\lceil\log N\rceil$, so that it does not depend on the entry law (Appendix~\ref{app:cold-rounding-transfer}). Each partial coloring fixes a constant fraction of the free coordinates at the price of slightly lowering the row margins, so the rounding stage reaches a target margin $\kappa$ as long as the surplus margin $\kappa_K-\kappa$ covers its total cost. This cost is an explicit function $\mathcal D_\alpha(q_K)$ of the free fraction, of order $\sqrt{q_K\log(1/q_K)}$ for small $q_K$, which we define in~\eqref{eq:rounding-cost}. Since our theorems determine $q_K$, they yield the following guarantee for the whole algorithm.

\begin{theorem}[Algorithm for the binary perceptron]\label{thm:intro-algorithm}
  Fix a density $\alpha$ and an LP margin schedule $(\kappa_t)_{t=1}^K$ satisfying the hypotheses of Theorem~\ref{thm:intro-state-evolution}, and let $A$ be an i.i.d.\ subgaussian matrix as in Theorem~\ref{thm:intro-universality}. Let $q_K$ be computed by Algorithm~\ref{alg:state-evolution}, and suppose that the target margin $\kappa$ satisfies $\kappa_K-\kappa>\mathcal D_\alpha(q_K)$. Then the $K$ rounds of \IterLP\ followed by the rounding stage form a randomized polynomial-time algorithm $\Xi$ that finds a solution of the binary perceptron, that is, with high probability,
  \begin{equation*}
    \Xi(A)\in\{\pm1\}^N
    \qquad\text{and}\qquad
    A\,\Xi(A)\ge\kappa\1.
  \end{equation*}
\end{theorem}

We evaluate the condition of Theorem~\ref{thm:intro-algorithm} numerically. Like known algorithmic thresholds for the binary perceptron, it is characterized implicitly, and its value requires numerical evaluation. For example, the threshold of~\cite{JoLee2026} for online algorithms is located by a computer-assisted proof, and the value of the threshold of~\cite{HuangSellkeSun2026} at zero margin has not yet been evaluated, which would require a substantial numerical study of the associated stochastic control problem. In Section~\ref{sec:numerical}, we evaluate our recursion along a quadratic margin schedule at density $\alpha=0.60$. The evaluation provides numerical evidence that our algorithm succeeds at this density with zero margin, beyond the current best guarantee.

\subsection{Further related work}\label{subsec:related-work}

\paragraph{Capacity and the algorithmic gap.}
At zero margin, solutions exist up to the capacity $\alpha_\star\approx0.833$ predicted by Krauth and M\'ezard~\cite{KrauthMezard1989}. After early bounds~\cite{KimRoche1998,Talagrand1999}, a sharp threshold~\cite{Xu2021}, the Gardner formula at small densities~\cite{BolthausenEtAl2022}, and universality of the threshold sequence~\cite{NakajimaSun2023}, matching lower and upper bounds were proved subject to numerical conditions~\cite{DingSun2025,Huang2024}, which were recently verified with computer assistance~\cite{Shmalo2026}. The capacity, sharp threshold, and critical window of the symmetric model are also well understood~\cite{AubinEtAl2019,PerkinsXu2021,AbbeLiSlyContiguity2022,SahSawhney2023,Altschuler2023}. Provable algorithmic guarantees remain far below the capacity~\cite{KimRoche1998,AbbeLiSly2022,LSZ2025,JoLee2026}. Closing this gap requires understanding concrete algorithms at constant density, which is the aim of our analysis of iterated linear programming.

\paragraph{Solution space structure and algorithmic barriers.}
Remarkably, although all but a vanishing fraction of the solutions are isolated, the solution space is not entirely frozen. Statistical physics predicted rare, well-connected clusters of solutions that explain the success of simple learning algorithms~\cite{BaldassiEtAl2015}, and~\cite{AbbeLiSly2022} proved that such well-connected clusters exist at low density. We now review further algorithmic barriers, beyond those of Section~\ref{subsec:one-lp-to-many}. Stable algorithms cannot reliably find isolated solutions~\cite{GongEtAl2026}. For the symmetric model, the overlap-gap property rules out stable and online algorithms above certain densities~\cite{GamarnikEtAl2022,GamarnikEtAl2023}, lattice-based reductions give evidence of hardness~\cite{VafaVaikuntanathan2025}, and sampling solutions is hard~\cite{ElAlaouiGamarnik2025}. The Lipschitz threshold builds on work on spherical perceptrons and projection pursuit~\cite{ElAlaouiSellke2022,MontanariZhou2024}, online discrepancy thresholds are universal~\cite{JoLee2026b}, and related thresholds are known for online balancing~\cite{BansalSpencer2020,FiedlerEtAl2026,AltschulerTikhomirov2026} and random matrix discrepancy~\cite{KizildagLi2026,ChenElAlaoui2026}; see~\cite{AchlioptasCojaOghlan2008,Gamarnik2021,LiSchramm2024} for discussions of the overlap-gap property and related barriers.

\paragraph{Discrepancy and integer programming.}
\IterLP\ follows the LP and random-walk methods of algorithmic discrepancy theory~\cite{Spencer1985,Bansal2010,LovettMeka2015,BansalDadushGarg2016,Rothvoss2017,EldanSingh2018,BansalEtAl2019,JainSahSawhney2023}, which recently led to an announced resolution of the Koml\'os conjecture~\cite{GuoFangLu2026b,GuoFangLu2026a}. These methods are analyzed for worst-case inputs, where their guarantees are tight only up to constant factors. Discrepancy~\cite{Costello2009,EzraLovett2019,HobergRothvoss2019,BansalMeka2019,FranksSaks2020,TurnerEtAl2020,AltschulerNilesWeed2022} and integer and linear programs~\cite{ChandrasekaranVempala2014,BorstEtAl2023,AltschulerEtAl2026,AbdallaVershyninZou2026} with random inputs have been studied directly.

\paragraph{Exact asymptotics and universality.}\label{subsec:universality-amp}
Our proof builds on two lines of work on exact high-dimensional asymptotics, namely Gaussian comparison and state evolution. Gaussian comparison gives sharp analyses of a single convex program~\cite{Gordon1985,Stojnic2013,ThrampoulidisOymakHassibi2015,ThrampoulidisEtAl2018}, including linear programs for the negative perceptron and max-margin classification~\cite{MontanariEtAl2024,LiangSur2022}. State evolution describes iterative algorithms such as approximate message passing~\cite{DonohoEtAl2009,Bolthausen2014,BayatiMontanari2011,JavanmardMontanari2013,BerthierMontanariNguyen2020,FengEtAl2022}. It was recently connected to Gaussian comparison~\cite{MalikHassibi2026}, extended to free-probabilistic limits~\cite{ChenElAlaoui2026}, and developed for optimization-based iterations with fresh samples in each step~\cite{ChandrasekherEtAl2023}. The closest work to ours is~\cite{CelentanoEtAl2025}, which derives state evolution for saddle-point updates that reuse one Gaussian matrix, with smooth, strongly convex penalties and unconstrained variables. Each round of \IterLP\ instead solves an exact linear program with hard constraints. Its optimal vertex places positive fractions of the coordinates at $\pm1$ and of the rows at the margin, which produces the atoms of our state evolution, and conditioning on the coordinates and rows selected by earlier vertices makes the law of the matrix non-Gaussian. \IterLP\ therefore needs both an exact analysis of a nonsmooth convex program and a state evolution on reused data, and its memory terms come from conditioning on the optimization history rather than from an Onsager correction. Our universality argument follows the Lindeberg principle for convex estimators~\cite{Chatterjee2006,KoradaMontanari2011,OymakTropp2018,MontanariSaeed2022,HanShen2023,MontanariEtAl2023} and complements universality results for message passing and first-order methods~\cite{ChenLam2021,Han2025}. The closest result for linear programs is the universality of the phase transition of $\ell_1$ minimization, which Bayati, Lelarge, and Montanari proved through message passing~\cite{BayatiLelargeMontanari2015}.

\subsection{Organization and notation}\label{subsec:organization}

\paragraph{Organization.}
In Section~\ref{sec:technical-overview}, we outline the proof and state the state-evolution recursion in full (Algorithm~\ref{alg:state-evolution}). In Section~\ref{sec:preliminaries}, we collect the Gaussian tools we use, namely the Gaussian min--max comparison theorem, Gaussian regression, and relative-entropy bounds. In Section~\ref{sec:model}, we set up the iterated LP, its optimization histories, and their KKT structure. In Section~\ref{sec:optimizer-coupling}, we couple the matrix conditioned on the optimization history with a Gaussian matrix. In Section~\ref{sec:finite-po}, we analyze a single round on a fixed Gaussian history and derive the population variational problem and the update laws of Algorithm~\ref{alg:state-evolution}. In Section~\ref{sec:closing-gaussian-recursion}, we close the induction over rounds and prove Theorem~\ref{thm:intro-state-evolution}. In Section~\ref{sec:numerical}, we evaluate the recursion numerically. In Section~\ref{sec:universality}, we prove Theorem~\ref{thm:intro-universality} and transfer rounding certificates to subgaussian matrices. In Appendix~\ref{app:deferred-proofs}, we give the deferred proofs, including the rounding guarantee behind Theorem~\ref{thm:intro-algorithm} (Appendix~\ref{app:cold-rounding-transfer}).

\paragraph{Notation.}
We write $[N]=\{1,\ldots,N\}$, $\1$ for the all-ones vector, $I_N$ for the $N\times N$ identity matrix, and $e_p$ for the $p$-th standard basis vector. For a vector $x$ and a set $S$, $x_S$ is the restriction of $x$ to the coordinates in $S$. For a matrix $A$ and sets $I$ and $S$, $A_{I,S}$ is the submatrix with rows in $I$ and columns in $S$, and $A_{:,S}$ keeps all rows. We write $\|\cdot\|$ for the Euclidean norm, $\|\cdot\|_{\op}$ for the operator norm, and $z_+=\max\{z,0\}$. For a coordinate trajectory, $x_j^{(1:t)}=(x_j^{(1)},\ldots,x_j^{(t)})$. We write $\law(X)$ for the law of a random vector $X$, $\delta_x$ for the point mass at $x$, and $\Wtwo$ for the quadratic Wasserstein distance, $\Wtwo(\mu,\nu)^2=\inf\E\|X-Y\|^2$, where the infimum is over all couplings of $X\sim\mu$ and $Y\sim\nu$. We write $\xrightarrow{\Pp}$ for convergence in probability as $N\to\infty$ and $o_{\Pp}(1)$ for a quantity that converges to zero in probability. Finally, $P_N$ denotes a polylogarithmic factor in $N$ whose exponent may change between occurrences.

\section{Technical overview}\label{sec:technical-overview}

We first give a brief overview of the argument. Sections~\ref{subsec:overview-conditional-law} and~\ref{subsec:overview-gaussian-recursion} explain the exact analysis behind Theorem~\ref{thm:intro-state-evolution}, Section~\ref{subsec:state-evolution} states the state-evolution recursion in full, and Section~\ref{subsec:overview-universality} explains universality (Theorem~\ref{thm:intro-universality}).

Our analysis of the first round refines that of~\cite{LSZ2025}. Instead of tracking only the empirical law of the row margins $Ax$, we track jointly the empirical laws of the primal quantities $x$ and $Ax$ and of the dual quantities $\lambda$ and $A^\top\lambda$, and we determine the exact limit of the frozen fraction (Sections~\ref{sec:finite-po} and~\ref{sec:closing-gaussian-recursion}, with an empty history). These joint laws are precisely the information that later rounds require.

After the first round, conditioning on the optimization history changes the law of the reused matrix $A$, and this conditional law is no longer Gaussian. Drawing a fresh objective independently of $A$ and the earlier objectives does not undo this conditioning, so Gaussian comparison cannot be applied directly~\cite{LSZ2025}. We record the complete optimization history and compute explicitly the likelihood ratio between the law of the matrix given this history and the Gaussian law conditioned only on the linear constraints that the history imposes. This ratio is proportional to a product of squared determinants. Relative-entropy and transportation bounds then give, outside a set of histories of vanishing probability, a coupling under which the difference between the two matrices vanishes in operator norm in probability. This allows us to apply Gaussian comparison again (Section~\ref{sec:optimizer-coupling}).

A general round follows the same principle with a longer history. The auxiliary problem from Gaussian comparison then reduces to a variational problem over $2t-1$ overlap parameters, and its value gaps identify the limiting laws of the next round (Sections~\ref{sec:finite-po} and~\ref{sec:closing-gaussian-recursion}).

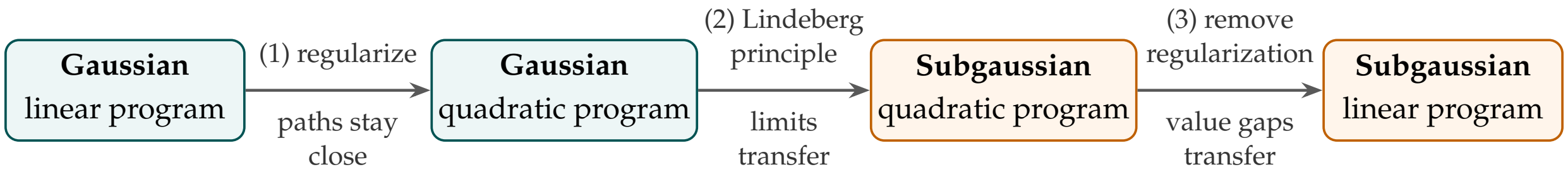
\begin{figure}[htbp]
\centering
\begin{tikzpicture}[
  box/.style={draw,rounded corners=4pt,align=center,minimum width=2.5cm,minimum height=1.05cm,inner sep=2pt,line width=.7pt,font=\footnotesize},
  gbox/.style={box,draw=teal!65!black,fill=teal!7},
  ybox/.style={box,draw=orange!75!black,fill=orange!8},
  arr/.style={-{Stealth[length=2.2mm]},line width=.85pt,black!65},
  lab/.style={font=\scriptsize,align=center,text=black!80}]
\node[gbox] (GL) at (0,0) {\textbf{Gaussian}\\[1pt] linear program};
\node[gbox] (GQ) at (4.6,0) {\textbf{Gaussian}\\[1pt] quadratic program};
\node[ybox] (YQ) at (9.2,0) {\textbf{Subgaussian}\\[1pt] quadratic program};
\node[ybox] (YL) at (13.8,0) {\textbf{Subgaussian}\\[1pt] linear program};
\draw[arr] (GL) -- node[lab,above=2pt] {(1) regularize} node[lab,below=2pt] {paths stay\\close} (GQ);
\draw[arr] (GQ) -- node[lab,above=2pt] {(2) Lindeberg\\principle} node[lab,below=2pt] {limits\\transfer} (YQ);
\draw[arr] (YQ) -- node[lab,above=2pt] {(3) remove\\regularization} node[lab,below=2pt] {value gaps\\transfer} (YL);
\end{tikzpicture}
\caption{The chain of comparisons in the proof of universality.}
\label{fig:overview-universality-chain}
\end{figure}

Finally, we transfer the conclusions to subgaussian matrices through the Lindeberg principle. The exact linear program is not stable enough for this purpose, since its optimal vertex can jump under small perturbations of the matrix and its feasible set depends on $A$. We therefore pass through a regularized quadratic program with a smooth freezing rule, along the chain of comparisons in Figure~\ref{fig:overview-universality-chain} (Section~\ref{sec:universality}).

\subsection{Conditioning on the optimization history}\label{subsec:overview-conditional-law}

Fix a depth $p$. The complete history after $p$ rounds records, for every round $r\le p$, the optimizer $x^{(r)}$ and its row margins $Ax^{(r)}$, the row multipliers $\lambda^{(r)}$ and the vector $A^\top\lambda^{(r)}$, the active row set $I_r=\{i:\lambda_i^{(r)}>0\}$, and the free and frozen sets $F_r,B_r$. We work with \emph{regular} histories (Definition~\ref{def:regular-history}), in which the KKT conditions of every round are nondegenerate. Every row constraint is either active with a positive multiplier or strictly slack, every newly frozen coordinate has a reduced cost of the same sign as its value, the active block $A_{I_r,F_r}$ is square and invertible, and both $x^{(1)},\ldots,x^{(p)}$ and $\lambda^{(1)},\ldots,\lambda^{(p)}$ are linearly independent. For Gaussian matrices, the history is regular at every fixed depth with high probability.

Let $P_h$ be the conditional law of $A$ given that the first $p$ rounds produce the history $h$, and let $Q_h$ be the Gaussian law of $A$ conditioned only on the affine constraints that fix the recorded values of $Ax^{(r)}$ and $A^\top\lambda^{(r)}$ for $r\le p$. By Gaussian regression, which is the conditioning technique of~\cite{Bolthausen2014,BayatiMontanari2011}, $Q_h$ is explicit. Under $Q_h$, the matrix is a deterministic function of $h$ plus an isotropic Gaussian matrix supported on the orthogonal complement of the history directions. The two laws differ because $P_h$ also conditions on the event that the recorded points are optimizers. We show that $P_h$ is absolutely continuous with respect to $Q_h$, with Radon--Nikodym derivative
\[
  \frac{dP_h}{dQ_h}(A)
  =\frac{\prod_{r\le p}\left|\det A_{I_r,F_r}\right|^2}{\E_{Q_h}\prod_{r\le p}\left|\det A_{I_r,F_r}\right|^2}.
\]
The squared determinants are the Jacobians of the KKT equations with respect to the primal and dual optimizer coordinates. Moment bounds for these determinants show that the relative entropy of $P_h$ with respect to $Q_h$, averaged over $h$, is $O(\log N)$ at fixed depth. Since $Q_h$ is Gaussian, Talagrand's transportation inequality~\cite{Talagrand1996} then yields, for all $h$ outside a set of histories of vanishing probability, a coupling $(A^*,A^G)$ of $P_h$ and $Q_h$ with $\|A^*-A^G\|_{\op}\to0$ in probability.

This coupling suffices for the analysis of the next round. In the min--max formulation of the next LP, the matrix enters only through the bilinear form $\lambda^\top Ax$. When $x$ and $\lambda$ are restricted to balls of radius $O(\sqrt N)$, replacing $A^*$ by $A^G$ changes this form, and hence the restricted min--max values, by $o_{\Pp}(N)$. Consequently, every gap of order $N$ between restricted values that identifies the optimizer under the Gaussian law $Q_h$ remains valid under the true conditional law $P_h$ (Section~\ref{sec:optimizer-coupling}).

\subsection{Reduction of each round to a variational problem}\label{subsec:overview-gaussian-recursion}

We now describe how we analyze a single round $t=p+1$ once the history is fixed. By Section~\ref{subsec:overview-conditional-law}, we may replace the conditional law of the matrix by the Gaussian law $Q_h$. Under $Q_h$, Gaussian regression decomposes the matrix into a part determined by the history and an independent Gaussian part acting on the directions orthogonal to the history. The new round therefore sees the history only through a deterministic shift and fresh Gaussian noise.

We then apply Gaussian comparison to the LP of round $t$. The resulting auxiliary problem becomes tractable once we fix finitely many order parameters. Write $q=q_p$, let $Y\in[-1,1]$ be the new value of a currently free coordinate, and let $(X,R)\in\R^p\times\R^p$ be its primal and dual history, with last primal coordinate $X_p$. The order parameters are the second moment of the new coordinate and the overlaps of its increment with the history,
\[
  s=q\E Y^2,\qquad a=q\E[X(Y-X_p)],\qquad b=q\E[R(Y-X_p)],
\]
which form a vector $\omega=(s,a,b)\in\R^{2t-1}$. With $\omega$ fixed, the auxiliary problem splits into a column problem and a row problem. The column problem has value $\phi_t(\omega)$, a Gaussian width, namely the largest correlation with a fresh Gaussian that a coordinate update with these moments can achieve. The row problem has value $d_t(\omega)$, a projection cost, namely the smallest normalized $L^2$ cost of correcting the regressed row margins so that they are at least $\kappa_t$ and remain consistent with the old dual variables. The limiting normalized value of the LP is
\[
  V_t=\max_{\omega:\,\phi_t(\omega)\ge d_t(\omega)}
  \sqrt{\phi_t(\omega)^2-d_t(\omega)^2},
\]
and its unique maximizer $\omega_t^\star$ produces the explicit update laws~\eqref{eq:intro-recursion}, namely Gaussians projected onto $[-1,1]$ for the coordinates and onto $[\kappa_t,\infty)$ for the margins.

To show that the actual optimizer follows these laws, we prove quadratic value gaps. Strong concavity on the column side and the Pythagorean inequality for the projection on the row side show that a candidate whose empirical law is at $W_2$ distance $\varepsilon$ from the prediction loses an amount of order $\varepsilon^2$ in normalized value. We transfer this gap from the population problem to the finite auxiliary problem, and then, through Gaussian comparison and the coupling of Section~\ref{subsec:overview-conditional-law}, to the actual conditional LP, which identifies the limiting empirical laws of the coordinates and the row margins.

Since $\Wtwo$ convergence alone does not determine the masses of the atoms, we obtain the free fraction by combining the population identity $\alpha\,\Pp(U_t=\kappa_t)=q_t$ with a count of the active constraints at the optimal vertex, and analogous value gaps on the dual side identify the laws of the multipliers. Finally, the new population history is again strict in the sense of Definition~\ref{def:strict-history}, so the argument applies to the next round (Sections~\ref{sec:finite-po} and~\ref{sec:closing-gaussian-recursion}).

\subsection{The state-evolution recursion in full}\label{subsec:state-evolution}

We now state precisely the recursion that the preceding discussion derives. Algorithm~\ref{alg:state-evolution} acts on probability laws in the same way that \IterLP\ acts on the matrix, and its outputs are the limits $q_t$, $\mu_t^{\rm col}$, and $\mu_t^{\rm row}$ in Theorems~\ref{thm:intro-state-evolution} and~\ref{thm:intro-universality}.

Entering round $t$, write $p=t-1$ and $q=q_p$. The state consists of the free fraction $q$, the free-column history $(X,R)\in\R^p\times\R^p$, the row history $(U,\Lambda)\in\R^p\times\R^p$, and the full coordinate trajectory $\bar X\in[-1,1]^p$, whose law $\mu_p^{\rm col}$ also records the frozen coordinates. We write $X_p$ for the last coordinate of $X$ and $e_p$ for the last standard basis vector of $\R^p$. For $t=1$, the histories are empty, and every term that involves them is omitted.

\begin{algorithm}[!tp]
\caption{State evolution for \IterLP}
\label{alg:state-evolution}
\small
\textbf{Input:} a density $\alpha$ and margins $\kappa_1>\cdots>\kappa_K$ such that $(\alpha,\kappa_1)$ satisfies~\eqref{eq:stage1-feasibility}. Set $q_0=1$, and start with empty histories.\\
\textbf{For $t=1,\ldots,K$,} with $p=t-1$ and $q=q_p$:
\begin{enumerate}[label=\arabic*.,leftmargin=1.8em,itemsep=.15em,topsep=.2em]
  \item \emph{Moments.} Compute $Q^x=\E[\bar X\bar X^\top]$, $Q_F^x=q\E[XX^\top]$, $Q_F^{rx}=q\E[RX^\top]$, $Q^\lambda=\alpha\E[\Lambda\Lambda^\top]$, and $\Gamma=\alpha\E[\Lambda U^\top]$.
  \item \emph{Column value.} Let $G\sim N(0,1)$ be independent of the state. For $\omega=(s,a,b)\in\R\times\R^p\times\R^p$, let
  \[
    \phi_t(\omega)=\sup\bigl\{q\E[GY]:\ Y\in[-1,1],\ q\E Y^2=s,\ q\E[X(Y-X_p)]=a,\ q\E[R(Y-X_p)]=b\bigr\},
  \]
  where $Y$ may depend on $(X,R,G)$ and on independent randomness.
  \item \emph{Row cost.} Let $G_{\rm row}\sim N(0,1)$ be independent of the state and of $G$, and let
  \begin{gather*}
    \rho_\omega^2=s-(Q_F^x)_{pp}-2e_p^\top a-a^\top(Q^x)^{-1}a,\\
    Z_\omega=\bigl(e_p+(Q^x)^{-1}a\bigr)^\top U+\bigl((Q^\lambda)^{-1}(b-\Gamma(Q^x)^{-1}a)\bigr)^\top\Lambda+\rho_\omega G_{\rm row},\\
    d_t(\omega)^2=\alpha\min\bigl\{\E(\widetilde U-Z_\omega)^2:\ \widetilde U\ge\kappa_t,\ \alpha\E[\Lambda\widetilde U]=\Gamma e_p+b\bigr\}.
  \end{gather*}
  \item \emph{Value.} Call $\omega$ admissible if $\rho_\omega^2\ge0$ and the sets in Steps~2 and~3 are nonempty. Compute
  \[
    V_t=\max\Bigl\{\sqrt{\phi_t(\omega)^2-d_t(\omega)^2}:\ \omega\text{ admissible},\ \phi_t(\omega)\ge d_t(\omega)\Bigr\},
  \]
  let $\omega_t^\star=(s_t^\star,a_t^\star,b_t^\star)$ be the maximizer, and set $\rho_t=\rho_{\omega_t^\star}$ and $\gamma_t=d_t(\omega_t^\star)/V_t$.
  \item \emph{Coefficients.} Let $(\tau_t,\eta_{x,t},\eta_{r,t})\in\R\times\R^p\times\R^p$ minimize the dual of the column problem at $\omega_t^\star$,
  \[
    \frac{\tau s_t^\star}{2}-\eta_x^\top(Q_F^xe_p+a_t^\star)-\eta_r^\top(Q_F^{rx}e_p+b_t^\star)+q\E\,\ell(G+\eta_x^\top X+\eta_r^\top R;\tau),
  \]
  where $\ell(z;\tau)=\max_{|y|\le1}\{zy-\tau y^2/2\}$. Let $Z_t=Z_{\omega_t^\star}$, and let $\eta_{\lambda,t}\in\R^p$ be the multiplier for which $U_t=\max\{\kappa_t,Z_t-\eta_{\lambda,t}^\top\Lambda\}$ satisfies $\alpha\E[\Lambda U_t]=\Gamma e_p+b_t^\star$. Then $U_t$ attains the minimum in Step~3.
  \item \emph{Update.} Let $V\sim N(0,1)$, which represents the objective coordinate $v_{t,j}$, be independent of $G$, $G_{\rm row}$, and the state, and let
  \[
    S_t=\frac{V+\gamma_tG}{\sqrt{1+\gamma_t^2}},\qquad
    Y_t=\max\Bigl\{-1,\min\Bigl\{1,\frac{S_t+\eta_{x,t}^\top X+\eta_{r,t}^\top R}{\tau_t}\Bigr\}\Bigr\},\qquad
    \Lambda_t=\frac{(\kappa_t-Z_t+\eta_{\lambda,t}^\top\Lambda)_+}{V_t}.
  \]
  Since $S_t$ is again standard Gaussian and independent of the state, $Y_t$ is the optimizer of Step~2 at $\omega_t^\star$ with $G$ replaced by $S_t$.
  \item \emph{New state.} Set $q_t=q\,\Pp(|Y_t|<1)$. The new free-column history has law $\law\bigl((X,Y_t),(R,-V)\,\big|\,|Y_t|<1\bigr)$, since stationarity gives $r_j^{(t)}=-v_{t,j}$ on free coordinates, the new row history has law $\law\bigl((U,U_t),(\Lambda,\Lambda_t)\bigr)$, and the full trajectory keeps its frozen signs,
  \[
    \mu_t^{\rm col}(\cdot)=\Pp\bigl((\bar X,\bar X_p)\in\cdot,\ |\bar X_p|=1\bigr)+q\,\Pp\bigl((X,Y_t)\in\cdot\bigr).
  \]
\end{enumerate}
\textbf{Output:} for $t=1,\ldots,K$, the free fraction $q_t$, the law $\mu_t^{\rm col}$ of the coordinate trajectory $(\bar X_1,\ldots,\bar X_t)$, and the law $\mu_t^{\rm row}=\law(U_1,\ldots,U_t)$ of the margin trajectory.
\end{algorithm}

Steps~2--4 are the population limit of the auxiliary problem of Section~\ref{subsec:overview-gaussian-recursion}, and Steps~5 and~6 describe its optimizer. A free coordinate becomes a Gaussian projected onto $[-1,1]$, and a row margin becomes a Gaussian projected onto $[\kappa_t,\infty)$, as in the update laws~\eqref{eq:intro-recursion}.

Under the first-round feasibility condition~\eqref{eq:stage1-feasibility}, every step is well defined. Lemma~\ref{lem:positive-branch} shows that $V_t>0$, Appendix~\ref{app:population-multipliers} shows that the maximum in Step~4 is attained and that the multipliers in Step~5 exist with $\tau_t>0$, and Proposition~\ref{prop:population-optimizer-unique} shows that $\omega_t^\star$ is unique. Theorem~\ref{thm:explicit-population-recursion} shows that the new state is again a strict history in the sense of Definition~\ref{def:strict-history}, so that $Q^x$ and $Q^\lambda$ are invertible in the next round. We derive each step in Sections~\ref{subsec:population-geometry} and~\ref{subsec:canonical-kkt-update}. In Section~\ref{sec:numerical}, we evaluate the algorithm numerically by representing each law by particles.

\subsection{Universality through a surrogate path}
\label{subsec:overview-universality}

A Lindeberg-type comparison lemma (Lemma~\ref{lem:repeated-design-comparison}), in the spirit of~\cite{HanShen2023}, shows that the infimum of a cost over a fixed set of candidates has asymptotically the same law under Gaussian and subgaussian matrices if the candidates lie in an $\ell_\infty$ ball of polylogarithmic radius and the cost is smooth in the matrix entries and Lipschitz in the candidate. The LP fails these conditions, since its feasible set depends on $A$, its optimal vertex can jump with $A$, and nothing controls $Ax$, $\lambda$, and $A^\top\lambda$ in $\ell_\infty$. We therefore replace round $t$ by the surrogate quadratic program
\[
  -v_t^\top x+\frac{\tau_{q,t}}{2}\|x\|^2+\frac{1}{2\eta_t}\bigl\|(\kappa_t\1-Ax)_+\bigr\|^2+\frac{\beta_t}{2}\,\|m\odot x-c\|^2
  \qquad\text{over }x\in[-1,1]^N.
\]
A quadratic penalty replaces the row constraints, a ridge term makes the objective strongly convex, and the memory $(m,c)$ from the earlier rounds replaces hard freezing by a smooth pull toward the frozen values (Section~\ref{subsec:univ-paths}). Hence the minimizer is unique and Lipschitz in $A$ and in the earlier rounds, and leave-one-out estimates bound its row margins and multipliers in $\ell_\infty$.

Since the lemma compares infima rather than optimizers, we encode each statement about an optimizer as a lower bound on the infimum over the candidates that violate it, and we follow the three steps of Figure~\ref{fig:overview-universality-chain}. First, for Gaussian matrices, suitable regularization keeps the LP path within normalized squared distance $O(\varepsilon)$ of the surrogate path (Corollary~\ref{cor:univ-gaussian-proxy}). The proof uses an objective $F^\ell$ with an $\ell_1$ penalty in place of the quadratic one, which keeps the localization gap of the LP and has bounded slope. Second, the lemma transfers a Gaussian lower bound on a residual energy, which vanishes exactly on the surrogate path and grows quadratically away from it, over candidate paths whose empirical laws are far from the prediction. Third, the value gap of $F^\ell$ transfers in the same way, with the residual energy of the earlier rounds as an extra penalty because the memory depends on the matrix, so the two subgaussian paths stay close and have the same limits.

\section{Preliminaries}\label{sec:preliminaries}

\subsection{Gaussian min--max comparison}
\label{subsec:prelim-cgmt}

Let $C\subset\R^n$ and $D\subset\R^m$ be nonempty compact sets,
let $\psi:C\times D\to\R$ be continuous, and let
$G\in\R^{m\times n}$, $g\in\R^m$, and $h\in\R^n$ have mutually
independent standard Gaussian entries. Define the primary and
auxiliary values
\begin{align}
  \Phi(G)
  &=
  \min_{w\in C}\max_{u\in D}
  \bigl\{u^\top Gw+\psi(w,u)\bigr\},
  \label{eq:prelim-po}\\
  \phi(g,h)
  &=
  \min_{w\in C}\max_{u\in D}
  \bigl\{\|w\|g^\top u+\|u\|h^\top w+\psi(w,u)\bigr\}.
  \label{eq:prelim-ao}
\end{align}

\begin{theorem}[Gaussian min--max comparison
{\cite{Gordon1985,ThrampoulidisOymakHassibi2015}}]
\label{thm:prelim-cgmt}
For every $c\in\R$,
\begin{equation}\label{eq:prelim-gordon-tail}
  \Pp(\Phi<c)\le 2\Pp(\phi\le c).
\end{equation}
If, in addition, $C,D$ are convex and $\psi$ is convex in $w$ and
concave in $u$, then for every $v\in\R$ and $\eps>0$,
\begin{equation*}
  \Pp(|\Phi-v|>\eps)
  \le 2\Pp(|\phi-v|\ge\eps).
\end{equation*}
\end{theorem}

The one-sided comparison does not require convexity. Under the
convexity--concavity assumptions, the opposite tail follows from
Sion's minimax theorem~\cite{Sion1958} and the same comparison,
applied after negating the payoff and exchanging the roles of the two
variables.

The theorem also holds, with the same proof, in the following slightly more general form, which we use in Section~\ref{sec:finite-po}. The minimizing variable is $z$ in a compact set $C\subset\R^{n'}$, the bilinear term is $u^\top G\,w(z)$ for a continuous map $w:C\to\R^n$, the payoff is $\psi(z,u)$, and $\|w\|$ and $h^\top w$ in~\eqref{eq:prelim-ao} are replaced by $\|w(z)\|$ and $h^\top w(z)$. Indeed, Gordon's comparison involves the two Gaussian processes only through the pairs $(w(z),u)$. In particular, $z$ may contain coordinates that do not enter the bilinear term, and $C$ may couple them to the other coordinates. The two-sided bound additionally requires $C$ and $D$ to be convex, $w$ to be affine, and $\psi$ to be convex in $z$ and concave in $u$.

\begin{corollary}[Localization of optimizers]
\label{cor:prelim-cgmt-localization}
Consider a sequence of problems as in Theorem~\ref{thm:prelim-cgmt}, indexed by $N$, with primary values $\Phi_N$ and constraint sets $C_N$ for the minimizing variable. Let $B_N\subset C_N$ be nonempty deterministic compact sets.
\begin{enumerate}[label=\textnormal{(\roman*)}]
\item Let $\Phi_{B_N}$ and $\phi_{B_N}$ denote the primary and auxiliary values when the minimizing variable is restricted to $B_N$. If $\Phi_N\xrightarrow{\Pp}v$ and $\Pp(\phi_{B_N}\ge v+\delta)\to1$ for some $\delta>0$, then, with high probability, no primary minimizer belongs to $B_N$.
\item For problems written in max--min order, $\max_{w\in C_N}\min_{u\in D}\{u^\top Gw+\psi(w,u)\}$, let $\Phi_N$ denote the primary value, and let $\Phi_{B_N}$ and $\phi_{B_N}=\max_{w\in B_N}\min_{u\in D}\{\|w\|g^\top u+\|u\|h^\top w+\psi(w,u)\}$ denote the primary and max--min auxiliary values when the maximizing variable is restricted to $B_N$. If $\Phi_N\xrightarrow{\Pp}v$ and $\Pp(\phi_{B_N}\le v-\delta)\to1$ for some $\delta>0$, then, with high probability, no primary maximizer belongs to $B_N$.
\end{enumerate}
\end{corollary}

\begin{proof}
(i) By \eqref{eq:prelim-gordon-tail},
$\Phi_{B_N}\ge v+\delta/2$ with high probability, while
$\Phi_N=v+o_{\Pp}(1)$. If a primary minimizer belonged to $B_N$, these
two values would coincide.

(ii) Negating the payoff turns the max--min problem into a min--max problem of the form~\eqref{eq:prelim-po} with $G$ replaced by $-G$, which has the same law. By the symmetry $(g,h)\mapsto(-g,-h)$, the auxiliary value of the negated problem has the law of $-\phi_{B_N}$. Part~(i), applied to the negated problem, gives the claim.
\end{proof}

Neither part requires the restricted sets to be convex, since both use only the one-sided bound~\eqref{eq:prelim-gordon-tail}. Both parts also hold conditionally when the sets and the
deterministic part of the payoff are determined by randomness
independent of the fresh Gaussian inputs.

\subsection{Gaussian regression and whitening}
\label{subsec:prelim-gaussian}

We recall the Gaussian regression formulas used below.

\begin{proposition}[Gaussian regression]
\label{prop:prelim-gaussian-regression}
Let $Z\sim N(m,\Sigma)$ in $\R^d$, where $\Sigma\succ0$, and let
$T:\R^d\to\R^k$ be a linear map of full row rank. Then
\begin{equation*}
Z\mid\{TZ=b\}
\sim
N\!\left(
m+\Sigma T^\top(T\Sigma T^\top)^{-1}(b-Tm),
\,
\Sigma-\Sigma T^\top(T\Sigma T^\top)^{-1}T\Sigma
\right).
\end{equation*}
In particular, the regression residual
$Z-\E[Z\mid TZ]$ is Gaussian and independent of $TZ$.
\end{proposition}

When $\Sigma=\sigma^2I_d$, let $z_b$ be the point of the affine subspace
$\{z:Tz=b\}$ closest to $m$. Then
$\law(Z\mid TZ=b) = \law\!(z_b+\sigma P_{\ker T}G), \allowbreak\quad G\sim N(0,I_d)$.

For the matrix form, let $A=N^{-1/2}G\in\R^{M\times N}, $ $G_{ij}\stackrel{\rm iid}{\sim}N(0,1),$ and let $X\in\R^{N\times p}$ and $L\in\R^{M\times p}$ have full column rank. These will be the primal and row-dual history matrices. Set $P_X=X(X^\top X)^{-1}X^\top, $ $P_L=L(L^\top L)^{-1}L^\top.$

\begin{corollary}[Two-sided Gaussian matrix regression]
\label{cor:prelim-matrix-regression}
Suppose that $U\in\R^{M\times p}$ and $R\in\R^{N\times p}$ satisfy
the consistency condition $L^\top U=R^\top X$. Conditionally on $AX=U,$ $A^\top L=R,$ we have
\begin{equation*}
A
\stackrel{d}{=}
A_0+\frac1{\sqrt N}P_L^\perp G_0P_X^\perp,
\end{equation*}
where $G_0$ is an independent matrix with i.i.d.\ standard Gaussian
entries and
\begin{equation}
A_0 ={} U(X^\top X)^{-1}X^\top +L(L^\top L)^{-1}R^\top - L(L^\top L)^{-1}(L^\top U)(X^\top X)^{-1}X^\top . \label{eq:prelim-matrix-mean}
\end{equation}
\end{corollary}

\begin{proof}
Let $Q_X\in\R^{N\times(N-p)}$ have orthonormal columns spanning $\operatorname{span}(X)^\perp$. On $\{AX=U\}$, the consistency condition gives $X^\top A^\top L=U^\top L=X^\top R$, so the constraints are equivalent to $AX=U$ and $Q_X^\top A^\top L=Q_X^\top R$. The linear map $A\mapsto(AX,Q_X^\top A^\top L)$ is onto, since $(U',R')\mapsto U'(X^\top X)^{-1}X^\top+L(L^\top L)^{-1}R'^\top Q_X^\top$ is a right inverse, and its kernel is $\{H:HX=0,\ H^\top L=0\}=P_L^\perp\R^{M\times N}P_X^\perp$. The matrix $A_0$ satisfies both constraints and is orthogonal to this kernel in the trace inner product, so it is the point of the constraint set closest to the mean $0$. The isotropic case of Proposition~\ref{prop:prelim-gaussian-regression}, with $\sigma^2=1/N$, gives the claim.
\end{proof}

The residual lies in $\{H:HX=0,\ H^\top L=0\} = P_L^\perp\R^{M\times N}P_X^\perp.$ To write the residual in terms of a standard Gaussian matrix, let
$F\subset[N]$, let $Q_L$
have orthonormal columns spanning $\operatorname{span}(L)^\perp$, and
define $\Sigma_F = I_F-X_F(X^\top X)^{-1}X_F^\top \succeq0 .$ Then, in distribution,
\begin{equation*}
  A_{:,F}
  =
  A_{0,:,F}
  +
  \frac1{\sqrt N}Q_LG_F\Sigma_F^{1/2},
\end{equation*}
where $G_F$ has i.i.d.\ standard Gaussian entries. Hence, if
$\mu=Q_L\nu+Lc$, then $\mu^\top(A_{:,F}-A_{0,:,F})w = \frac1{\sqrt N}\nu^\top G_F\Sigma_F^{1/2}w.$ Thus, after conditioning, the bilinear form splits into a
deterministic regression term and a standard Gaussian bilinear form
in the coordinates transverse to the history.

Finally, we use Gaussian integration by parts. For $G\sim N(0,I_d)$
and Lipschitz $f:\R^d\to\R$, $\E[G_jf(G)] = \E[\partial_jf(G)],$ where the derivative is understood almost everywhere. The identity
also holds conditionally on independent marks, provided that the
terms are integrable.

\subsection{Relative entropy and Gaussian transport}
\label{subsec:prelim-entropy}

For probability measures $P,Q$ on a common measurable space, we
define the relative entropy $\Ent(P\mid Q) = \int \log\!\left(\frac{dP}{dQ}\right)dP$ if $P\ll Q$, and $+\infty$ otherwise.

Let $P_{XY}$ and $Q_{XY}$ be two joint laws on
$\mathcal X\times\mathcal Y$. Write $P_X,Q_X$ for the marginal laws
of $X$, and $P_{Y\mid X=x},Q_{Y\mid X=x}$ for the corresponding
conditional laws.

\begin{proposition}[Chain rule and data processing
{\cite[Section~2.5]{PolyanskiyWu2022}}]
\label{prop:prelim-entropy}
We have
\begin{equation*}
\Ent(P_{XY}\mid Q_{XY})
=
\Ent(P_X\mid Q_X)
+
\int
\Ent(P_{Y\mid X=x}\mid Q_{Y\mid X=x})\,P_X(dx).
\end{equation*}
Consequently, $\int \Ent(P_{Y\mid X=x}\mid Q_{Y\mid X=x})\,P_X(dx) \le \Ent(P_{XY}\mid Q_{XY}).$ If $F$ is measurable and $P_F,Q_F$ are the laws of $F(X,Y)$ under
$P_{XY},Q_{XY}$, respectively, then $\Ent(P_F\mid Q_F) \le \Ent(P_{XY}\mid Q_{XY}).$ Adjoining the same independent random variable to both models leaves
their relative entropy unchanged.
\end{proposition}

We also use the Donsker--Varadhan variational inequality
$\E_P f \le \Ent(P\mid Q)+\log\E_Qe^f,$
which holds whenever $P\ll Q$, $f$ is $P$-integrable, and
$\E_Qe^f<\infty$; see~\cite{DonskerVaradhan1975}. In particular, if
$P\ll S\ll Q$ and $W=dS/dQ$, then $\Ent(P\mid Q) \le 2\Ent(P\mid S)+\log\E_QW^2.$

\begin{theorem}[Gaussian transportation inequality {\cite{Talagrand1996}}]
\label{thm:prelim-gaussian-transport}
Let $Q=N(m,\sigma^2I_E)$ be the isotropic Gaussian law on the affine
subspace $m+E$ of a Euclidean space. Then every probability measure
$P$ on this subspace satisfies $\Wtwo^2(P,Q) \le 2\sigma^2\Ent(P\mid Q).$ \end{theorem}

For matrix-valued Gaussian variables, we take the Euclidean norm to be
the Frobenius norm. Thus, for a Gaussian law with covariance
$N^{-1}I_E$ on an affine subspace of matrices, we obtain $\Wtwo^2(P,Q) \le \frac2N\Ent(P\mid Q).$

\section{The iterated LP and its optimization histories}\label{sec:model}

\subsection{The iterated LP}\label{subsec:iterated-replacement-lp}

Let $v_1,\ldots,v_K\stackrel{\mathrm{iid}}{\sim}N(0,I_N)$ be independent of $A$.
Set $B_0=\varnothing$, $F_0=[N]$, and $x^{(0)}=0$. In every round
$t=1,\ldots,K$, keep the previously frozen coordinates fixed and solve
\begin{align}
  x_{B_{t-1}}^{(t)} & = x_{B_{t-1}}^{(t-1)}, \label{eq:freeze-old} \\
  x_{F_{t-1}}^{(t)} & \in\argmax_{y\in[-1, 1]^{F_{t-1}}}
  \left\{v_{t, F_{t-1}}^\top y:
  A_{:, B_{t-1}}x_{B_{t-1}}^{(t-1)}+A_{:, F_{t-1}}y\geq\kappa_t\1\right\}.
  \label{eq:roundt}
\end{align}
Set
$C_t = \{j\in F_{t-1}:\abs{x_j^{(t)}} = 1\}, \quad B_t = B_{t-1}\cup C_t, \quad F_t = F_{t-1}\setminus C_t.$
For $t\ge2$, the previous iterate is strictly feasible for the row constraints, since
$Ax^{(t-1)}\geq\kappa_{t-1}\1 = (\kappa_t+\Delta_t)\1, \allowbreak\quad \Delta_t := \kappa_{t-1}-\kappa_t>0.$
Write $\Phi_{N,t}:=v_t^\top x^{(t)}$ for the full objective value.
Let $\lambda^{(t)}\ge0$ be a row-constraint multiplier, and set
$u^{(t)}=Ax^{(t)}$ and $r^{(t)}=A^\top\lambda^{(t)}$.
The regular KKT structure is recorded in Section~\ref{subsec:regular-kkt}.
Initial feasibility is addressed in Section~\ref{sec:finite-po}.

\subsection{Complete histories}\label{subsec:complete-histories}

At depth $p$, we define
$X_p = [x^{(1)}|\cdots|x^{(p)}], \quad U_p = AX_p,\allowbreak\quad L_p = [\lambda^{(1)}|\cdots|\lambda^{(p)}], \quad R_p = A^\top L_p$.
The normalized history matrices are
\begin{equation}\label{eq:history-matrices}
  Q_p^x = \frac1N X_p^\top X_p,
  \qquad Q_p^\lambda = \frac1N L_p^\top L_p,
  \qquad \Gamma_p = \frac1N L_p^\top U_p.
\end{equation}
On the current free set $F_p$, set
\begin{equation*}
  Q_{p, F}^x = \frac1N X_{p, F_p}^\top X_{p, F_p},\qquad Q_{p, F}^{rx} = \frac{1}{N}R_{p, F_p}^\top X_{p, F_p},
  \qquad D_p^x = Q_p^x-Q_{p, F}^x.
\end{equation*}
The frozen-coordinate contribution $D_p^x$ is positive definite precisely when the next round's column covariance has a spectral gap, up to the conditioning of $Q_p^x$.

The complete empirical history consists of the matrices in~\eqref{eq:history-matrices}, the free fraction $q_p = |F_p|/N$, the free-column empirical law of
$\bigl(x_j^{(1)}, \ldots, x_j^{(p)}, r_j^{(1)}, \ldots, r_j^{(p)}\bigr), \allowbreak\quad j\in F_p,$
and the row empirical law of
$\bigl(u_i^{(1)}, \ldots, u_i^{(p)}, \lambda_i^{(1)}, \ldots, \lambda_i^{(p)}\bigr), \allowbreak\quad i\in[M].$
We also record the full primal trajectory law $\widehat\mu_{N,p}^{\rm col}=N^{-1}\sum_{j=1}^N\delta_{x_j^{(1:p)}}$, which includes the coordinates frozen before round $p$.

We use the following conventions for the empty history. Set $q_0=1$;
$X_0,R_0$ have size $N\times0$, and $U_0,L_0$ have size $M\times0$.
All old Gram and mixed-moment matrices are empty, products through
the zero-dimensional space vanish, and its unique linear map is its
own inverse. Set $u^{(0)}=0$ and let $e_0$ be the empty vector;
an old-moment entry indexed by $(p,p)$ is zero at $p=0$.
A Schur complement over an empty block is the remaining block.
These conventions impose no feasibility requirement on $x^{(0)}$.

\subsection{Regular KKT pairs}\label{subsec:regular-kkt}

For the reduced LP at round $t$, row multipliers satisfy
$u^{(t)}\ge\kappa_t\1,\allowbreak\quad \lambda^{(t)}\ge0,\allowbreak\quad \lambda_i^{(t)}(u_i^{(t)}-\kappa_t)=0$.
On $F_{t-1}$, box multipliers $\lambda_{\rm box}^+,\lambda_{\rm box}^-\ge0$
satisfy
\[
  v_{t,F_{t-1}}+r_{F_{t-1}}^{(t)}
  -\lambda_{\rm box}^++\lambda_{\rm box}^-=0,
  \qquad
  \lambda_{{\rm box},j}^+(1-x_j^{(t)})
  =\lambda_{{\rm box},j}^-(1+x_j^{(t)})=0.
\]
In particular, $r_j^{(t)}=-v_{t,j}$ on $F_t$.

\begin{definition}[Regular KKT pair]\label{def:regular-kkt}
Consider the round-$t$ LP over the coordinates $F_{t-1}$ that enter this round free. Let $(x^{(t)},\lambda^{(t)})$ be a primal--dual KKT pair, and write $u^{(t)}=Ax^{(t)}$, $r^{(t)}=A^\top\lambda^{(t)}$, and $I_t=\{i:\lambda_i^{(t)}>0\}$. The pair is \emph{regular} if:
\begin{enumerate}[label=\textnormal{(\roman*)}]
  \item The row constraints are strictly complementary, that is,
  \[
    u_i^{(t)}=\kappa_t,\quad \lambda_i^{(t)}>0\quad(i\in I_t),
    \qquad
    u_i^{(t)}>\kappa_t,\quad \lambda_i^{(t)}=0\quad(i\notin I_t).
  \]
  \item The free and newly frozen coordinates satisfy
  \[
    |x_j^{(t)}|<1,\quad (v_t+r^{(t)})_j=0\quad(j\in F_t),
    \qquad
    x_j^{(t)}(v_t+r^{(t)})_j>0\quad(j\in C_t),
  \]
  where $C_t=F_{t-1}\setminus F_t$ and $x_j^{(t)}\in\{\pm1\}$ on $C_t$. The coordinates in $B_{t-1}$ remain fixed by \eqref{eq:freeze-old}.
  \item $|I_t|=|F_t|$ and $A_{I_t,F_t}$ is invertible, with the convention that the determinant of the empty block is $1$.
\end{enumerate}
A regular KKT cell is obtained by fixing the active sets and the frozen signs in these conditions.
\end{definition}

For a regular pair, the active normals form a basis on $F_{t-1}$. Strict complementarity forces every optimizer to satisfy the same active equalities, giving a unique simple vertex. Stationarity on $F_t$ then determines the row and reduced box multipliers uniquely.

Throughout the Gaussian analysis, $A=G/\sqrt N$, where the entries of
$G\in\R^{M\times N}$ are independent standard Gaussians,
$M/N\to\alpha\in(0,\infty)$, and the objective vectors are independent
of $A$.

\begin{lemma}[Gaussian general position]\label{lem:s1-general-position}
Outside a null set of the Gaussian matrix and objective vectors, every
first-round KKT pair is regular whenever the optimal value
$\Phi_{N,1}$ is positive. On the event $B_1\ne\varnothing$, every later
round $t\le K$ also has a regular KKT pair. In particular, the primal
optimizer and the row multipliers of every round are unique,
$|I_t|=|F_t|$, and $A_{I_t,F_t}$ is invertible.
\end{lemma}

\begin{proof}
We first fix a deterministic frozen set $B\subset[N]$ and signs $s\in\{\pm1\}^B$, write $F=B^c$, and consider the reduced LP
\[
  \max\bigl\{v_{t,F}^\top y:\ y\in[-1,1]^F,\ A_{:,F}y\ge b\bigr\},
  \qquad b=\kappa_t\1-A_{:,B}s.
\]
For $B\ne\varnothing$, the vector $A_{:,B}s\sim N(0,|B|N^{-1}I_M)$ is independent of $A_{:,F}$, so $b$ has a density and is independent of $A_{:,F}$. For the first round, $B=\varnothing$ and $b=\kappa_1\1$. Regularity can fail in two ways.

\emph{Primal degeneracy.} Some feasible point has $|F|+1$ active constraints. Then there are $I\subset[M]$, $C\subset F$, and signs $\sigma\in\{\pm1\}^C$ with $|I|+|C|=|F|+1$ such that $A_{I,F\setminus C}y=b_I-A_{I,C}\sigma$ is solvable, so $\det[A_{I,F\setminus C}\mid b_I-A_{I,C}\sigma]=0$. For $B\ne\varnothing$, this determinant is a nonzero polynomial in $(A_{:,F},b)$, because $b$ has an independent density. For the first round, it is a nonzero polynomial in $A$ unless $\kappa_1=0$ and $C=\varnothing$. In that case the only solution is $y=0$, whose objective value is $0$, which is excluded by $\Phi_{N,1}>0$.

\emph{Dual degeneracy.} At a simple vertex with active rows $I$ and newly frozen coordinates $C$, we have $|I|=|F\setminus C|$, the block $A_{I,F\setminus C}$ is invertible, and the KKT conditions give $\lambda_I=-A_{I,F\setminus C}^{-\top}v_{t,F\setminus C}$ and $(v_t+r)_C=v_{t,C}+A_{I,C}^\top\lambda_I$. After multiplication by $\det A_{I,F\setminus C}$, the events $\lambda_i=0$ for some $i\in I$ and $(v_t+r)_j=0$ for some $j\in C$ are zero sets of nonzero linear forms in $v_t$. They are null because $v_t$ has a density and is independent of $A$.

Outside these null sets, the optimal vertex is simple and its multipliers are nonzero, which gives conditions~(i)--(iii) of Definition~\ref{def:regular-kkt} and the uniqueness claims. Taking the finite union over $(B,s)$, the patterns $(I,C,\sigma)$, and the rounds $t\le K$, the conclusion holds simultaneously for the realized frozen sets $B_{t-1}$ and signs.
\end{proof}

\section{Coupling the conditioned matrix with a Gaussian matrix}\label{sec:optimizer-coupling}

We compute the change in the Gaussian matrix law caused by conditioning on LP optimizers and active sets, then couple the conditional law to a Gaussian law with vanishing operator-norm error. This supplies the matrix used in Sections~\ref{sec:finite-po} and~\ref{sec:closing-gaussian-recursion}.

\begin{definition}[Regular complete history]\label{def:regular-history}
A depth-$p$ complete history
$h = (X_p,\allowbreak U_p,\allowbreak L_p,\allowbreak R_p,\allowbreak \{I_r, F_r, B_r\}_{r\leq p})$
is \emph{regular} if there exist a matrix $A_0$ and objective vectors $v_1,\ldots,v_p$ such that the prescribed LP updates produce exactly this history, every round has a regular KKT pair (Definition~\ref{def:regular-kkt}), and $\operatorname{rank}X_p=\operatorname{rank}L_p=p.$ In particular, $A_0X_p=U_p,$ $A_0^\top L_p=R_p.$ We denote this class by $\mathcal R_{p,N}$, and we let $\mathcal R_{0,N}$ consist of the empty history.
\end{definition}

Every prefix is regular, and $\mathcal A_{h,p} :=\{A\in\R^{M\times N}:AX_p=U_p,\ A^\top L_p=R_p\}$ contains $A_0$ with invertible active blocks. Hence $L_p^\top U_p=R_p^\top X_p$, and the active-block determinants and their product do not vanish identically on $\mathcal A_{h,p}$. Regularity imposes no uniform spectral lower bound. The Gaussian induction propagates full rank through Gram-matrix convergence and the positive definite limits in Theorem~\ref{thm:explicit-population-recursion}.

Fix $h\in\mathcal R_{p,N}$ and write $F=F_p$. The density formulas concern this deterministic history; the transport estimate will give uniform coupling on a set of histories of high probability.

\subsection{The Gaussian law conditioned on the affine history}\label{subsec:gaussian-reference-law}

Condition first on the affine observations alone:
\begin{equation}\label{eq:history-linear-constraints}
  AX_p = U_p, \qquad A^\top L_p = R_p.
\end{equation}
Let $Q_{h, p} := \law\!\left( A \, \middle|\, AX_p = U_p, \ A^\top L_p = R_p \right).$

\begin{lemma}[Gaussian regression on the affine history]\label{lem:history-gaussian-regression}
  Under $Q_{h, p}$,
  \begin{equation}\label{eq:history-gaussian-regression}
    A
    =
    A_{h, p}
    +
    \frac1{\sqrt N}
    P_{L_p}^{\perp}G_0P_{X_p}^{\perp},
  \end{equation}
  where $G_0$ is an i.i.d.\ standard Gaussian matrix and
  \begin{equation}
A_{h, p} = U_p(X_p^\top X_p)^{-1}X_p^\top +L_p(L_p^\top L_p)^{-1}R_p^\top \quad -L_p(L_p^\top L_p)^{-1} (L_p^\top U_p) (X_p^\top X_p)^{-1}X_p^\top . \label{eq:history-conditional-mean}
\end{equation}
\end{lemma}

\begin{proof}
Apply Corollary~\ref{cor:prelim-matrix-regression} with $(X,L,U,R)=(X_p,L_p,U_p,R_p)$; regularity gives full rank and $L_p^\top U_p=R_p^\top X_p$. The mean $A_{h,p}$ satisfies the two affine constraints and is orthogonal to their tangent space $\mathcal T_{h,p}=P_{L_p}^{\perp}\R^{M\times N}P_{X_p}^{\perp}$. The residual is therefore isotropic Gaussian with covariance $N^{-1}I$ on this space, giving \eqref{eq:history-gaussian-regression}.

\end{proof}

We now restrict \eqref{eq:history-gaussian-regression} to the current free columns. Let $E_F$ denote the coordinate embedding of $\R^{F}$ into $\R^N$, and let $Q_{L, p}$ be a matrix whose orthonormal columns span $\spanop(L_p)^\perp$. Define $\Sigma_{p, F} = I_F - X_{p, F}(X_p^\top X_p)^{-1}X_{p, F}^\top .$ Then
\begin{equation}\label{eq:free-column-gaussian-representation}
  A_{:, F}
  =
  A_{h, p, :, F}
  +
  \frac1{\sqrt N}
  Q_{L, p}G_{p, F}\Sigma_{p, F}^{1/2},
\end{equation}
where $G_{p, F}$ is an i.i.d.\ standard Gaussian matrix.

For the empty history, $A_{h,0}=0$,
$P_{X_0}^{\perp}=\Sigma_{0,F_0}=I_N$, and
$P_{L_0}^{\perp}=Q_{L,0}=I_M$; hence $Q_{h,0}$ is the original
Gaussian law.

\subsection{From KKT coordinates to the Radon--Nikodym derivative}\label{subsec:kkt-coordinate-density}

The Jacobian between optimizer coordinates and KKT residuals determines the change from affine conditioning to optimizer conditioning.

For the first round, fix a regular KKT cell (Definition~\ref{def:regular-kkt}) with active rows $I$, free columns $F$, and frozen signs $s = x_{F^c}\in\{\pm1\}^{F^c}$. The continuous optimizer coordinates are $y = x_F, $ $\ell = \lambda_I,$ with $y\in(-1,1)^F$ and $\ell\in(0,\infty)^I$. Their equality residuals are
\begin{align}
  \mathsf q(v, A;y, \ell)&=A_{I,F}y+A_{I,F^c}s-\kappa_1\1_I,
  \label{eq:kkt-coordinate-q}\\
  \mathsf p(v, A;y, \ell)&=v_F+A_{I, F}^{\top}\ell.
  \label{eq:kkt-coordinate-p}
\end{align}
Thus $(y, \ell)$ is the optimizer on this cell exactly when $\mathsf q(v, A;y, \ell) = 0$, $\mathsf p(v, A;y, \ell) = 0$, and the inactive constraints satisfy the strict cell inequalities. Let $x(y,s)$ equal $y$ on $F$ and $s$ on $F^c$, and let $\lambda(\ell,I)$ equal $\ell$ on $I$ and zero on $I^c$. The strict cell event is
\begin{equation*}
  \mathcal E_{I,F,s}(y,\ell)
  :=\left\{(Ax(y,s))_{I^c}>\kappa_1\1,
  \quad s\odot\bigl(v+A^\top\lambda(\ell,I)\bigr)_{F^c}>0\right\}.
\end{equation*}

Conditioning on optimizer coordinates or only on the KKT equalities gives, respectively,
\begin{align*}
 P_{y,\ell}&=\law\bigl((v,A)\mid(\hat x_F,\hat\lambda_I)=(y,\ell),
                         \ \text{KKT cell }(I,F,s)\bigr),\\
 Q_{y,\ell}&=\law\bigl((v,A)\mid\mathsf q(v,A;y,\ell)=0,\ \mathsf p(v,A;y,\ell)=0\bigr),
\end{align*}
where $(\hat x,\hat\lambda)$ is the first-round KKT pair, which is unique almost surely by Lemma~\ref{lem:s1-general-position}. Both are conditional laws given continuous variables, and the formulas below define versions of them, which we use throughout.

We compare the two laws through the coarea formula~\cite[3.2.12 and 3.2.22]{Federer1969}; see also~\cite{AzaisWschebor2009}. On the open cell, the optimizer coordinates are explicit smooth functions of the input,
\[
  \Pi(v,A):=(\hat x_F,\hat\lambda_I)
  =\bigl(A_{I,F}^{-1}(\kappa_1\1_I-A_{I,F^c}s),\,-A_{I,F}^{-\top}v_F\bigr).
\]
For fixed $(y,\ell)$, the residual map $\Psi_{y,\ell}:=(\mathsf q,\mathsf p)(\cdot\,;y,\ell)$ is affine in $(v,A)$, and its differential is surjective, since $\mathsf p$ contains $v_F$ with the identity coefficient and $\mathsf q$ is surjective in $A_{I,:}$ because $x(y,s)\ne0$. On the cell, the fiber $\{\Pi=(y,\ell)\}$ is the part of $\{\Psi_{y,\ell}=0\}$ on which $\mathcal E_{I,F,s}(y,\ell)$ holds. Differentiating the identity $\Psi_{\Pi(v,A)}(v,A)=0$ shows that $D\Psi_{y,\ell}=-J\,D\Pi$ on this fiber, where
\begin{equation*}
  J=\frac{\partial(\mathsf q,\mathsf p)}{\partial(y,\ell)}
  =\begin{pmatrix}
    A_{I, F} & 0               \\
    0        & A_{I, F}^{\top}
  \end{pmatrix},
  \qquad
  |\det J|=|\det A_{I,F}|^2.
\end{equation*}
Hence the normal Jacobians satisfy $J_\Psi=|\det J|\,J_\Pi$. By the coarea formula, the two conditional laws have densities proportional to $f/J_\Pi$ and $f/J_\Psi$ with respect to the same surface measure on the fiber, where $f$ is the Gaussian density of $(v,A)$. Therefore
\begin{equation*}
  \frac{dP_{y, \ell}}{dQ_{y, \ell}}(v, A)
  =\frac{|\det A_{I, F}|^2\1_{\mathcal E_{I,F,s}(y,\ell)}}
  {\E_{Q_{y, \ell}}\left[|\det A_{I, F}|^2\1_{\mathcal E_{I,F,s}(y,\ell)}\right]}.
\end{equation*}

Condition both laws further on the complete first-round history $h = (x, u, \lambda, r)$.

The equations $\mathsf q=0$ and $\mathsf p=0$ determine
\begin{equation}\label{eq:active-free-history}
  (Ax)_I=\kappa_1\1_I,
  \qquad
  (A^\top\lambda)_F = -v_F.
\end{equation}
To specify the complete affine history, fix
$u_J=(Ax)_J,\allowbreak\quad r_B=(A^\top\lambda)_B,\allowbreak\quad r_F=(A^\top\lambda)_F, \allowbreak\quad J=I^c,\quad B=F^c$.
Under the free-stationarity constraint $\mathsf p=0$, fixing $r_F$ also fixes $v_F=-r_F$. Together with \eqref{eq:active-free-history}, these conditions give $Ax = u, $ $A^\top\lambda = r.$

After this conditioning, write $Q_h=\law(A\mid Ax=u,\ A^\top\lambda=r)$ for the affine-conditioned law and $P_h$ for the law conditioned on the exact first-round KKT history, including its active and free sets and frozen signs.

Given $h$, the inactive-row inequalities either hold or fail for all matrices in the support, $v_F=-r_F$ is fixed, and $v_{F^c}$ is independent of $A$ under both laws. Integrating the frozen-coordinate sign conditions against $v_{F^c}$ therefore gives a factor that depends only on $h$, since $r_{F^c}$ is fixed. These factors cancel on normalization, leaving
\begin{equation*}
  {
    \frac{dP_h}{dQ_h}(A)
    =
    \frac{
      |\det A_{I, F}|^2
    }{
      \E_{Q_h}|\det A_{I, F}|^2
    }.
  }
\end{equation*}

\subsection{The selection density along a finite path}\label{subsec:finite-path-coordinate-density}

For $h\in\mathcal R_{p,N}$, let $P_{h,p}$ denote the law of the matrix conditioned on the event that the first $p$ rounds of \IterLP\ produce the history $h$, and recall the affine-conditioned Gaussian law $Q_{h,p}$ of Section~\ref{subsec:gaussian-reference-law}.

\begin{proposition}[Selection density]\label{prop:selection-density}
For every $p\le K$ and $h\in\mathcal R_{p,N}$,
\begin{equation}\label{eq:finite-path-selection-density}
  \frac{dP_{h, p}}{dQ_{h, p}}(A)
  =
  \frac{
    \displaystyle
    \prod_{r = 1}^p
    |\det A_{I_r, F_r}|^2
  }{
    \displaystyle
    \E_{Q_{h, p}}
    \prod_{r = 1}^p
    |\det A_{I_r, F_r}|^2
  }
\end{equation}
defines a version of $P_{h,p}$, and we use this version throughout.
\end{proposition}

\begin{proof}
At round $r$, put $y_r=x_{F_r}^{(r)}$ and $\ell_r=\lambda_{I_r}^{(r)}$, and let $s_{B_r}=x_{B_r}^{(r)}\in\{\pm1\}^{B_r}$ be the frozen signs. The active-primal and free-stationarity residuals are
\[
  \mathsf q_r=A_{I_r,F_r}y_r+A_{I_r,B_r}s_{B_r}-\kappa_r\1_{I_r},
  \qquad
  \mathsf p_r=v_{r,F_r}+A_{I_r,F_r}^\top\ell_r.
\]
Since the frozen values are constant on the cell, $(\mathsf q_r,\mathsf p_r)$ depends on the optimizer coordinates only through $(y_r,\ell_r)$. Hence the Jacobian of the residuals with respect to $(y_1,\ell_1,\ldots,y_p,\ell_p)$ is block diagonal with blocks $\diag(A_{I_r,F_r},A_{I_r,F_r}^\top)$, and
\[
 \left|\det\frac{\partial(\mathsf q_1,\mathsf p_1,\ldots,\mathsf q_p,\mathsf p_p)}
 {\partial(y_1,\ell_1,\ldots,y_p,\ell_p)}\right|
 =\prod_{r=1}^p|\det A_{I_r,F_r}|^2.
\]
The residual map is a submersion. Full column rank of $X_p$ makes $A\mapsto AX_p$ surjective, which gives full row rank to the active-primal residuals, and the free-stationarity residuals have full rank in the independent objective coordinates $v_{r,F_r}$. The coarea argument of Section~\ref{subsec:kkt-coordinate-density}, applied to the optimizer map and to this residual map, therefore gives the density $\prod_{r\le p}|\det A_{I_r,F_r}|^2$, restricted to the strict cell inequalities, between the optimizer-conditioned and the residual-conditioned laws.

Conditioning further on the complete affine history $AX_p = U_p, $ $A^{\top}L_p = R_p$ fixes the inactive-row inequalities and the free-stationarity objective coordinates at every round.  Integrating the box-sign indicators on the newly frozen coordinates $F_{r-1}\setminus F_r$, with $F_0=[N]$, against the remaining independent objective coordinates yields a scalar that depends only on the complete path history.  As in the first-round calculation, these scalars cancel after normalization.  The resulting residual-conditioned law is the law $Q_{h,p}$ defined by the constraints \eqref{eq:history-linear-constraints}, which proves~\eqref{eq:finite-path-selection-density}.
\end{proof}

\subsection{Recursive relative-entropy control and conditional coupling}
\label{subsec:transport-coupling}

We remove the squared-determinant selection bias recursively. Fix $p\le K$.
For $0\le r\le p$, let $H^r$ denote the complete depth-$r$ history
whenever the first $r$ rounds are defined and regular. Undefined or
nonregular outputs are sent to a cemetery state and are not used as
inputs to the next step. Let $\mathsf H_r$ be the restriction of the
unconditioned law of $H^r$ to $\mathcal R_{r,N}$; it is a subprobability
measure. All history integrals below are over this regular set.
For a regular history $h$, $P_{h,r}=\law(A\mid H^r=h)$ is the
optimizer-conditioned law of Proposition~\ref{prop:selection-density}, and
$Q_{h,r}=\law\!\left(A\,\middle|\,AX_r=U_r,\ A^\top L_r=R_r\right)$
is the Gaussian law conditioned only on the affine history. These
conditional laws do not condition on regularity at any later depth.
At depth zero the two laws coincide. We define the averaged conditional
relative entropy $\mathcal D_r:=\int\Ent(P_{h,r}\mid Q_{h,r})\,\mathsf H_r(dh),$
$\mathcal D_0=0$.

For the one-step reference experiment, keep the actual depth-$(r-1)$ history, replace its conditional matrix law $P_{h,r-1}$ by $Q_{h,r-1}$, draw the same fresh Gaussian objective $v_r$, and perform the round-$r$ LP update. Let $S_{h,r}$ be the resulting conditional matrix law given the depth-$r$ history.

The two input laws differ only in the matrix law, with relative entropy averaging to $\mathcal D_{r-1}$ against $\mathsf H_{r-1}$. Apply the same measurable update, data processing, and the entropy chain rule. For the subprobability input, first normalize its common mass and multiply back afterward; zero mass is immediate. Discarding the nonnegative history-marginal entropy and the conditional entropies of nonregular outputs gives
\begin{equation}\label{eq:one-step-conditional-entropy}
  \int\Ent(P_{h,r}\mid S_{h,r})\,\mathsf H_r(dh)\le\mathcal D_{r-1}.
\end{equation}

It remains to compare $S_{h,r}$ with the Gaussian reference law
$Q_{h,r}$. For $h\in\mathcal R_{r,N}$, set $T_{<r}(A) = \prod_{s<r}|\det A_{I_s,F_s}|^2, $ $T_r(A) = |\det A_{I_r,F_r}|^2.$ By~\eqref{eq:finite-path-selection-density} at depth $r-1$, the true and reference joint laws of $(H^{r-1},A,v_r)$ differ by the density $T_{<r}(A)/\E_{Q_{H^{r-1},r-1}}T_{<r}$. Since $H^r$ is a function of $(H^{r-1},A,v_r)$ that determines $H^{r-1}$, conditioning on $H^r=h$ gives $\frac{dP_{h,r}}{dS_{h,r}} = \frac{T_{<r}} {\E_{S_{h,r}}T_{<r}},$ and $T_{<r}>0$ holds $S_{h,r}$-almost surely on regular histories, whereas the depth-$r$ formula
\eqref{eq:finite-path-selection-density} gives $\frac{dP_{h,r}}{dQ_{h,r}} \propto T_{<r}T_r.$ Hence
\begin{equation*}
  \frac{dS_{h,r}}{dQ_{h,r}}(A)
  =
  W_{h,r}(A),
  \qquad
  W_{h,r}(A)
  :=
  \frac{|\det A_{I_r,F_r}|^2}
       {\E_{Q_{h,r}}|\det A_{I_r,F_r}|^2}.
\end{equation*}
Only the current round's squared-determinant factor remains.

We have the decomposition $\Ent(P_{h,r}\mid Q_{h,r}) = \Ent(P_{h,r}\mid S_{h,r}) + \E_{P_{h,r}}\log W_{h,r},$ and the Donsker--Varadhan variational inequality gives
$\E_{P_{h,r}}\log W_{h,r} \le \Ent(P_{h,r}\mid S_{h,r}) + \log\E_{S_{h,r}}W_{h,r}.$
Since $dS_{h,r}/dQ_{h,r}=W_{h,r}$, $\E_{S_{h,r}}W_{h,r} = \E_{Q_{h,r}}W_{h,r}^2.$ Combining these bounds with \eqref{eq:one-step-conditional-entropy} yields the
closed recursion
\begin{equation*}
  \mathcal D_r
  \le
  2\mathcal D_{r-1}
  +
  \int
  \log\E_{Q_{h,r}}W_{h,r}^2\,
  \mathsf H_r(dh).
\end{equation*}

We control the remaining term by the following determinant estimate,
which we prove in Appendix~\ref{app:determinant-moment}.

\begin{lemma}[Ratio of determinant moments]
\label{lem:determinant-fourth-moment}
For every $r\le K$ and every $h\in\mathcal R_{r,N}$,
\begin{equation}
\label{eq:determinant-fourth-moment}
  \frac{
  \E_{Q_{h,r}}|\det A_{I_r,F_r}|^4
  }{
  \bigl(
  \E_{Q_{h,r}}|\det A_{I_r,F_r}|^2
  \bigr)^2
  }
  \le N^{C_K},
\end{equation}
where $C_K$ depends only on the fixed depth $K$.
\end{lemma}

Therefore $\log\E_{Q_{h,r}}W_{h,r}^2 \le C_K\log N,$ uniformly over $h\in\mathcal R_{r,N}$, and $\mathcal D_r \le 2\mathcal D_{r-1}+C_K\log N \le C_K2^r\log N.$

Under $Q_{h,r}$ the residual is isotropic Gaussian with covariance $N^{-1}I$ on the history's affine space. Talagrand's inequality gives
\begin{equation*}
  \int
  W_{2,F}^2(P_{h,r},Q_{h,r})\,\mathsf H_r(dh)
  \le
  \frac{2}{N}\mathcal D_r
  \le
  C_K\frac{\log N}{N},
\end{equation*}
where $W_{2,F}$ denotes the quadratic Wasserstein distance with respect
to the Frobenius norm.

Set
$a_N=\left(\frac{\log N}{N}\right)^{1/4}, \allowbreak\quad \mathcal G_{r,N} = \left\{ h: W_{2,F}(P_{h,r},Q_{h,r})\le a_N \right\}$.
By Markov's inequality, $\mathsf H_r(\mathcal G_{r,N}^c) \le C_K a_N^2 \longrightarrow0.$ For every $h\in\mathcal G_{r,N}$, we choose a conditional coupling
$(A^*,A^G)$ of $P_{h,r}$ and $Q_{h,r}$ with $\E_h\|A^*-A^G\|_F^2\le 2a_N^2.$ Then, for every fixed $\eps>0$,
\begin{equation*}
  \sup_{h\in\mathcal G_{r,N}}
  \Pp_h\!\left(
  \|A^*-A^G\|_{\op}>\eps
  \right)
  \le
  \frac{2a_N^2}{\eps^2}
  \longrightarrow0.
\end{equation*}
Fix $C_A>1+\sqrt\alpha$ and set
\begin{equation}\label{eq:conditional-norm-control}
  \mathcal E_N^{\rm op}=\{\|A\|_{\op}\le C_A\},\qquad
  \mathcal N_{p,N}=\{h:P_{h,p}((\mathcal E_N^{\rm op})^c)\le a_N\}.
\end{equation}
Gaussian norm concentration and Markov's inequality give
$\mathsf H_p(\mathcal N_{p,N}^c)\le\Pp((\mathcal E_N^{\rm op})^c)/a_N=o(1)$.
For $h\in\mathcal G_{p,N}\cap\mathcal N_{p,N}$, the coupling above also gives $Q_{h,p}(\|A\|_{\op}>C_A+1)\le a_N+2a_N^2=o(1)$ uniformly in such $h$. By a union bound, all these history events can be intersected over the finitely many depths.

At $p=0$, use the identity coupling and set
$\mathcal G_{0,N}=\mathcal N_{0,N}=\{\varnothing\}$;
the norm bounds follow directly from Gaussian concentration.

\section{One Gaussian step with a fixed history}
\label{sec:finite-po}

In this section, we analyze one round $t=p+1$, $p\ge0$, with the history of the first $p$ rounds fixed. On the histories $h\in\mathcal G_{p,N}\cap\mathcal N_{p,N}$ of Section~\ref{subsec:transport-coupling}, the coupling lets us replace the conditional law of the matrix by the Gaussian law $Q_{h,p}$ conditioned on the affine history, at the cost of an $o_{\Pp_h}(N)$ error in restricted saddle values. Gaussian regression (Proposition~\ref{prop:prelim-gaussian-regression}) and the Gaussian comparison of Theorem~\ref{thm:prelim-cgmt} then reduce the LP to a column width and a row projection cost, both indexed by the overlap parameters $\omega=(s,a,b)$, and their limits $\phi_t(\omega)$ and $d_t(\omega)$ determine the laws of the next round. In Section~\ref{subsec:fixed-history-problem}, we formulate the finite problem on a fixed history, and in Section~\ref{subsec:population-geometry}, we formulate the population variational problem. In Section~\ref{subsec:canonical-kkt-update}, we derive the update of the laws. In Sections~\ref{subsec:population-uniqueness} and~\ref{subsec:population-gaps}, we prove that the population optimizer is unique and that the population problem has quantitative value gaps, which we use in Section~\ref{sec:closing-gaussian-recursion} to localize the LP optimizer.

For the first round, we use the Gaussian width $\varphi_1$ and the feasibility condition~\eqref{eq:stage1-feasibility} of Section~\ref{subsec:intro-main-results}.
Choose $\rho_0\in(0,1]$ with $\varphi_1(\rho_0)^2>\alpha\E(\kappa_1-\rho_0G)_+^2$. Dominated convergence gives continuity in the margin on bounded intervals, so the strict inequality persists at $\kappa_1+\Delta_0$ for some $\Delta_0>0$. The feasibility result of~\cite{LSZ2025} then supplies, with high probability, $x\in[-1,1]^N$ with $Ax\ge(\kappa_1+\Delta_0)\1$. This is the initial reserve; later reserves come from the margin drops in Section~\ref{subsec:iterated-replacement-lp}.

Throughout the population analysis, the input for round $t=p+1$
is strict when $p\ge1$, in the sense of Definition~\ref{def:strict-history}.
At $p=0$, we use the empty history under \eqref{eq:stage1-feasibility},
with empty population marks and $X_p=U_p=0$.

\subsection{The finite problem on a fixed Gaussian history}\label{subsec:fixed-history-problem}

Fix $h\in\mathcal R_{p,N}$ and write $F=F_p$, $B=F^c$, and
$t=p+1$. 

The new iterate agrees with $x^{(p)}$ on $B$ and ranges over the box on
$F$. Write
$y=x_F^{(p)}+w\in[-1,1]^F,\allowbreak\quad x=x^{(p)}+E_Fw,\allowbreak\quad u^{(p)}=U_pe_p.$
Introducing the row margin $m$ as a separate variable splits the
constraint into $m\ge\kappa_t\1$ and $m=Ax$. The fixed-history PO is

\begin{equation}\label{eq:conditional-po}
  \Phi_{N, h}^{\mathrm{free}}
  = \max_{\substack{y\in[-1, 1]^F\\m\geq\kappa_t\1}}
  \inf_{\mu\in\R^M}
  \left\{
  v_{t, F}^\top y+
  \mu^\top\bigl(u^{(p)}+A_{:, F}w-m\bigr)
  \right\}.
\end{equation}
The omitted frozen contribution is centered Gaussian conditionally on the history: $B,x_B^{(p)}$ are fixed, $|x_j^{(p)}|=1$ on $B$, and $v_t$ is fresh. Hence
\[
 \E_h\left(\frac{v_{t,B}^\top x_B^{(p)}}N\right)^2
 =\frac{|B|}{N^2}\le\frac1N,
 \qquad \sup_h\Pp_h(|v_{t,B}^\top x_B^{(p)}|>\eps N)
 \le\frac1{\eps^2N}.
\]
It is therefore $o_{\Pp_h}(N)$ uniformly, and the free and full values have the same normalized limit.

We split the equality multiplier $\mu$ into a component along the old dual history and a transverse component. Eliminating the old component turns the earlier dual KKT relations into finitely many exact moment constraints on the new margin.

\begin{lemma}[Eliminating the old dual directions]
\label{lem:finite-compatibility}
Decompose $\mu=Q_{L,p}\nu+L_pc$, where
$\nu\in\R^{M-p}$ and $c\in\R^p$, and set
$a_N(w):=\frac1N X_{p,F}^\top w,\allowbreak\quad b_N(w):=\frac1N R_{p,F}^\top w.$
Taking the infimum over $c$ in \eqref{eq:conditional-po} imposes the
compatibility constraint
\begin{equation}
\label{eq:finite-row-compatibility}
  \frac1N L_p^\top m
  =
  \Gamma_pe_p+b_N(w).
\end{equation}
Moreover, the conditional mean of the new row field is
\begin{align}
  z_{N,h}(a_N,b_N)
  &:=
  u^{(p)}+A_{h,p,:,F}w \notag\\*
  &=
  U_p\!\left\{e_p+(Q_p^x)^{-1}a_N\right\}
  +L_p(Q_p^\lambda)^{-1}
  \left\{b_N-\Gamma_p(Q_p^x)^{-1}a_N\right\}.
  \label{eq:finite-row-baseline}
\end{align}
It satisfies the same compatibility relation, $\frac1N L_p^\top z_{N,h}(a_N,b_N) = \Gamma_pe_p+b_N.$ \end{lemma}

\begin{proof}
Under the affine history, $L_p^\top A_{:,F}=R_{p,F}^\top$ and $L_p^\top u^{(p)}=L_p^\top U_pe_p=N\Gamma_pe_p$. Hence the $c$-component of the payoff in~\eqref{eq:conditional-po} is $Nc^\top\{\Gamma_pe_p+b_N(w)-\frac1NL_p^\top m\}$, and the infimum over $c\in\R^p$ is $-\infty$ unless~\eqref{eq:finite-row-compatibility} holds. Substituting~\eqref{eq:history-conditional-mean} into $A_{h,p,:,F}w$ and using $X_p^\top X_p=NQ_p^x$, $L_p^\top L_p=NQ_p^\lambda$, and $L_p^\top U_p=N\Gamma_p$ gives~\eqref{eq:finite-row-baseline}. Finally, $L_p^\top A_{h,p}=R_p^\top$, so $\frac1NL_p^\top z_{N,h}=\Gamma_pe_p+b_N$.
\end{proof}

We impose fixed finite cutoffs $\|m\|\le R_m\sqrt N$ and
$\|\nu\|\le R_{\rm dual}\sqrt N$, choosing
$R_m>C_A+1,\allowbreak\quad R_m^2>\alpha\E U_t^2$
with uniform positive slack over the prescribed compact input family.
Here $U_t$ is the population row update of
Section~\ref{subsec:canonical-kkt-update}. On
$\{\|A\|_{\op}\le C_A+1\}$, every box vector satisfies
$\|Ax\|\le R_m\sqrt N$. This event has conditional probability tending
to one uniformly on $\mathcal G_{p,N}\cap\mathcal N_{p,N}$ under both
matrix laws. We choose $R_{\rm dual}$ above the limiting scalar optimizer
and remove it in Section~\ref{subsec:conditional-optimizer}.

For $b\in\R^p$, define the finite feasible row-margin set
\begin{equation*}
  \cC_{N, h}^{(R_m)}(b) =
  \left\{
  m\in\R^M:
  m\geq\kappa_t\1, \quad
  \norm m\leq R_m\sqrt N, \quad
  \frac1N L_p^\top m = \Gamma_pe_p+b
  \right\}.
\end{equation*}
By~\eqref{eq:free-column-gaussian-representation} and Lemma~\ref{lem:finite-compatibility}, the compactified PO becomes
\begin{align}
  \Phi_{N, h}^{(R_{\rm dual}, R_m)}
  = \max_{\substack{y\in[-1, 1]^F                                  \\
    m\in\cC_{N, h}^{(R_m)}(b_N(w))}}
  \inf_{\norm\nu\leq R_{\rm dual}\sqrt N}
  \biggl\{ & v_{t, F}^\top y +
  \nu^\top Q_{L, p}^\top\{z_{N, h}(a_N, b_N)-m\}\notag             \\
  & +\frac1{\sqrt N}\nu^\top G_{p, F}\Sigma_{p, F}^{1/2}w
  \biggr\}.
  \label{eq:compact-po}
\end{align}
The random part of~\eqref{eq:compact-po} is a single bilinear form in a standard Gaussian matrix, optimized over compact convex sets.

We apply the Gaussian comparison of Theorem~\ref{thm:prelim-cgmt} with the history fixed. Let $g\in\R^{|F|}$ and $G_{\rm row}\in\R^{M-p}$ be independent standard Gaussian vectors, independent of the history and $v_t$. Write
$c_v=N^{-1}v_{t,F}^\top y$ and $c_g=N^{-1}g^\top\Sigma_{p,F}^{1/2}w$.
The auxiliary optimization (AO), normalized by $N$, is
\begin{equation}\label{eq:conditional-ao}
 \frac{\phi_{N,h}^{(R_{\rm dual},R_m)}}N
 =\max_{\substack{y\in[-1,1]^F\\m\in\cC_{N,h}^{(R_m)}(b_N(w))}}
 \inf_{\|\nu\|\le R_{\rm dual}\sqrt N}
 \left\{c_v+\frac{\nu^\top Q_{L,p}^\top(Z_{N,h}-m)}N
                +\frac{\|\nu\|}{\sqrt N}c_g\right\}.
\end{equation}
Here the order parameters
$s_N=\|y\|^2/N$, $a_N=N^{-1}X_{p,F}^\top w$, $b_N=N^{-1}R_{p,F}^\top w$, and $\omega_N=(s_N,a_N,b_N)$ record the new squared norm and the overlaps of the increment with the old histories. They determine the transverse variance
\begin{equation}\label{eq:finite-rho}
 \rho_{N,h}(\omega_N)^2=\frac1Nw^\top\Sigma_{p,F}w
 =s_N-(Q_{p,F}^x)_{pp}-2e_p^\top a_N-a_N^\top(Q_p^x)^{-1}a_N
\end{equation}
and the row field $Z_{N,h}=z_{N,h}(a_N,b_N)+\rho_{N,h}(\omega_N)Q_{L,p}G_{\rm row}$.
Thus the two Gaussian terms replacing the matrix bilinear form are $\|\nu\|c_g/\sqrt N$ and $\rho_{N,h}G_{\rm row}^\top\nu/N$ in the normalized payoff.
Both $Z_{N,h}$ and $m$ satisfy the old-dual compatibility relation, so $Z_{N,h}-m$ is transverse. Minimizing over the direction and norm of $\nu$ gives
\begin{equation}\label{eq:conditional-ao-reduced}
 \frac{\phi_{N,h}^{(R_{\rm dual},R_m)}}N
 =\max_{\substack{y\in[-1,1]^F\\m\in\cC_{N,h}^{(R_m)}(b_N(w))}}
 \left\{c_v-R_{\rm dual}
       \left(\frac{\|Z_{N,h}-m\|}{\sqrt N}-c_g\right)_+\right\}.
\end{equation}
For each $\omega_N$, let $d_{N,h}^{(R_m)}(\omega_N)=N^{-1/2}\dist(Z_{N,h},\cC_{N,h}^{(R_m)}(b_N))$, and let $\cA_{N,h}(\omega_N)$ be the compact set of pairs $(c_v,c_g)$ attained by box vectors with these order parameters. Minimizing the row distance then yields

\begin{equation}\label{eq:finite-hinge}
  \frac1N\phi_{N, h}^{(R_{\rm dual}, R_m)}
  = \sup_{\omega_N}
  \sup_{(c_v,c_g)\in\cA_{N,h}(\omega_N)}
  \left\{
  c_v-R_{\rm dual}\bigl(d_{N,h}^{(R_m)}(\omega_N)-c_g\bigr)_+
  \right\}.
\end{equation}

This expresses the next-round problem in terms of a column attainable set $\cA_{N, h}(\omega_N)$ and a row repair distance $d_{N, h}^{(R_m)}(\omega_N)$. The population problem below is defined through the limits of the support functions of the column sets and the projection costs of the row sets.

\subsection{The population variational problem}
\label{subsec:population-geometry}

This subsection and the next derive the steps of Algorithm~\ref{alg:state-evolution}. Write $q=q_p$ for the free fraction at depth $p$, and let
$(X,R)\sim\pi_p^{\rm col},\allowbreak\quad (U,\Lambda)\sim\pi_p^{\rm row},\allowbreak\quad \bar X=(\bar X_1,\ldots,\bar X_p)\sim\mu_p^{\rm col}.$
The marks $(X,R)$ describe the free columns on which the next update acts; $\bar X$ also records previously frozen coordinates. Their moments are
\begin{align*}
  Q^x&=\E[\bar X\bar X^\top],
  &Q_F^x&=q\E[XX^\top],
  &D^x&=Q^x-Q_F^x,\\
  Q_F^{rx}&=q\E[RX^\top],
  &Q^\lambda&=\alpha\E[\Lambda\Lambda^\top],
  &\Gamma&=\alpha\E[\Lambda U^\top].
\end{align*}

\begin{definition}[Strict population history]\label{def:strict-history}
For $p\ge1$, a depth-$p$ population history is \emph{strict} if its three laws have finite second moments and satisfy the following conditions:
\begin{enumerate}[label=\textnormal{(\roman*)},leftmargin=*]
\item $X\in(-1,1)^p$ and $\bar X\in[-1,1]^p$ almost surely. Frozen coordinates stay frozen, that is, $\bar X_{r+1}=\bar X_r$ on $\{|\bar X_r|=1\}$ for $r<p$. The full and free laws are related by $q=\Pp(|\bar X_p|<1), $ $\law(\bar X\mid |\bar X_p|<1)=\law(X).$ \item For every $r\le p$, the row marks satisfy $U_r\ge\kappa_r,$ $\Lambda_r\ge0,$ $\Lambda_r(U_r-\kappa_r)=0 \quad\text{almost surely}.$ \item $q>0$, $Q^x\succ0$, $Q^\lambda\succ0$, and $D^x\succ0$.
\end{enumerate}
Such a history is a \emph{strict input for round $t=p+1$} if
$\Delta_t:=\kappa_p-\kappa_t>0$.
\end{definition}

Compact input families have fixed depth, density, and margins. Positivity and finite
dual attainment follow from Lemma~\ref{lem:positive-branch} and
Appendix~\ref{app:population-multipliers}.

The finite order parameters $(s_N,a_N,b_N)$ introduced in the preceding subsection have the following population counterparts.

\begin{definition}[Population order parameters and transverse variance]
  \label{def:column-order-parameters}
  A parameter $\omega=(s,a,b)\in\R\times\R^p\times\R^p$ is column-feasible if there exists a random variable $Y\in[-1,1]$ such that
  \begin{equation}
    \label{eq:population-column-moments}
    q\E Y^2=s,
    \qquad
    q\E[X(Y-X_p)]=a,
    \qquad
    q\E[R(Y-X_p)]=b.
  \end{equation}
  For a feasible parameter, define its transverse variance by
  \begin{equation}
    \label{eq:population-rho}
    \rho_\omega^2
    =
    s-(Q_F^x)_{pp}
    -2e_p^\top a
    -a^\top(Q^x)^{-1}a.
  \end{equation}
  The feasible parameter domain consists of the triples satisfying \eqref{eq:population-column-moments} and $\rho_\omega^2\ge0$.
\end{definition}

\subsubsection{Column width}
Let $G\sim N(0,1)$ be independent of the old column marks. For a column-feasible $\omega$, define
\begin{equation}
  \label{eq:population-width}
  \phi_t(\omega)
  =
  \sup\left\{
  q\E[GY]:
  Y\in[-1,1]\text{ satisfies }
  \eqref{eq:population-column-moments}
  \right\}.
\end{equation}
Here the supremum is taken over randomized kernels, that is, over functions of $(X,R,G)$ and of an independent auxiliary random seed.

Define $\ell(z;\tau) = \max_{|y|\le1} \left\{ zy-\frac{\tau}{2}y^2 \right\}.$ Dualizing the moment constraints in \eqref{eq:population-width}, we obtain the finite-dimensional dual representation
\begin{align}
  \phi_t(\omega)
  =
  \inf_{\substack{\tau\in\R\\
      \eta_x,\eta_r\in\R^p}}
  \biggl\{
  &\frac{\tau s}{2}
  -\eta_x^\top(Q_F^xe_p+a)
  -\eta_r^\top(Q_F^{rx}e_p+b)
  \notag\\
  &\qquad
  +q\E\ell(G+\eta_x^\top X+\eta_r^\top R;\tau)
  \biggr\}.
  \label{eq:population-column-dual}
\end{align}
Appendix~\ref{app:population-multipliers} proves that, at every population maximizer, the column and row dual optima are attained at finite multipliers, with $\tau_\omega>0$. The pointwise column optimizer is then unique and given by
\begin{equation*}
  Y_\omega
  =
  \clip\left(
  \frac{
    G+\eta_{x,\omega}^\top X+\eta_{r,\omega}^\top R
  }{\tau_\omega},
  -1,1
  \right).
\end{equation*}
Here $\clip(z,-1,1)=\max\{-1,\min\{z,1\}\}$ denotes the projection onto $[-1,1]$.
\subsubsection{Row repair cost}
The parameter $\omega$ also determines the population limit of the finite row pre-projection margin. Let $G_{\rm row}\sim N(0,1)$ be independent of the old history and of $G$, and define
\begin{equation}
Z_\omega ={} \{e_p+(Q^x)^{-1}a\}^\top U + \left[ (Q^\lambda)^{-1} \{b-\Gamma(Q^x)^{-1}a\} \right]^\top\Lambda +\rho_\omega G_{\rm row}. \label{eq:population-row-baseline}
\end{equation}
The admissible row margins form the closed convex set
\begin{equation*}
  \cC_\omega
  =
  \left\{
  \widetilde U\in L^2:
  \widetilde U\ge\kappa_t,
  \quad
  \alpha\E[\Lambda\widetilde U]
  =\Gamma e_p+b
  \right\}.
\end{equation*}
If $\cC_\omega$ is nonempty, define $d_t(\omega)^2 = \alpha\,\dist_{L^2}(Z_\omega,\cC_\omega)^2, $ $U_\omega=P_{\cC_\omega}Z_\omega.$ If $\cC_\omega=\varnothing$, we set $d_t(\omega)=+\infty$ and assign the value $-\infty$ to the corresponding population branch.

By construction, $\alpha\E[\Lambda Z_\omega]=\Gamma e_p+b,$ so the pre-projection margin already satisfies the old-dual moment constraint. If the row dual is attained, the projection admits the KKT representation
\begin{equation}
  \label{eq:canonical-row}
  U_\omega
  =
  \max\{
  \kappa_t,
  Z_\omega-\eta_{\lambda,\omega}^\top\Lambda
  \},
  \qquad
  \alpha\E[\Lambda U_\omega]
  =
  \Gamma e_p+b,
\end{equation}
for some multiplier $\eta_{\lambda,\omega}\in\R^p$.  Consequently,
\begin{equation*}
  d_t(\omega)^2
  =
  \alpha\E
  \left[
    \left(
    (\kappa_t-Z_\omega
    +\eta_{\lambda,\omega}^\top\Lambda)_+
    -\eta_{\lambda,\omega}^\top\Lambda
    \right)^2
    \right].
\end{equation*}
Moreover, under the nondegeneracy condition
$\alpha\E\left[ \Lambda\Lambda^\top \1_{\{ Z_\omega-\eta_{\lambda,\omega}^\top\Lambda>\kappa_t \}} \right] \succ0$
the multiplier $\eta_{\lambda,\omega}$ is unique.

\subsubsection{Scalar reduction and the population objective}

Let $V\sim N(0,1)$ be independent of $(G,G_{\rm row})$ and of the old
population marks.  For fixed $\omega$ and $\gamma\ge0$,
\begin{equation}
  \label{eq:rotated-column-width}
  \sup_Y q\E[(V+\gamma G)Y]
  =
  \sqrt{1+\gamma^2}\,\phi_t(\omega),
\end{equation}
where the supremum is over $Y\in[-1,1]$ satisfying
\eqref{eq:population-column-moments}.  Indeed,
$(V+\gamma G)/\sqrt{1+\gamma^2}$ is again standard Gaussian, and this
orthogonal rotation of the fresh Gaussian variables leaves the
admissible class unchanged.

For $\cC_\omega\ne\varnothing$, use
$-R_{\rm dual}(d-c)_+=\inf_{0\le\gamma\le R_{\rm dual}}\gamma(c-d)$
to write the population hinge objective as
\begin{align*}
 &\sup_Y\{q\E[VY]-R_{\rm dual}(d_t(\omega)-q\E[GY])_+\}\\
 &\qquad=\sup_Y\inf_{0\le\gamma\le R_{\rm dual}}
       \{q\E[(V+\gamma G)Y]-\gamma d_t(\omega)\}.
\end{align*}

Let $\mathcal A_\omega$ contain the joint laws of $(X,R,V,G,Y)$ with the prescribed $(X,R,V,G)$ marginal, $|Y|\le1$, and moments \eqref{eq:population-column-moments}. Independent randomization and the affine constraints make this set convex. Minimax therefore interchanges the supremum and infimum; \eqref{eq:rotated-column-width} gives
\begin{equation}
  \label{eq:truncated-branch-definition}
  \Psi_{t,R_{\rm dual}}(\omega)
  =
  \inf_{0\le\gamma\le R_{\rm dual}}
  \left\{
    \sqrt{1+\gamma^2}\,\phi_t(\omega)
    -\gamma d_t(\omega)
  \right\}.
\end{equation}

The remaining minimization is one-dimensional. Write
$f_\omega(\gamma)=\phi_t(\omega)\sqrt{1+\gamma^2}-\gamma d_t(\omega)$.
On the strict branch $\phi_t(\omega)>d_t(\omega)$,
\[
  f_\omega'(\gamma)
  =\frac{\gamma\phi_t(\omega)}{\sqrt{1+\gamma^2}}-d_t(\omega),
  \qquad
  f_\omega''(\gamma)
  =\frac{\phi_t(\omega)}{(1+\gamma^2)^{3/2}}>0.
\]
Solving $f_\omega'(\gamma)=0$ and substituting the solution gives
$\gamma=d_t(\omega)/\sqrt{\phi_t(\omega)^2-d_t(\omega)^2}$ and
$f_\omega(\gamma)=\sqrt{\phi_t(\omega)^2-d_t(\omega)^2}$.
If $\phi_t(\omega)=d_t(\omega)$, the infimum is zero, approached as
$\gamma\to\infty$ unless both quantities vanish; if
$\phi_t(\omega)<d_t(\omega)$, the negative linear term dominates.
Thus the untruncated population value is
\begin{equation}
  \label{eq:positive-recursion-value}
  \Psi_t(\omega)
  =\inf_{\gamma\ge0}f_\omega(\gamma)
  =
  \begin{cases}
    \sqrt{\phi_t(\omega)^2-d_t(\omega)^2},
      & \phi_t(\omega)\ge d_t(\omega),\\[1mm]
    -\infty,
      & \phi_t(\omega)<d_t(\omega),
  \end{cases}
\end{equation}
and $V_t=\max_\omega\Psi_t(\omega), $ $\Omega_t=\argmax_\omega\Psi_t(\omega).$ On the strict branch $\phi_t(\omega)>d_t(\omega)$, the scalar minimizer
is unique and equals $\gamma_\omega = \frac{d_t(\omega)}{\Psi_t(\omega)}.$

The next lemma gives $V_t>0$.

\begin{lemma}[A feasible point with positive value]
  \label{lem:positive-branch}
  Fix a depth-$p$ input for round $t=p+1$ satisfying the assumptions of Section~\ref{sec:finite-po}.
  Then there exists a column-feasible $\omega$ with
  $\cC_\omega\ne\varnothing$ such that $\phi_t(\omega)>d_t(\omega).$ Consequently, $\Psi_t(\omega)>0$ and hence $V_t>0$. Moreover, on every compact family of such inputs, with the product $\Wtwo$ topology on the laws and the Euclidean topology on $q$, there is $v_*>0$ such that $V_t\ge v_*$ throughout the family.
\end{lemma}

\begin{proof}
  See Appendix~\ref{app:positive-value-point}.
\end{proof}

For the empty input, $\omega=s\in[0,1]$ and $\rho^2=s$, so
\begin{equation}\label{eq:empty-history-functionals}
  \phi_1(s)=\varphi_1(\sqrt s),\qquad
  Z_s=\sqrt s\,G_{\rm row},\qquad
  d_1(s)^2=\alpha\E(\kappa_1-\sqrt s\,G_{\rm row})_+^2.
\end{equation}
The projection $U_s=\max\{\kappa_1,Z_s\}$ makes every row set nonempty. Substitution in \eqref{eq:positive-recursion-value} gives $\Psi_1(\rho^2)=\inf_{\gamma\ge0}\{\sqrt{1+\gamma^2}\,\varphi_1(\rho)-\gamma d_1(\rho^2)\}$. Thus \eqref{eq:stage1-feasibility} supplies a positive branch, and the column and row gaps below apply with no old moment constraints.

\subsubsection{Removal of the dual cutoff}
On a compact family of such inputs, the preceding lemma yields a uniform lower bound $V_t\ge v_*>0.$ On the other hand, every branch whose scalar optimizer is affected by the cutoff has value $O(R_{\rm dual}^{-1})$, uniformly over the same family. Indeed, if $\phi_t(\omega)\le d_t(\omega)$, then evaluating \eqref{eq:truncated-branch-definition} at $\gamma=R_{\rm dual}$ gives $\Psi_{t,R_{\rm dual}}(\omega) \le \frac{\E|G|}{2R_{\rm dual}}.$ If instead $\phi_t(\omega)>d_t(\omega)$ but $\gamma_\omega>R_{\rm dual}$, then
$\Psi_t(\omega) < \frac{d_t(\omega)}{R_{\rm dual}}, \allowbreak\quad \phi_t(\omega)-d_t(\omega) < \frac{\E|G|}{R_{\rm dual}^2}$,
and hence $\Psi_{t,R_{\rm dual}}(\omega) \le \frac{3\E|G|}{2R_{\rm dual}}.$ Thus, for $R_{\rm dual}=C/v_*$ with $C$ sufficiently large, no branch affected by the cutoff maximizes the population problem, and hence the truncated and untruncated population optima coincide.

For the finite problem choose $R_{\rm dual}>\gamma_t$. The norm localization proved in Section~\ref{subsec:conditional-optimizer} makes the dual cutoff inactive. First-order optimality in the transverse directions and the old-dual compatibility constraint then give $m=Ax$; the operator-norm bound also makes the row cutoff inactive. The proof there shows that the compactified and original LPs coincide with high probability.

\subsection{The population law update}
\label{subsec:canonical-kkt-update}

Let $\omega_t^\star=(s_t^\star,a_t^\star,b_t^\star)$ be the population maximizer, whose uniqueness is proved in Proposition~\ref{prop:population-optimizer-unique}, and write
$\phi_t^\star = \phi_t(\omega_t^\star), \allowbreak\quad d_t^\star = d_t(\omega_t^\star), \allowbreak\quad \rho_t = \rho_{\omega_t^\star}, \allowbreak\quad V_t = \sqrt{(\phi_t^\star)^2-(d_t^\star)^2}.$
The scalar minimizer is $\gamma_t = \frac{d_t^\star}{V_t}.$

\subsubsection{Column update}
Let $V,G$ be independent standard Gaussian variables, independent of the previous population history, and set $S_t = \frac{V+\gamma_tG}{\sqrt{1+\gamma_t^2}}.$ Let $(\tau_t,\eta_{x,t},\eta_{r,t})$ be a column-dual minimizer in \eqref{eq:population-column-dual} at $\omega_t^\star$.  The update of a previously free coordinate is
\begin{equation}
  \label{eq:canonical-primal-update}
  Y_t
  =
  \clip\left(
  \frac{
    S_t+\eta_{x,t}^\top X+\eta_{r,t}^\top R
  }{\tau_t},
  -1,1
  \right).
\end{equation}

Define the new free mass by $q_t = q\,\Pp(|Y_t|<1).$ Since $S_t$ is again a standard Gaussian independent of the old column marks and $Y_t$ attains the constrained width, $\phi_t^\star = q\E[S_tY_t].$ Gaussian integration by parts therefore gives
\begin{equation}
  \label{eq:stein-free-mass}
  q_t
  =
  q\,\Pp(|Y_t|<1)
  =
  \tau_t\phi_t^\star.
\end{equation}
Thus Stein's identity relates the column width to the free-coordinate mass of the next population state.

\subsubsection{Row update}
Set $Z_t=Z_{\omega_t^\star}$, $U_t=U_{\omega_t^\star}$, and $\eta_{\lambda,t}=\eta_{\lambda,\omega_t^\star}$. Equation~\eqref{eq:canonical-row} gives $U_t=\max\{\kappa_t,Z_t-\eta_{\lambda,t}^\top\Lambda\}$ and $\alpha\E[\Lambda U_t]=\Gamma e_p+b_t^\star$. The obstacle residual and new row multiplier are
\begin{align}
 \zeta_t&=(\kappa_t-Z_t+\eta_{\lambda,t}^\top\Lambda)_+,
 \label{eq:canonical-row-obstacle}\\
 U_t-Z_t&=\zeta_t-\eta_{\lambda,t}^\top\Lambda,
 \label{eq:canonical-row-residual}\\
 \Lambda_t&=\frac{\gamma_t}{d_t^\star}\zeta_t.
 \label{eq:canonical-dual-update}
\end{align}

Here $V$ represents the objective coordinate $v_{t,j}$ of a free coordinate. Since stationarity gives $r_j^{(t)}=-v_{t,j}$ on the free coordinates (Section~\ref{subsec:regular-kkt}), we set $R_t=-V$ on $\{|Y_t|<1\}$. The multiplier update~\eqref{eq:canonical-dual-update} is the population form of complementary slackness. In the auxiliary problem, the transverse dual variable has the empirical law of $(\gamma_t/d_t^\star)(U_t-Z_t)$, and rows with $U_t>\kappa_t$ carry no multiplier. We derive~\eqref{eq:canonical-dual-update} from the finite LP in the proof of Theorem~\ref{thm:conditional-localization}, and $d_t^\star>0$ by Appendix~\ref{app:population-multipliers}.

\begin{lemma}[Free-coordinate/active-row balance]
  \label{lem:free-row-balance}
  At a population optimizer with $V_t>0$, the updated free fraction satisfies
  \begin{equation*}
    q_t
    =
    \alpha\,\Pp(U_t=\kappa_t)
    =
    \alpha\,\Pp(\Lambda_t>0).
  \end{equation*}
\end{lemma}

\begin{proof}
  By Lemma~\ref{lem:app-population-multipliers} and its proof, $\rho_t>0$, $d_t^\star>0$, the column multiplier $\tau_t$ is unique, and $\phi_t$ is differentiable in $\rho$ at the maximizer with $a,b$ fixed. With $a,b$ fixed, \eqref{eq:population-rho} gives
  $s=(Q_F^x)_{pp}+2e_p^\top a+a^\top(Q^x)^{-1}a+\rho^2.$
  Applying the envelope theorem to \eqref{eq:population-column-dual}
  and the row projection, together with Gaussian integration by parts
  in $Z_t=z_t+\rho G_{\rm row}$, where $z_t$ does not depend on $G_{\rm row}$, gives
  $\partial_\rho\phi_t=\tau_t\rho,\allowbreak\quad \partial_\rho d_t= \frac{\alpha\rho\,\Pp(U_t=\kappa_t)}{d_t}.$
  Stationarity of $\Psi_t=(\phi_t^2-d_t^2)^{1/2}$ at $\rho_t>0$ yields
  $\tau_t\phi_t^\star=\alpha\Pp(U_t=\kappa_t).$
  Combining this with \eqref{eq:stein-free-mass} proves the first equality.

  Since $Z_t-\eta_{\lambda,t}^\top\Lambda$ has a continuous distribution,
  \eqref{eq:canonical-row}, \eqref{eq:canonical-row-obstacle}, and
  \eqref{eq:canonical-dual-update} give
  $\{U_t=\kappa_t\}=\{\Lambda_t>0\}$ almost surely, proving the second.
\end{proof}

For the empty input, the update becomes
\begin{equation}\label{eq:empty-history-update}
\begin{aligned}
  Y_1&=\clip(S_1/\tau_1,-1,1),
  &U_1&=\max\{\kappa_1,\rho_1G_{\rm row}\},\\
  \Lambda_1&=\frac{(\kappa_1-\rho_1G_{\rm row})_+}{V_1},
  &q_1&=\Pp(|S_1|<\tau_1).
\end{aligned}
\end{equation}
\subsubsection{Updated population laws and moments}

For the next population state, set
$\widetilde X_t=(X^\top,Y_t)^\top, \allowbreak\quad \widetilde R_t=(R^\top,R_t)^\top,\allowbreak\quad \text{and}\allowbreak\quad \widetilde U_t=(U^\top,U_t)^\top, \allowbreak\quad \widetilde\Lambda_t=(\Lambda^\top,\Lambda_t)^\top$.

The new free-column law is
$\pi_t^{\rm col} = \law\!\left( \widetilde X_t,\widetilde R_t \,\middle|\, |Y_t|<1 \right),$
while the updated row law is
$\pi_t^{\rm row} = \law\!\left( \widetilde U_t,\widetilde\Lambda_t \right).$
In particular, \eqref{eq:empty-history-update} gives
$\pi_1^{\rm col}=\law(Y_1,-V\mid |Y_1|<1)$ and
$\pi_1^{\rm row}=\law(U_1,\Lambda_1)$, with
$S_1=(V+\gamma_1G)/\sqrt{1+\gamma_1^2}$.

For the full primal trajectory, retain the previously frozen coordinates. Writing $\mu_{t-1}^{\rm col}$ for its old law, define $\mu_t^{\rm col}$ by
\begin{equation}
\int f(z)\,\mu_t^{\rm col}(dz) ={} \int_{\{|z_{t-1}|=1\}} f(z_1,\ldots,z_{t-1},z_{t-1})\, \mu_{t-1}^{\rm col}(dz) + q_{t-1}\, \E f(X,Y_t) \label{eq:full-column-law-update}
\end{equation}
for every bounded measurable $f$, with the first integral omitted at $t=1$.
The terms retain old signs and update entering free coordinates, respectively.

In particular,
\begin{equation*}
  \Pp(|\bar X_t|<1)=q_t,
  \qquad
  \law(\bar X_1,\ldots,\bar X_t\mid |\bar X_t|<1)
  =
  \law(\widetilde X_t\mid |Y_t|<1),
\end{equation*}
for $(\bar X_1,\ldots,\bar X_t)\sim\mu_t^{\rm col}$.

The free-column moments needed in the next round are therefore
\begin{equation*}
  Q_{t,F}^x
  =
  q_t\E\!\left[
    \widetilde X_t\widetilde X_t^\top
    \,\middle|\,
    |Y_t|<1
  \right],
  \qquad
  Q_{t,F}^{rx}
  =
  q_t\E\!\left[
    \widetilde R_t\widetilde X_t^\top
    \,\middle|\,
    |Y_t|<1
  \right].
\end{equation*}
Similarly, the updated row moments are $Q_t^\lambda = \alpha\E[ \widetilde\Lambda_t\widetilde\Lambda_t^\top ], $ $\Gamma_t = \alpha\E[ \widetilde\Lambda_t\widetilde U_t^\top ].$

The full primal Gram matrix $Q_t^x = \E[ \bar X_{1:t}\bar X_{1:t}^\top ]$ is given by
\begin{equation}
  \label{eq:primal-gram-update}
  Q_t^x
  =
  \begin{pmatrix}
    Q^x
    &
    Q^xe_p+a_t^\star
    \\
    e_p^\top Q^x+(a_t^\star)^\top
    &
    s_t^\star+(D^x)_{pp}
  \end{pmatrix},
\end{equation}
and the frozen contribution is $D_t^x = Q_t^x-Q_{t,F}^x.$

Together with $q_t$, these laws and moments define the transition
\begin{equation}
  \label{eq:population-transition}
  \mathfrak h_{t-1}\longmapsto\mathfrak h_t.
\end{equation}

\subsection{Uniqueness of the population optimizer}
\label{subsec:population-uniqueness}

For uniqueness we use $(\rho,a,b)$: the column width is concave and the row field affine in these coordinates.

For the fixed input, abbreviate $Q:=Q_p^x$,
$Q_F:=Q_{p,F}^x$, and $D:=Q-Q_F$. Set
$h(a):=(Q_F)_{pp}+2e_p^\top a+a^\top Q^{-1}a,\allowbreak\quad s=h(a)+\rho^2.$
For $\theta=(\rho,a,b)$, define
$\widehat\phi_t(\theta):=\phi_t(h(a)+\rho^2,a,b),\allowbreak\quad \widehat d_t(\theta):=d_t(h(a)+\rho^2,a,b),$
and
\[
  \widehat\Psi_t(\theta)
  :=
  \inf_{\gamma\ge0}
  \left\{
    \sqrt{1+\gamma^2}\,\widehat\phi_t(\theta)
    -\gamma\widehat d_t(\theta)
  \right\}.
\]
Thus $\widehat\Psi_t(\rho,a,b)=\Psi_t(s,a,b)$.

We will use mean-preserving randomization: if $Y\in[-1,1]$ and $q\E Y^2\le\widetilde s\le q$, mix $Y$ with a conditionally $\{\pm1\}$-valued variable of mean $Y$. This gives $\widetilde Y\in[-1,1]$ with $\E[\widetilde Y\mid\mathcal F]=Y$ and $q\E\widetilde Y^2=\widetilde s$, preserving all correlations with $\mathcal F$-measurable variables.

\begin{proposition}[Uniqueness of the positive population optimizer]
\label{prop:population-optimizer-unique}
Assume $V_t:=\max_\omega\Psi_t(\omega)>0.$ Suppose that at every maximizing parameter the column and row dual
optima are attained at finite multipliers, with positive column
second-moment multiplier.  Then
$\operatorname*{argmax}_\omega\Psi_t(\omega) = \{\omega_t^\star\}.$
\end{proposition}

\begin{proof}
We first prove concavity in $(\rho,a,b)$ and then analyze the equality
case for two maximizers.

\medskip
\noindent\textbf{Concavity.}
For a column-feasible $Y$, write
$W=Y-X_p$, $a(Y)=q\E[XW]$, $b(Y)=q\E[RW]$, and
$\beta(Y)=Q^{-1}a(Y)$.  Define
\begin{equation}
  \label{eq:uniqueness-T}
  T(Y)
  :=
  \left(
    \sqrt q\,[W-\beta(Y)^\top X],
    D^{1/2}\beta(Y)
  \right).
\end{equation}
A direct expansion gives
\[
  \|T(Y)\|^2
  =
  q\E Y^2-(Q_F)_{pp}
  -2e_p^\top a(Y)-a(Y)^\top Q^{-1}a(Y)
  =\rho(Y)^2.
\]
Thus $T$ is affine in $Y$ and its squared norm is the transverse variance.
With no old coordinates, \eqref{eq:uniqueness-T} simply gives $T(Y)=Y$
and $\rho(Y)^2=\E Y^2$.

Take two feasible points
$\theta_i=(\rho_i,a_i,b_i)$ with column optimizers $Y_i$, and let
$\theta_\lambda=\lambda\theta_1+(1-\lambda)\theta_2$ and
$\bar Y=\lambda Y_1+(1-\lambda)Y_2$.  Then
$a(\bar Y)=a_\lambda$, $b(\bar Y)=b_\lambda$, and
\[
  \rho(\bar Y)
  =
  \|\lambda T(Y_1)+(1-\lambda)T(Y_2)\|
  \le
  \lambda\rho_1+(1-\lambda)\rho_2.
\]
If $\bar Y$ has too small a second moment for $\theta_\lambda$, the
mean-preserving randomization above raises it to the required value
without changing $a_\lambda$, $b_\lambda$, or $q\E[G\bar Y]$.
Therefore the feasible set in $(\rho,a,b)$ is convex and
$\widehat\phi_t$ is jointly concave.

The pre-projection margin $Z_\theta$ in \eqref{eq:population-row-baseline}, with $\rho$ treated as an independent coordinate, is affine in $\theta$.  The admissible set $\mathcal C_b = \{\widetilde U\ge\kappa_t: \alpha\E[\Lambda\widetilde U]=\Gamma e_p+b\}$ is stable under convex interpolation of the parameter.  Interpolating the two row
projections therefore gives $\widehat d_t(\theta_\lambda) \le \lambda\widehat d_t(\theta_1) +(1-\lambda)\widehat d_t(\theta_2),$ so $\widehat d_t$ is convex.  Hence, for every $\gamma\ge0$, $F_\gamma(\theta) := \sqrt{1+\gamma^2}\,\widehat\phi_t(\theta) -\gamma\widehat d_t(\theta)$ is concave, and therefore so is
$\widehat\Psi_t=\inf_{\gamma\ge0}F_\gamma$.

\medskip
\noindent\textbf{Equality at two maximizers.}
Suppose that two distinct points $\theta_i=(\rho_i,a_i,b_i)$ both maximize
$\widehat\Psi_t$, with value $V_t>0$, and let
$\theta_0=(\theta_1+\theta_2)/2$.  Concavity gives
$\widehat\Psi_t(\theta_0)=V_t$.

At a positive maximizer, $\rho_i>0$ and
$d_i:=\widehat d_t(\theta_i)>0$ by Lemma~\ref{lem:app-population-multipliers}.  Let $\gamma_0$ minimize the scalar
profile at $\theta_0$.  Then
\[
  V_t
  =
  F_{\gamma_0}(\theta_0)
  \ge
  \frac{F_{\gamma_0}(\theta_1)+F_{\gamma_0}(\theta_2)}2
  \ge V_t.
\]
Thus equality holds throughout, and $\gamma_0$ minimizes both endpoint
profiles.  Since
$\gamma_i=d_i/V_t$ and $V_t^2=\phi_i^2-d_i^2$ on the positive branch, we obtain $d_1=d_2=:d, $ $\phi_1=\phi_2=:\phi.$ Writing $\phi_0=\widehat\phi_t(\theta_0)$ and
$d_0=\widehat d_t(\theta_0)$, concavity and convexity give
$\phi_0\ge\phi$ and $d_0\le d$.  Equality in the preceding chain and
$\gamma_0>0$ therefore force $\phi_0=\phi$ and $d_0=d$.

Let $\widetilde U_i=P_{\mathcal C_{b_i}}Z_{\theta_i}$ and
$E_i=\widetilde U_i-Z_{\theta_i}$.  Since
$\|E_i\|_2=d/\sqrt\alpha$ and
$(\widetilde U_1+\widetilde U_2)/2$ is midpoint-feasible,
\[
  \frac d{\sqrt\alpha}
  \le
  \left\|\frac{E_1+E_2}{2}\right\|_2
  \le
  \frac{\|E_1\|_2+\|E_2\|_2}{2}
  =
  \frac d{\sqrt\alpha}.
\]
Hence $E_1=E_2$ in $L^2$.

Write $Z_{\theta_i}=z_i+\rho_iG_{\rm row}$.  By the row KKT formula, $E_i = \max\{ \kappa_t-z_i-\rho_iG_{\rm row}, -\eta_i^\top\Lambda \}.$ For almost every value of the old row marks, the two resulting continuous
piecewise-affine functions of $g\in\R$ agree everywhere. For
sufficiently negative $g$, the first branch is active in both functions,
and comparing slopes and intercepts gives
\begin{equation}
  \label{eq:uniqueness-rho-z}
  \rho_1=\rho_2=:\rho,
  \qquad
  z_1=z_2
  \quad\text{a.s.}
\end{equation}

\medskip
\noindent\textbf{Identification of $a$ and $b$.}
It remains to identify the overlap vectors, when present.
Suppose that $a_1\ne a_2$. Since
$\nabla^2h=2Q^{-1}\succ0$, the strict convexity of $h$ and
\eqref{eq:uniqueness-rho-z} imply $h\!\left(\frac{a_1+a_2}{2}\right)+\rho^2 < \frac{s_1+s_2}{2} \le q.$ Since the midpoint is itself a maximizer, its column second-moment
multiplier $\tau_0$ is positive by hypothesis.

Realize $Y_1,Y_2$ on the same copy of $(X,R,G)$ and set
$\bar Y=(Y_1+Y_2)/2$.  Then $a(\bar Y)=a_0$, $b(\bar Y)=b_0$,
$q\E[G\bar Y]=\phi$, and
$\bar\rho := \rho(\bar Y) = \left\|\frac{T(Y_1)+T(Y_2)}2\right\| \le\rho$.
If $\bar\rho<\rho$, mean-preserving randomization produces another
midpoint-feasible optimizer with the same width value.  This contradicts
the $\tau_0$-strong concavity of the column problem, $\widehat\phi_t(\theta_0)-q\E[GY] \ge \frac{q\tau_0}{2}\E(Y-Y_0)^2.$ Hence $\bar\rho=\rho$.  Equality in the Hilbert-space triangle
inequality then gives $T(Y_1)=T(Y_2)$.  Comparing the second components
of \eqref{eq:uniqueness-T} and using $D\succ0$ and $Q\succ0$ yields
$a_1=a_2$, a contradiction.

Finally, since $a_1=a_2$, the identity \eqref{eq:uniqueness-rho-z} gives $\bigl[(Q^\lambda)^{-1}(b_1-b_2)\bigr]^\top\Lambda=0 $ $\text{a.s.}$ Setting $c=(Q^\lambda)^{-1}(b_1-b_2)$, we obtain
$0=\alpha\E(c^\top\Lambda)^2=c^\top Q^\lambda c$. Since
$Q^\lambda\succ0$, it follows that $b_1=b_2$.

Thus $\theta_1=\theta_2$, contradicting the assumption that the two points are distinct.  Hence
$\widehat\Psi_t$ has a unique maximizer, and since $s=h(a)+\rho^2$,
so does the original objective $\Psi_t$.
\end{proof}

\subsection{Quantitative population gaps}
\label{subsec:population-gaps}

The gaps below localize the finite optimizer: restrictions on maximizing variables lower the value, while restrictions on minimizing variables raise it. Retain the population optimizer and update notation, and set
\[
  \Xi_t:=S_t+\eta_{x,t}^\top X+\eta_{r,t}^\top R,
  \qquad
  \mathcal C_t
  :=
  \left\{
    \widetilde U\ge\kappa_t:
    \alpha\E[\Lambda\widetilde U]=\Gamma e_p+b_t^\star
  \right\},
\]
so that
$Y_t=\clip(\Xi_t/\tau_t,-1,1), \allowbreak\quad U_t=P_{\mathcal C_t}Z_t, \allowbreak\quad d_t^\star = \sqrt{\alpha\E(U_t-Z_t)^2}.$
The transverse row multiplier satisfies
\[
  \ell_t
  :=
  \frac{\gamma_t}{d_t^\star}(U_t-Z_t),
  \qquad
  \alpha\E[\Lambda\ell_t]=0,
  \qquad
  \alpha\E\ell_t^2=\gamma_t^2.
\]

We fix the dual cutoff $R_{\rm dual}$ chosen above, so that the truncated
and untruncated population problems have the same unique maximizer
$\omega_t^\star$ and value $V_t$.

\begin{proposition}[Population gaps]
\label{prop:population-restricted-gaps}
For every $\eps>0$, the following hold.

\begin{enumerate}[label=\textnormal{(\roman*)}]

\item \textbf{Parameter gap.}
There exists $\Delta_t^{\rm par}(\eps)>0$ such that
\[
  \sup_{\substack{\omega\text{ column-feasible}\\
                  \dist(\omega,\omega_t^\star)\ge\eps}}
  \Psi_{t,R_{\rm dual}}(\omega)
  \le
  V_t-\Delta_t^{\rm par}(\eps).
\]

\item \textbf{Column gap.}
Every $Y$ satisfying the moment constraints of $\omega_t^\star$ satisfies
\begin{equation}
  \label{eq:population-column-gap}
  \phi_t^\star-q\E[S_tY]
  \ge
  \frac{q\tau_t}{2}\E(Y-Y_t)^2.
\end{equation}
Consequently, if $\Wtwo\!\left( \law(X,R,V,Y), \law(X,R,V,Y_t) \right)\ge\eps,$ then
\[
  q\E[(V+\gamma_tG)Y]-\gamma_td_t^\star
  \le
  V_t-
  \frac{q\tau_t\sqrt{1+\gamma_t^2}}{2}\eps^2.
\]

\item \textbf{Row gap.}
Every $\widetilde U\in\mathcal C_t$ satisfies
\begin{equation}
  \label{eq:population-row-gap}
  \alpha\E(\widetilde U-Z_t)^2-(d_t^\star)^2
  \ge
  \alpha\E(\widetilde U-U_t)^2.
\end{equation}
Hence, if $\Wtwo\!\left( \law(U,\Lambda,\widetilde U), \law(U,\Lambda,U_t) \right)\ge\eps,$ then, with
$d_t(\widetilde U):=\sqrt{\alpha\E(\widetilde U-Z_t)^2}$,
\[
  \sqrt{1+\gamma_t^2}\,\phi_t^\star
  -\gamma_td_t(\widetilde U)
  \le
  V_t-\gamma_t
  \left(
    \sqrt{(d_t^\star)^2+\alpha\eps^2}-d_t^\star
  \right).
\]

\item \textbf{Dual gaps.}
For the scalar profile $f_t(\gamma) := \phi_t^\star\sqrt{1+\gamma^2}-d_t^\star\gamma,$ we have, uniformly for $\gamma\in[0,R_{\rm dual}]$,
\begin{equation}
  \label{eq:population-scalar-gap}
  f_t(\gamma)-V_t
  \ge
  \frac{c_\gamma}{2}|\gamma-\gamma_t|^2,
  \qquad
  c_\gamma
  :=
  \frac{\phi_t^\star}{(1+R_{\rm dual}^2)^{3/2}}>0.
\end{equation}
At the optimal dual radius $\gamma_t$, every $\ell\in L^2$ satisfying
$\alpha\E[\Lambda\ell]=0$ and $\alpha\E\ell^2=\gamma_t^2$ obeys
\begin{equation}
  \label{eq:population-transverse-angular-gap}
  \alpha\E[\ell(Z_t-U_t)]+\gamma_td_t^\star
  =
  \frac{\alpha d_t^\star}{2\gamma_t}
  \E(\ell-\ell_t)^2.
\end{equation}
In particular, if $\Wtwo\!\left( \law(U,\Lambda,\ell), \law(U,\Lambda,\ell_t) \right)\ge\eps,$ then
\[
  \sqrt{1+\gamma_t^2}\,\phi_t^\star
  +\alpha\E[\ell(Z_t-U_t)]
  \ge
  V_t+\frac{\alpha d_t^\star}{2\gamma_t}\eps^2.
\]
\end{enumerate}

All four gaps hold uniformly over compact families of such inputs.
\end{proposition}

\begin{proof}
For (i), note that the admissible parameter set is compact. The column width $\phi_t$ is upper semicontinuous, while weak compactness and lower semicontinuity of the row projection give lower semicontinuity of $d_t$. Hence $\Psi_{t,R_{\rm dual}}$, extended by $-\infty$ to
row-infeasible parameters, is upper semicontinuous. Its supremum over
the closed set
$\{\dist(\omega,\omega_t^\star)\ge\eps\}$ is therefore attained, and it is
strictly smaller than $V_t$ because $\omega_t^\star$ is the unique maximizer.

For (ii), the terms involving the affine moment multipliers cancel in
the column dual representation, which gives $\phi_t^\star-q\E[S_tY] = q\E\!\left[ \ell(\Xi_t;\tau_t) -\Xi_tY+\frac{\tau_t}{2}Y^2 \right].$ Since $Y_t$ maximizes
$y\mapsto\Xi_ty-\tau_ty^2/2$ on $[-1,1]$, its first-order condition
is $(\Xi_t-\tau_tY_t)(y-Y_t)\le0$ for every $y\in[-1,1]$. Hence
\[
  \ell(\Xi_t;\tau_t)-\Xi_tY+\frac{\tau_t}{2}Y^2
  =(\Xi_t-\tau_tY_t)(Y_t-Y)+\frac{\tau_t}{2}(Y-Y_t)^2
  \ge\frac{\tau_t}{2}(Y-Y_t)^2.
\]
Taking expectations gives \eqref{eq:population-column-gap}. Coupling the two laws through a common
realization of $(X,R,V)$ gives $\Wtwo^2\!\left( \law(X,R,V,Y), \law(X,R,V,Y_t) \right) \le \E(Y-Y_t)^2,$ and multiplying by $\sqrt{1+\gamma_t^2}$ yields the stated value gap.

For (iii), the projection variational inequality gives
$\E[(U_t-Z_t)(\widetilde U-U_t)]\ge0$ for every
$\widetilde U\in\mathcal C_t$. Expanding the square therefore gives
\[
  \E(\widetilde U-Z_t)^2-\E(U_t-Z_t)^2
  =\E(\widetilde U-U_t)^2
     +2\E[(U_t-Z_t)(\widetilde U-U_t)]
  \ge\E(\widetilde U-U_t)^2.
\]
This proves \eqref{eq:population-row-gap}, and hence
$d_t(\widetilde U)\ge
\sqrt{(d_t^\star)^2+\alpha\E(\widetilde U-U_t)^2}$.
Coupling the two laws with the same old row marks gives
$\Wtwo^2(\law(U,\Lambda,\widetilde U),\law(U,\Lambda,U_t))
\le\E(\widetilde U-U_t)^2$, which yields the stated value gap.

For (iv), $f_t'(\gamma_t)=0, $ $f_t''(\gamma) = \frac{\phi_t^\star}{(1+\gamma^2)^{3/2}} \ge c_\gamma,$ which proves \eqref{eq:population-scalar-gap}.  Moreover,
$U_t-Z_t=(d_t^\star/\gamma_t)\ell_t$ and
$\alpha\E\ell^2=\alpha\E\ell_t^2=\gamma_t^2$, so $\alpha\E[\ell(Z_t-U_t)]+\gamma_td_t^\star = \frac{\alpha d_t^\star}{2\gamma_t} \E(\ell-\ell_t)^2,$ proving \eqref{eq:population-transverse-angular-gap} and its
$\Wtwo$ consequence.

The bounds in Lemma~\ref{lem:app-population-multipliers} make the estimates (ii)--(iv) uniform over compact families of such inputs. Uniformity in (i) follows from Lemma~\ref{lem:app-input-continuity}.
\end{proof}

\section{Closing the Gaussian recursion}\label{sec:closing-gaussian-recursion}

In this section, we close the Gaussian recursion. In Section~\ref{subsec:finite-ao-gaps}, we transfer the population value gaps of Section~\ref{subsec:population-gaps} to the finite auxiliary problem. In Section~\ref{subsec:conditional-optimizer}, we use Gaussian comparison and the matrix coupling to localize the LP optimizer and to identify the conditional laws of the next round. In Section~\ref{subsec:new-free-set}, we show by an active-constraint count that the fraction of frozen coordinates converges as well. In Section~\ref{subsec:gaussian-recursion-closure}, we verify that the updated moments remain nondegenerate, so that the recursion can be iterated to any fixed depth, and we prove Theorem~\ref{thm:intro-state-evolution}.

Fix $t\ge1$, put $p=t-1$, and fix an input $\mathfrak h_p$ satisfying the assumptions of Section~\ref{sec:finite-po}. As in the preceding section, we write $q=q_p$, denote by $\omega_t^\star=(s_t^\star,a_t^\star,b_t^\star)$ the population optimizer and by $V_t$ the optimal value, and use the updated laws from \eqref{eq:population-transition}.

\begin{definition}[Sets of finite histories]\label{def:typical-history}
Given a deterministic sequence $\delta_N\downarrow0$, let $\mathcal T_{p,N}(\mathfrak h_p)$ be the set of histories $h\in\mathcal R_{p,N}\cap\mathcal G_{p,N}\cap\mathcal N_{p,N}$ whose empirical laws, defined in Section~\ref{subsec:complete-histories}, satisfy
\begin{equation*}
\left|\frac{|F_p|}{N}-q_p\right| +\Wtwo(\widehat\pi_{N,p}^{\rm col},\pi_p^{\rm col}) +\Wtwo(\widehat\pi_{N,p}^{\rm row},\pi_p^{\rm row}) +\Wtwo(\widehat\mu_{N,p}^{\rm col},\mu_p^{\rm col})\le\delta_N.
\end{equation*}
For $t=1$, set $\mathcal T_{0,N}=\{\varnothing\}$.
\end{definition}

Once convergence at depth $p$ is established, we choose $\delta_N$ to decrease slowly enough that $\mathsf H_p(\mathcal T_{p,N}(\mathfrak h_p))\to1$.

The probability $\Pp_h$ includes the next round's fresh variables. All conditional convergences are uniform over $h\in\mathcal T_{p,N}(\mathfrak h_p)$, up to $\mathsf H_p$-null sets, and over compact input families of fixed depth, density, and margins. Compactness uses the product $\Wtwo$ topology on the three laws and the Euclidean topology on $q$. At depth zero, $\Pp_h$ is unconditional.

\subsection{Finite gaps for the auxiliary problem}\label{subsec:finite-ao-gaps}

Fix the scalar cutoff $R_{\rm dual}<\infty$ chosen in the preceding section, so that the truncated and untruncated population optima coincide. We suppress the fixed row cutoff $R_m$ from the notation and write $d_{N,h}=d_{N,h}^{(R_m)}$. Let $\mathsf{AO}_{N,h}^{(R_{\rm dual})}$ denote the finite auxiliary value and $\mathsf{PO}_{N,h}^{(R_{\rm dual})}$ the corresponding primary value for the fixed history.

The finite hinge formula \eqref{eq:finite-hinge} has population counterpart
$\mathcal J_t(Y,\omega)=q\E[VY]-R_{\rm dual}(d_t(\omega)-q\E[GY])_+$, with $V_t=\sup_{\omega,Y}\mathcal J_t(Y,\omega)=\max_\omega\Psi_{t,R_{\rm dual}}(\omega)$ by \eqref{eq:truncated-branch-definition}.

Appendix~\ref{app:finite-population} proves the matching approximation bounds: subsequential compactness and lower semicontinuity give the upper bound, while Lemma~\ref{lem:app-finite-recovery} supplies recovery states with exact row compatibility for the lower bound. Applying compactness to restricted states transfers the population gaps.

\begin{proposition}[Finite AO approximation and localization]\label{prop:finite-ao-gaps}
  Uniformly over $h\in\mathcal T_{p,N}(\mathfrak h_p)$,
  $\frac1N\mathsf{AO}_{N, h}^{(R_{\rm dual})} \xrightarrow{\Pp_h} V_t.$

  Let $\mathcal B_{N,\eps}$ be the finite analogue of any of the parameter, column-law, or row-law bad sets in Proposition~\ref{prop:population-restricted-gaps}. For every fixed $0<\eps'<\eps$, there is a constant $c_t^{\mathrm{bad}}(\eps,\eps')>0$, uniform over the prescribed family of finite histories, such that
  \begin{equation}\label{eq:app-finite-restricted-gap}
    \Pp_h\left(
    \frac1N
    \mathsf{AO}_{N,h}^{(R_{\rm dual}),\mathrm{bad}}
    >V_t-c_t^{\mathrm{bad}}(\eps,\eps')
    \right)
    \longrightarrow0.
  \end{equation}
\end{proposition}

\subsection{From finite gaps to the conditional optimizer}
\label{subsec:conditional-optimizer}

Proposition~\ref{prop:finite-ao-gaps} supplies the downward gaps for maximizing variables. For dual variables we use finite primal test points in the dual-first AO to obtain upward gaps, remove the cutoffs, and identify both the transverse and old-history components of the multiplier.

\begin{theorem}[One-step conditional localization]
\label{thm:conditional-localization}
For the round-$t$ iterated LP,
\begin{align}
  \frac1N\Phi_{N,t}
  &\xrightarrow{\Pp_h}V_t,
  \label{eq:conditional-primary-value}\\*[-1mm]
  \omega_{N,t}
  &\xrightarrow{\Pp_h}\omega_t^\star,
  \label{eq:conditional-order-parameter}\\*[-1mm]
  \gamma_{N,t}
  :=
  \frac{\|P_{L_p}^\perp\lambda^{(t)}\|}{\sqrt N}
  &\xrightarrow{\Pp_h}\gamma_t.
  \label{eq:conditional-dual-norm}
\end{align}
Before restriction to the new free set,

\begin{gather}
\label{eq:conditional-column-law}
  \Wtwo\left(
    \frac1{|F_p|}
    \sum_{j\in F_p}
    \delta_{(x_j^{(1:p)},r_j^{(1:p)},v_{t,j},x_j^{(t)})},
    \law(X,R,V,Y_t)
  \right)
  \xrightarrow{\Pp_h}0,
 \\
\label{eq:conditional-row-law}
  \Wtwo\left(
    \frac1M\sum_{i=1}^M
    \delta_{(u_i^{(1:t)},\lambda_i^{(1:t)})},
    \law(U_{1:t},\Lambda_{1:t})
  \right)
  \xrightarrow{\Pp_h}0.
\end{gather}

These convergences are uniform over the histories of Definition~\ref{def:typical-history}, in the sense specified after that definition.
\end{theorem}

\begin{proof}
We first work with the compactified saddle problem
\eqref{eq:compact-po}, with the cutoffs fixed before $N\to\infty$.
Write $(\widetilde y_N,\widetilde m_N,\widetilde\nu_N)$ for a compact
saddle point and $\widetilde\omega_N$ for its order parameter. Let
$\widetilde x_N$ agree with $\widetilde y_N$ on $F_p$ and with the
old iterate on $B_p$.
\begin{enumerate}[label=\textnormal{(\roman*)}, leftmargin=*, itemsep=0.8em]

\item
We begin with the maximizing variables.  For the parameter,
column-law, and row-margin restrictions,
Proposition~\ref{prop:finite-ao-gaps} gives auxiliary values strictly
below $V_t$.  Gaussian comparison (Corollary~\ref{cor:prelim-cgmt-localization}(ii)) transfers these downward gaps to the
corresponding restricted primary problems, and the conditional matrix
coupling transfers them to the actual conditional law of the matrix.
The unrestricted comparison, using both tails and the finite recovery
sequence, gives $N^{-1}\Phi_{N,h}^{(R_{\rm dual},R_m)}\to V_t$.
An optimizer in a bad set would make the restricted and unrestricted
primary values equal, contradicting the corresponding gap.
Thus $\widetilde\omega_N\to\omega_t^\star$, and the compact column
law satisfies \eqref{eq:conditional-column-law} with
$x_{F_p}^{(t)}$ replaced by $\widetilde y_N$. The same argument gives
\[
  \Wtwo\left(
    \frac1M\sum_{i=1}^M
      \delta_{(U_{p,i},L_{p,i},\widetilde m_{N,i})},
    \law(U,\Lambda,U_t)\right)
  \xrightarrow{\Pp_h}0.
\]
These convergences hold in conditional probability, uniformly over
the prescribed histories.

\item
We next localize the norm of the transverse dual variable. Fix $\eps>0$ and let $J$ be
either nonempty interval
$[0,\gamma_t-\eps]\allowbreak\quad\text{or}\allowbreak\quad [\gamma_t+\eps,R_{\rm dual}].$
Let $a$ be its endpoint closest to $\gamma_t$.

With the rotated Gaussian
$S_N^a=\frac{v_t+a\widetilde g}{\sqrt{1+a^2}},\allowbreak\quad \widetilde g=\Sigma_{p,F_p}^{1/2}g,$
we repeat the finite construction of
Lemma~\ref{lem:app-finite-recovery} to obtain a primal test point
$y_N^a$ and its exact row projection $m_N^a$.  Their normalized
correlations and row distance satisfy
$c_{v,N}^a\to\frac{\phi_t^\star}{\sqrt{1+a^2}}, \allowbreak\quad c_{g,N}^a\to\frac{a\phi_t^\star}{\sqrt{1+a^2}}, \allowbreak\quad d_N^a\to d_t^\star.$
Thus, uniformly for $\gamma\in J$,
\[
  H_N^a(\gamma)
  :=
  c_{v,N}^a+\gamma(c_{g,N}^a-d_N^a)
  =
  \frac{\phi_t^\star}{\sqrt{1+a^2}}
  +
  \gamma\left(
    \frac{a\phi_t^\star}{\sqrt{1+a^2}}-d_t^\star
  \right)
  +o_{\Pp_h}(1).
\]
Let $\mathcal A_{N,h}^{\mathrm{dual}}(J)$ denote the normalized
dual-first AO restricted to $\|\nu\|/\sqrt N\in J$. The primal test
pair $(y_N^a,m_N^a)$ gives a lower bound on its inner supremum, and
hence on the restricted dual-first value:
$\mathcal A_{N,h}^{\mathrm{dual}}(J) \ge\inf_{\gamma\in J}H_N^a(\gamma).$
The first-order condition
$d_t^\star=\gamma_t\phi_t^\star/\sqrt{1+\gamma_t^2}$ determines the
slope of the limiting affine function $H_N^a$. Since
$\gamma/\sqrt{1+\gamma^2}$ is strictly increasing, this slope is
negative when $a<\gamma_t$ and positive when $a>\gamma_t$.
In either forbidden interval the infimum is therefore attained at
its endpoint $a$ nearest to $\gamma_t$, where the limiting value is
$f_t(a)$. Hence, by
\eqref{eq:population-scalar-gap},
$\mathcal A_{N,h}^{\mathrm{dual}}(J) \ge V_t+\frac{c_\gamma}{2}\eps^2-o_{\Pp_h}(1).$
The lower-tail Gaussian comparison and the matrix coupling exclude
both forbidden intervals for the compact minimizing variable, so
$\|\widetilde\nu_N\|/\sqrt N\xrightarrow{\Pp_h}\gamma_t$.

We use this norm localization to remove the compactification. Choose a
fixed $\eps<(R_{\rm dual}-\gamma_t)/2$. With high conditional
probability, $\widetilde\nu_N$ is in the interior of
the dual ball. First-order optimality there and the row compatibility
constraint give, respectively,
$Q_{L,p}^{\top}(A\widetilde x_N-\widetilde m_N)=0, \allowbreak\quad L_p^{\top}(A\widetilde x_N-\widetilde m_N)=0.$
These two sets of directions span $\R^M$, so
$\widetilde m_N=A\widetilde x_N$. On the operator-norm event,
$\|\widetilde m_N\|=\|A\widetilde x_N\| \le(C_A+1)\sqrt N<R_m\sqrt N,$
and the row cutoff is inactive as well.

Every exact LP feasible point, paired with $m=Ax$, is feasible for
the compact problem on this event, and its payoff is independent
of $\nu$. Conversely, the compact optimizer is now LP feasible,
and its saddle value is $v_{t,F_p}^{\top}\widetilde y_N$.
Together, these observations show that the compact and free LP
values are equal.

For $t=1$, the analysis of~\cite{LSZ2025} gives
$N^{-1}\Phi_{N,1}\xrightarrow{\Pp}V_1>0$ under
\eqref{eq:stage1-feasibility}, so Lemma~\ref{lem:s1-general-position}
gives KKT regularity and uniqueness of the primal and reduced dual
optimizers with high probability.

Restoring the multiplier of the compatibility equality in
the compact KKT conditions gives the reduced LP KKT conditions;
the row-norm constraint contributes no multiplier. In particular,
the transverse component of the row multiplier is
$Q_{L,p}\widetilde\nu_N$, which proves
\eqref{eq:conditional-dual-norm}. We may now identify the compact
primal optimizer with $x^{(t)}$ and its row variable with
$u^{(t)}=Ax^{(t)}$. Part (i) consequently proves
\eqref{eq:conditional-order-parameter},
\eqref{eq:conditional-column-law}, and
\begin{equation}
\label{eq:conditional-margin-law}
  \Wtwo\left(
    \frac1M\sum_{i=1}^M
    \delta_{(U_{p,i},L_{p,i},u_i^{(t)})},
    \law(U,\Lambda,U_t)
  \right)
  \xrightarrow{\Pp_h}0.
\end{equation}
Finally,
$\Phi_{N,t}=\Phi_{N,h}^{(R_{\rm dual},R_m)}
+v_{t,B_p}^{\top}x_{B_p}^{(p)}$ on this event.
The uniform estimate after \eqref{eq:conditional-po} now gives
\eqref{eq:conditional-primary-value}.

\item
We next identify the direction, as well as the norm, of the transverse
multiplier $\ell_{N,t}:=P_{L_p}^\perp\lambda^{(t)}$. Take the finite near-optimal
pair $(y_N^\star,m_N^\star)$ from
Lemma~\ref{lem:app-finite-recovery}, write
$Z_N^\star=Z_N(\omega_N^\star)$, and set
$d_N=\frac{\|m_N^\star-Z_N^\star\|}{\sqrt N},\allowbreak\quad \bar\ell_N=\frac{\gamma_t}{d_N}(m_N^\star-Z_N^\star)$.
Then $L_p^\top\bar\ell_N=0$ and
$\|\bar\ell_N\|^2/N=\gamma_t^2$. The marked empirical law of
$\bar\ell_N$ converges to that of
$\ell_t=\frac{\gamma_t}{d_t^\star}(U_t-Z_t)$.

For any transverse $\ell$, with
$\gamma=\|\ell\|/\sqrt N$, a direct expansion shows that the AO objective at
the same primal test pair equals
\begin{equation}
\label{eq:finite-transverse-gap}
  c_{v,N}+\gamma(c_{g,N}-d_N)
  +
  \frac{d_N}{2\gamma_t}
  \left\{
    \frac{\|\ell-\bar\ell_N\|^2}{N}
    -(\gamma-\gamma_t)^2
  \right\}.
\end{equation}
The first two terms equal $V_t+o_{\Pp_h}(1)$ for bounded $\gamma$.
If the marked law of $\ell$ remains at $\Wtwo$ distance at least
$\eps$ from $\law(U,\Lambda,\ell_t)$, then $\frac{\|\ell-\bar\ell_N\|^2}{N} \ge \frac MN(\eps-o_{\Pp_h}(1))^2.$ Choose $\delta>0$ with $\delta^2<\alpha\eps^2/4$.  On the restricted
set $|\gamma-\gamma_t|\le\delta$, the last term in
\eqref{eq:finite-transverse-gap} is bounded below by a positive
constant with high probability.

The lower-tail comparison therefore excludes the multiplier-law bad set in this region, while
\eqref{eq:conditional-dual-norm} excludes its complement.  Hence
\begin{equation}
\label{eq:conditional-transverse-multiplier-law}
  \Wtwo\left(
    \frac1M\sum_{i=1}^M
    \delta_{(U_{p,i},L_{p,i},\ell_{N,t,i})},
    \law(U,\Lambda,\ell_t)
  \right)
  \xrightarrow{\Pp_h}0.
\end{equation}

\item
For $p=0$, the full multiplier is already transverse. For $p\ge1$,
it remains to recover the component in $\operatorname{span}(L_p)$.
We first bound its coefficient and then identify its limit using dual
feasibility and complementarity.
Write $\lambda^{(t)}=\ell_{N,t}+L_pc_N$ and set
$c_\star:=\frac{\gamma_t}{d_t^\star}\eta_{\lambda,t},\allowbreak\quad \ell_t=\Lambda_t-\Lambda^\top c_\star.$
Let $\chi(s)=\min\{1,s_+\}$ and
$D_N=\diag(\chi(u_i^{(t)}-\kappa_t))$.  Since complementarity forces
the multiplier to vanish on rows with positive slack,
$D_N\lambda^{(t)}=0$.  Moreover,
\[
  B_N
  :=
  \frac1N L_p^\top D_NL_p
  \xrightarrow{\Pp_h}
  B
  :=
  \alpha\E[
    \chi(U_t-\kappa_t)\Lambda\Lambda^\top
  ]\succ0.
\]
Indeed, conditionally on the old row marks, the fresh Gaussian gives
the event $U_t>\kappa_t$ positive probability, while $Q^\lambda\succ0$.
It follows that $c_N^\top B_Nc_N \le \frac{\|\ell_{N,t}\|^2}{N},$ and hence $c_N=O_{\Pp_h}(1)$.

It remains to identify the limit of the coefficient. Along any convergent subsequence,
say $c_N\to c$, put
$\Delta_c:=\Lambda^\top(c-c_\star)$. Since
\[
  \ell_t=-\Lambda^\top c_\star+\frac{\gamma_t}{d_t^\star}\zeta_t,
  \qquad
  \zeta_t=(\kappa_t-Z_t+\eta_{\lambda,t}^\top\Lambda)_+\ge0,
\]
the limit of the full multiplier is
$\lambda=\Delta_c+\frac{\gamma_t}{d_t^\star}\zeta_t.$
Conditionally on $(U,\Lambda)$, the event $\{\zeta_t=0\}$ has positive
probability, so dual feasibility gives $\Delta_c\ge0$.  If
$\Delta_c>0$, then $\lambda>0$ for every realization of the fresh
Gaussian, and complementarity forces $U_t=\kappa_t$ almost surely
conditionally on $(U,\Lambda)$, contradicting
$\Pp(U_t>\kappa_t\mid U,\Lambda)>0$.  Hence $\Delta_c=0$ almost surely.
Since $Q^\lambda\succ0$, we obtain $c=c_\star$, and therefore
$c_N\xrightarrow{\Pp_h}c_\star$.

Combining this with
\eqref{eq:conditional-transverse-multiplier-law} yields
\begin{equation*}
  \Wtwo\left(
    \frac1M\sum_{i=1}^M
    \delta_{(U_{p,i},L_{p,i},\lambda_i^{(t)})},
    \law(U,\Lambda,\Lambda_t)
  \right)
  \xrightarrow{\Pp_h}0.
\end{equation*}

Finally, complementarity gives $\lambda=0$ or
$u=\kappa_t$. Hence, conditional on the old marks, the pair $(u,\lambda)$ lies on $\{\lambda=0,\ u>\kappa_t\}\cup\{u=\kappa_t,\ \lambda>0\}\cup\{(\kappa_t,0)\}$. The law of $u$ determines the mass on the first set, the law of $\lambda$ determines the mass on the second, and the remaining mass sits at $(\kappa_t,0)$. Thus the two marked marginals $(U,\Lambda,u)$ and
$(U,\Lambda,\lambda)$ determine the joint row law.  Convergence of
second moments upgrades this to $\Wtwo$ convergence, which proves
\eqref{eq:conditional-row-law}.

\end{enumerate}
\end{proof}

\subsection{Exact freezing and restriction to the new free set}
\label{subsec:new-free-set}

We now count the coordinates that reach the boundary and restrict the entering free-column law to those that remain free.

\begin{lemma}[Finite active-count inequality]
\label{lem:finite-active-count}
Let $\mathcal I_t^{\rm act} := \{i\in[M]:u_i^{(t)}=\kappa_t\}.$ On the strict-feasibility event, almost surely over $v_t$,
\begin{equation}
\label{eq:finite-active-count}
  |F_{t-1}|
  \le
  |C_t|+|\mathcal I_t^{\rm act}|.
\end{equation}
\end{lemma}

\begin{proof}
Perturbing a strictly row-feasible point into the relative interior of the entering box preserves the row inequalities, so the feasible polytope has dimension $|F_{t-1}|$. The absolutely continuous objective almost surely exposes a vertex. Its active box and row normals span $\R^{|F_{t-1}|}$ and are indexed by $C_t$ and $\mathcal I_t^{\rm act}$, giving \eqref{eq:finite-active-count}.
\end{proof}

\begin{proposition}[Exact one-step freezing]
\label{prop:one-step-exact-freezing}
\[
  \frac{|C_t|}{N}
  \xrightarrow{\Pp_h}
  q_{t-1}\Pp(|Y_t|=1)
  =
  q_{t-1}-q_t,
  \qquad
  \frac{|F_t|}{N}
  \xrightarrow{\Pp_h}
  q_t.
\]
Moreover,
\[
  \Wtwo\left(
    \frac1{|F_t|}
    \sum_{j\in F_t}
    \delta_{(x_j^{(1:t)},r_j^{(1:t)})},
    \pi_t^{\rm col}
  \right)
  \xrightarrow{\Pp_h}0.
\]
\end{proposition}

\begin{proof}
Write $q=q_{t-1}$ and $I_t=\mathcal I_t^{\rm act}$.  By
\eqref{eq:conditional-column-law} and
\eqref{eq:conditional-margin-law}, the portmanteau theorem applied to
the closed sets $\{|y|=1\}$ and $\{\kappa_t\}$ gives
\begin{equation*}
  \operatorname*{p-limsup}\frac{|C_t|}{N}
  \le
  q\,\Pp(|Y_t|=1)=q-q_t,
  \qquad
  \operatorname*{p-limsup}\frac{|I_t|}{N}
  \le
  \alpha\Pp(U_t=\kappa_t)=q_t,
\end{equation*}
where the last equality follows from Lemma~\ref{lem:free-row-balance}.
Together with \eqref{eq:finite-active-count} and
$|F_{t-1}|/N\to q$, these bounds give
\[
  \operatorname*{p-liminf}\frac{|C_t|}{N}
  \ge q-\operatorname*{p-limsup}\frac{|I_t|}{N}
  \ge q-q_t.
\]
Thus
$\frac{|C_t|}{N}\xrightarrow{\Pp_h}q-q_t,\allowbreak\quad \frac{|F_t|}{N}=\frac{|F_{t-1}|-|C_t|}{N} \xrightarrow{\Pp_h}q_t.$

It remains to restrict the column law to $F_t$.  Let
$\widehat\vartheta_{N,t}$ denote the empirical measure in
\eqref{eq:conditional-column-law}, and set $\vartheta_t:=\law(X,R,V,Y_t), $ $S:=\{|y|=1\}.$ Then
$\Wtwo(\widehat\vartheta_{N,t},\vartheta_t) \xrightarrow{\Pp_h}0, \allowbreak\quad \widehat\vartheta_{N,t}(S) = \frac{|C_t|}{|F_{t-1}|} \xrightarrow{\Pp_h} \vartheta_t(S)$.
Approximate $\1_{S^c}$ by continuous cutoffs vanishing near $S$. For bounded continuous tests, the error is bounded by the mass near $S$ minus the mass on $S$; it vanishes by weak convergence and convergence of boundary masses, first as $N\to\infty$ and then as the neighborhood shrinks. Thus the restrictions converge weakly. Their squared marks are uniformly integrable by the full $\Wtwo$ convergence, giving $\Wtwo$ convergence after normalization.

Moreover, $\tau_t>0$ and the update $Y_t$ contains a nondegenerate
fresh Gaussian component, so $\Pp(|Y_t|<1)>0, $ $q_t=q\,\Pp(|Y_t|<1)>0.$ We may therefore normalize the restrictions and obtain $\law(X,R,V,Y_t\mid |Y_t|<1)$ as the limiting law on $F_t$.  Finally, for $j\in F_t$ the box
constraint is inactive, and the KKT stationarity condition gives
$r_j^{(t)}=-v_{t,j}$.  Applying the continuous map $(X,R,V,Y_t)\longmapsto ((X,Y_t),(R,-V))$ yields the asserted convergence to $\pi_t^{\rm col}$.
\end{proof}

\subsection{Induction over rounds}
\label{subsec:gaussian-recursion-closure}

Theorem~\ref{thm:conditional-localization} and
Proposition~\ref{prop:one-step-exact-freezing} identify the updated
empirical laws. It remains to verify convergence of the updated
moments and preservation of the strict-history conditions.

\begin{theorem}[One-step Gaussian recursion]
\label{thm:explicit-population-recursion}
Let $t\ge1$, put $p=t-1$, and let $\mathfrak h_p$ be an
input for round $t$ satisfying the assumptions of Section~\ref{sec:finite-po}.
Let $\mathfrak h_t$ be its image under the
population transition \eqref{eq:population-transition}. Let
$h_{N,p}\in\mathcal T_{p,N}(\mathfrak h_p)$, with the history set of Definition~\ref{def:typical-history}.

Uniformly over this family and conditionally on $h_{N,p}$, the following convergences hold, with the last one in any fixed matrix norm:

\begin{gather}
\label{eq:gaussian-recursion-scalar-limits}
\left(
  \frac{\Phi_{N,t}}N,\,
  \omega_{N,t},\,
  \gamma_{N,t},\,
  \frac{|F_t|}{N}
\right)
\xrightarrow{\Pp_h}
\left(
  V_t,\,
  \omega_t^\star,\,
  \gamma_t,\,
  q_t
\right).
 \\
\label{eq:gaussian-recursion-law-limits}
\Wtwo(\widehat\pi_{N,t}^{\rm col},\pi_t^{\rm col})
+
\Wtwo(\widehat\pi_{N,t}^{\rm row},\pi_t^{\rm row})
+
\Wtwo(\widehat\mu_{N,t}^{\rm col},\mu_t^{\rm col})
\xrightarrow{\Pp_h}0,
 \\
\label{eq:gaussian-recursion-moment-limits}
\bigl(
Q_{N,t}^x,\,
Q_{N,t,F}^x,\,
Q_{N,t,F}^{rx},\,
Q_{N,t}^\lambda,\,
\Gamma_{N,t},\,
D_{N,t}^x
\bigr)
\xrightarrow{\Pp_h}
\bigl(
Q_t^x,\,
Q_{t,F}^x,\,
Q_{t,F}^{rx},\,
Q_t^\lambda,\,
\Gamma_t,\,
D_t^x
\bigr).
\end{gather}

Finally,
\begin{equation*}
q_t>0,\qquad
\operatorname{Schur}(Q_t^x\mid Q_{t-1}^x)=\rho_t^2>0,\qquad
\operatorname{Schur}(Q_t^\lambda\mid Q_{t-1}^\lambda)=\gamma_t^2>0,
\qquad
D_t^x\succ0.
\end{equation*}
Hence $Q_t^x\succ0$ and $Q_t^\lambda\succ0$, and, if $t<K$,
$\mathfrak h_t$ is a strict input for round $t+1$ in the sense of Definition~\ref{def:strict-history}.
\end{theorem}

\begin{proof}
\begin{enumerate}[label=\textnormal{(\roman*)}, leftmargin=*, itemsep=0.8em]

\item Theorem~\ref{thm:conditional-localization} and Proposition~\ref{prop:one-step-exact-freezing} give \eqref{eq:gaussian-recursion-scalar-limits} and the convergence of $\widehat\pi_{N,t}^{\rm col}$ and $\widehat\pi_{N,t}^{\rm row}$ in \eqref{eq:gaussian-recursion-law-limits}. It remains to treat the full column law, which is supported in $[-1,1]^t$, so that $\Wtwo$ convergence is equivalent to weak convergence. We split
\[
  \widehat\mu_{N,t}^{\rm col}
  =\frac1N\sum_{j\in B_p}\delta_{x_j^{(1:t)}}
  +\frac1N\sum_{j\in F_p}\delta_{x_j^{(1:t)}}.
\]
For $j\in B_p$, we have $x_j^{(t)}=x_j^{(p)}$ by~\eqref{eq:freeze-old}, so the first sum is the image of the restriction of $\widehat\mu_{N,p}^{\rm col}$ to the closed set $\{|z_p|=1\}$ under $z\mapsto(z,z_p)$. Since $h_{N,p}\in\mathcal T_{p,N}(\mathfrak h_p)$, we have $\widehat\mu_{N,p}^{\rm col}\to\mu_p^{\rm col}$ and $\widehat\mu_{N,p}^{\rm col}(\{|z_p|=1\})=|B_p|/N\to1-q_p=\mu_p^{\rm col}(\{|z_p|=1\})$. The restriction argument in the proof of Proposition~\ref{prop:one-step-exact-freezing} therefore shows that these restrictions converge. For $j\in F_p$, the trajectory $x_j^{(1:t)}$ consists of the old primal marks and $x_j^{(t)}$, so the second sum is $|F_p|/N$ times the image of the empirical measure in~\eqref{eq:conditional-column-law} under $(x,r,v,y)\mapsto(x,y)$. Since coordinate projections do not increase $\Wtwo$, it converges to $q_p\law(X,Y_t)$. Adding the two limits gives $\mu_t^{\rm col}$ by~\eqref{eq:full-column-law-update}.

\item
The marked $\Wtwo$ convergences, together with
$|F_t|/N\to q_t$ and $M/N\to\alpha$, give convergence of all second
moments, including the mixed ones.  In particular, $Q_{N,t,F}^x\to Q_{t,F}^x,$ $Q_{N,t,F}^{rx}\to Q_{t,F}^{rx},$ $Q_{N,t}^\lambda\to Q_t^\lambda,$ $\Gamma_{N,t}\to\Gamma_t.$ For the full primal Gram matrix, the finite order parameters give
$\frac1N X_p^\top x^{(t)} = Q_{N,p}^xe_p+a_{N,t}, \allowbreak\quad \frac1N\|x^{(t)}\|^2 = s_{N,t}+(D_{N,p}^x)_{pp}.$
Thus $\omega_{N,t}\to\omega_t^\star$ and the induction hypothesis
yield the update \eqref{eq:primal-gram-update} in the limit, and hence
$Q_{N,t}^x\to Q_t^x$.  Since
$D_{N,t}^x=Q_{N,t}^x-Q_{N,t,F}^x$, this proves
\eqref{eq:gaussian-recursion-moment-limits}.

\item
Finally, we verify that the updated population state satisfies the conditions of Definition~\ref{def:strict-history}.
Since $\tau_t>0$ and \eqref{eq:canonical-primal-update} contains a
nondegenerate fresh Gaussian, $q_t=q_{t-1}\Pp(|Y_t|<1)>0.$ Lemma~\ref{lem:app-population-multipliers} gives
$\rho_t>0$ and $d_t^\star>0$, and hence $\gamma_t>0$.
From \eqref{eq:primal-gram-update} and
\eqref{eq:population-rho}, $\operatorname{Schur}(Q_t^x\mid Q_{t-1}^x) = \rho_t^2>0.$

For the dual matrix, recall that $\ell_t = \frac{\gamma_t}{d_t^\star}(U_t-Z_t), $ $\alpha\E[\Lambda\ell_t]=0, $ $\alpha\E\ell_t^2=\gamma_t^2.$ Equations~\eqref{eq:canonical-row-residual} and
\eqref{eq:canonical-dual-update} give $\Lambda_t = \frac{\gamma_t}{d_t^\star}\eta_{\lambda,t}^\top\Lambda+\ell_t.$ Thus $\ell_t$ is the component of $\Lambda_t$ orthogonal to
the old dual history, and therefore $\operatorname{Schur}(Q_t^\lambda\mid Q_{t-1}^\lambda) = \alpha\E\ell_t^2 = \gamma_t^2>0.$

To prove $D_t^x\succ0$, use the increment coordinates
$(x_1,x_2-x_1,\ldots,x_t-x_{t-1})$. Previously frozen coordinates
contribute a positive-definite
depth-$p$ block and a zero last row and column; this block is empty
when $p=0$. The newly frozen coordinates contribute a
positive-semidefinite matrix whose last diagonal entry is
$q_{t-1}\E\!\left[(Y_t-X_p)^2\1_{\{|Y_t|=1\}}\right]>0.$
Indeed, $|X_p|<1$ on the entering free set, and the finite positive
clipping threshold and fresh Gaussian give
$\Pp(|Y_t|=1\mid X,R)>0$.

To check the kernel, write a vector in
increment coordinates as $z=(a,b)\in\R^p\times\R$. If its quadratic
form against the sum is zero, both nonnegative contributions vanish.
The old positive-definite block first gives $a=0$, and the new
contribution then equals
$b^2q_{t-1}\E[(Y_t-X_p)^2\1_{\{|Y_t|=1\}}]$, forcing $b=0$.
When $p=0$ only the latter step is needed. Thus $D_t^x\succ0$ for
every $t\ge1$. In particular, the initial output satisfies
$Q_1^x=\rho_1^2>0,\allowbreak\quad Q_1^\lambda=\gamma_1^2>0,\allowbreak\quad D_1^x=\E[Y_1^2\1_{\{|Y_1|=1\}}]=1-q_1>0.$

The update formulas preserve finite second moments.
Equation~\eqref{eq:full-column-law-update} preserves frozen coordinates
and consistency of the full and free laws. The row formulas
\eqref{eq:canonical-row} and \eqref{eq:canonical-dual-update} give
$U_t\ge\kappa_t,\allowbreak\quad \Lambda_t\ge0,\allowbreak\quad \Lambda_t(U_t-\kappa_t)=0\quad\text{almost surely}.$
Together with the positive free fraction and the matrix bounds above,
this verifies Definition~\ref{def:strict-history}. For $t<K$, the margin
drop $\kappa_t>\kappa_{t+1}$ makes $\mathfrak h_t$ a strict input for
round $t+1$.
\end{enumerate}
\end{proof}

\begin{proof}[Proof of Theorem~\ref{thm:intro-state-evolution}]
Start with $\mathfrak h_0$ and $\mathcal T_{0,N}=\{\varnothing\}$.
Lemma~\ref{lem:positive-branch} gives $V_1>0$, and the one-step theorem gives a positive frozen fraction, so
Lemma~\ref{lem:s1-general-position} gives regular KKT pairs along
the feasible path with high probability. At each fixed
depth, moment convergence and positive definiteness give full rank
of $X_p,L_p$, and the coupling bounds give
$\mathcal G_{p,N}\cap\mathcal N_{p,N}$ with high probability.
Choosing $\delta_N\downarrow0$ sufficiently slowly therefore places
the empirical history in $\mathcal T_{p,N}(\mathfrak h_p)$ with high
probability. Iterate the one-step theorem, integrate
the uniform conditional convergences, and take a union bound over
the fixed number $K$ of rounds. The row law $\widehat\mu_{N,t}^{\rm row}$ of Theorem~\ref{thm:intro-state-evolution} is the margin marginal of $\widehat\pi_{N,t}^{\rm row}$, and coordinate projections do not increase $\Wtwo$. Finally, $|B_t|=N-|F_t|$, so $|F_t|/N\to q_t$ gives $|B_t|/N\to1-q_t$.
\end{proof}

\section{Numerical evaluation}\label{sec:numerical}

In this section, we evaluate the recursion numerically at density $\alpha=0.60$. In Section~\ref{subsec:num-publication-finishing}, we make the condition of Theorem~\ref{thm:intro-algorithm} explicit by computing the margin that the rounding stage consumes. In Section~\ref{subsec:num-empirical-060}, we evaluate the recursion along a margin schedule, and the computed recursion satisfies this condition at zero margin at round $119$. In Section~\ref{subsec:num-finite-060}, we compare the recursion with \IterLP\ run on finite Gaussian matrices, and in Section~\ref{subsec:num-implementation}, we describe how we compute the recursion. Code and data are available at~\coderepo.

\subsection{The condition of Theorem~\ref{thm:intro-algorithm}}
\label{subsec:num-publication-finishing}

After the LP rounds, the rounding stage of~\cite{LSZ2025} applies the Gaussian-tail partial-coloring estimate of~\cite[Theorem~3.11]{LSZ2025} with constants $K_1=21/5$, $K_2=100$, and $K_3=4/3$. With these constants, each successful partial-coloring step makes at least three quarters of the remaining free coordinates nearly tight and lowers every row margin by at most $\sqrt{2Tr\log(200a/r)}$, where $r$ is the current free fraction. Summing these losses along the geometric sequence $r=q4^{-j}$ gives the total rounding cost
\begin{equation}\label{eq:rounding-cost}
 \mathcal D_{a,T}(q)=\sqrt{2Tq}
 \left(2\sqrt{\log(200a/q)}+
       \frac{\log4}{\sqrt{\log(200a/q)}}\right),
 \qquad T=\frac{21}{5},
\end{equation}
for $a\ge\max\{1,\alpha\}$ and $0<q\le1$. Since $\mathcal D_{a,T}$ increases with $a$, the best of these bounds is $\mathcal D_\alpha=\mathcal D_{\max\{1,\alpha\},21/5}$. If the LP rounds end at margin $\kappa_K$ with free fraction $q_K$ and $\kappa_K-\mathcal D_\alpha(q_K)>\kappa$, then the rounding stage reaches target margin $\kappa$ under every subgaussian entry law (Theorem~\ref{thm:intro-algorithm}, proved as Corollary~\ref{cor:cold-rounding-transfer}). At target margin $0$, we call $\kappa_t-\mathcal D_\alpha(q_t)$ the \emph{reserve} at round $t$.

\subsection{A margin schedule at density 0.60}
\label{subsec:num-empirical-060}

At density $\alpha=0.60$, we use the quadratic schedule
\begin{equation*}
 \kappa_t=\frac1{100}+c_0\left(\frac{150-t}{149}\right)^2,
 \qquad 1\le t\le140,
 \qquad c_0=0.5789153654719907.
\end{equation*}
Its first margin $\kappa_1\approx0.5889$ is $0.998$ times the numerically computed feasibility limit of the first LP. Lowering the margin in each round lets a further share of the free coordinates freeze, while the remaining margin must still cover the rounding cost. We fix the stopping rule in advance. A run stops at its first numerically resolved round (Section~\ref{subsec:num-implementation}) with
\begin{equation}\label{eq:num-cold-stop}
 \kappa_t-\mathcal D_\alpha(\widehat q_t)>10^{-3},
\end{equation}
where $\widehat q_t$ is the computed free fraction, and a run that has not stopped by round $140$ counts as a failure.

\begin{figure}[!htb]
\centering
\includegraphics[width=0.6\textwidth]{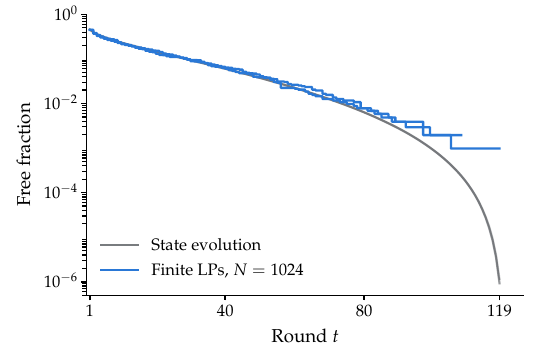}
\caption{The free fraction along the density-$0.60$ schedule, for the state-evolution recursion (gray, the three runs of Table~\ref{tab:num-060-stopping-replicas}) and for the two feasible finite LP paths with $N=1024$ (blue), drawn until every coordinate is frozen.}
\label{fig:num-060-trajectories}
\end{figure}

We evaluate the recursion with three random seeds, using $2^{14}$ column particles and $2^{10}$ or $2^{11}$ row particles per stratum (Section~\ref{subsec:num-implementation}). All three runs stop at round $119$, with smallest reserve $0.0068$ (Table~\ref{tab:num-060-stopping-replicas}). Their terminal free fractions agree within a factor of about $1.3$, and the run with twice as many row particles gives the smallest of the three. Figure~\ref{fig:num-060-trajectories} shows the free fraction decreasing to about $10^{-6}$. At round $119$, the stopping rule holds whenever the free fraction is below about $1.73\cdot10^{-6}$, while the largest computed value is $1.16\cdot10^{-6}$.

\begin{table}[htbp]
\centering
\caption{Population evaluations of the density-$0.60$ schedule. Each run stops at its first round satisfying~\eqref{eq:num-cold-stop}. Particle counts list column particles and row particles per stratum.}
\label{tab:num-060-stopping-replicas}
\begin{tabular}{rrrrrr}
\toprule
Seed & Particles & Stop $t$ & $\widehat q_t$ & $\kappa_t$ & Reserve\\
\midrule
2047 & $16\,384/1\,024$ & $119$ & $1.16306\cdot10^{-6}$ & $0.0350591$ & $0.0068420$\\
2718 & $16\,384/1\,024$ & $119$ & $9.49782\cdot10^{-7}$ & $0.0350591$ & $0.0094338$\\
3141 & $16\,384/2\,048$ & $119$ & $9.09999\cdot10^{-7}$ & $0.0350591$ & $0.0099502$\\
\bottomrule
\end{tabular}
\end{table}

\subsection{Comparison with finite linear programs}
\label{subsec:num-finite-060}

We also run \IterLP\ directly on Gaussian matrices with $N=1024$ and $M=\lfloor0.60N\rfloor=614$ for three random seeds, with the same schedule and a fresh Gaussian objective in each round. For one seed, the first LP is infeasible. The other two paths follow the recursion closely (Figure~\ref{fig:num-060-trajectories}). After rounds $1$, $2$, and $3$, they freeze on average $55.1\%$, $62.6\%$, and $66.4\%$ of the coordinates, compared with $54.3\%$, $60.1\%$, and $63.8\%$ for the recursion. On both paths, every coordinate is frozen at $\pm1$ by rounds $120$ and $109$, respectively, so \IterLP\ itself outputs a sign vector, with smallest row margins $0.0337$ and $0.0557$. The coordinatewise signs of their LP iterates first form a solution at zero margin already at rounds $44$ and $40$.

\subsection{Implementation}
\label{subsec:num-implementation}

We represent the column laws by weighted particles that carry the histories of the surviving coordinates, and the row laws by particles that carry the row histories, stratified by the last round in which the row was active. In each round, we reweight the column histories by their survival probability, resample them with scrambled Sobol points, and append a coordinate drawn from the corresponding truncated Gaussian law. Since the particles represent the law conditional on survival and the free fraction is the product of the conditional survival probabilities, free fractions far below the inverse number of particles remain resolved. We keep the Gram contributions of the frozen coordinates separately. We compute expectations over the fresh Gaussian variables in closed form through clipped and hinge moments, with a $12$-point Gauss--Legendre rule for very narrow column intervals ($\tau/\sigma<0.03$). We solve the variational problem of each round by bounded nonlinear least squares in coordinates whitened by Cholesky factors of the history Gram matrices, with analytic derivatives and continuation from the previous round.

We accept a round if the scaled residual is below $2\cdot10^{-6}$, the computed value is positive, and the free fraction decreases. For the stopping rule, we further require $0.01<\tau<10^3$, $\widehat q_t>10^{-10}$, a conditional survival probability above $10^{-5}$, and $\widehat q_t>100\,\varepsilon\,\widehat q_{t-1}$, where $\varepsilon$ is the largest scaled residual. These checks exclude unresolved scales.

The finite LPs use HiGHS~\cite{HuangfuHall2018} with primal and dual feasibility tolerances $10^{-9}$. We freeze coordinates with $|x_j|\ge1-10^{-9}$ at their signs and check row margins to tolerance $10^{-7}$. We save all iterates, and an independent script regenerates the matrices and objectives from their seeds and recomputes the row products, the freezing, and the sign margins.

\section{Universality for i.i.d.\ subgaussian matrices}\label{sec:universality}

In this section, we extend the Gaussian limits of the primal trajectories, row margins, free fractions, and LP values to i.i.d.\ subgaussian entries with the same first two moments. Since the LP optimizer and the frozen set can jump with the matrix, we cannot compare the two LP paths directly. Instead, we pass through the surrogate path of Section~\ref{subsec:overview-universality} in three steps (Figure~\ref{fig:overview-universality-chain}). We compare the Gaussian LP path with the Gaussian surrogate path, then the Gaussian surrogate path with the subgaussian surrogate path, and finally the subgaussian surrogate path with the subgaussian LP path. In Section~\ref{subsec:univ-paths}, we state the universality theorem, deduce Theorem~\ref{thm:intro-universality} from it, and define the surrogate path. In Sections~\ref{subsec:univ-inputs} and~\ref{subsec:univ-residual}, we collect auxiliary estimates and introduce a residual energy for the surrogate path, and in Section~\ref{subsec:univ-proof}, we prove the theorem. In Section~\ref{subsec:univ-stage2}, we transfer the rounding stage to subgaussian matrices, and in Appendix~\ref{app:cold-rounding-transfer}, we prove Theorem~\ref{thm:intro-algorithm}.

Fix $K<\infty$ and let $M/N\to\alpha\in(0,\infty)$. Write $A^G=G/\sqrt N$ and $A^Y=Y/\sqrt N$, where the entries $G_{ij}$ are i.i.d.\ standard Gaussian and the entries $Y_{ij}$ are i.i.d.\ copies of a random variable $\xi$ with $\E\xi=0$, $\E\xi^2=1$, and $\|\xi\|_{\psi_2}<\infty$. The two designs use the same independent standard Gaussian objectives $v_1,\ldots,v_K$, which are independent of the matrices. The margins satisfy $\kappa_{t-1}>\kappa_t$ for $2\le t\le K$. Unless stated otherwise, all limits are in probability.

\subsection{The universality theorem and the surrogate path}\label{subsec:univ-paths}

For design $A$, write $\widehat x_A^{(t)}$ for the exact iterates in \eqref{eq:freeze-old}--\eqref{eq:roundt} and set
$\widehat B_0=\varnothing$, $\widehat B_t=\widehat B_{t-1}\cup\{j:|\widehat x_{A,j}^{(t)}|=1\}$,
$\widehat F_t=[N]\setminus\widehat B_t$, and $\widehat u_A^{(t)}=A\widehat x_A^{(t)}$.
The sets depend on $A$, which is suppressed in their notation. Once a round is feasible, the margin drop makes its optimizer strictly row-feasible for the next round.

We retain the Gaussian limits $\mu_t^{\rm col}=\law(\bar X_1,\ldots,\bar X_t)$ from \eqref{eq:full-column-law-update} and $\mu_t^{\rm row}=\law(U_1,\ldots,U_t)$, as well as the free fraction $q_t$ and the normalized value $V_t$.

\begin{theorem}[Universality of the exact LP at finite depth]\label{thm:univ-exact}
Suppose that $(\alpha,\kappa_1)$ satisfies~\eqref{eq:stage1-feasibility} and that the Gaussian population histories through round $K$ are strict in the sense of Definition~\ref{def:strict-history}. Suppose also that, for some $\Delta_0>0$, the system $A^Gx\ge(\kappa_1+\Delta_0)\1$, $x\in[-1,1]^N$, is feasible with high probability. Then the exact path for $A^Y$ is well defined with high probability, and, jointly for $t\le K$,
\begin{align}
W_2\left(\frac1N\sum_j
\delta_{(\widehat x_{A^Y,j}^{(1)},\ldots,\widehat x_{A^Y,j}^{(t)})},
\mu_t^{\rm col}\right)&\longrightarrow0,\label{eq:univ-main-col}\\
W_2\left(\frac1M\sum_i
\delta_{(\widehat u_{A^Y,i}^{(1)},\ldots,\widehat u_{A^Y,i}^{(t)})},
\mu_t^{\rm row}\right)&\longrightarrow0,\label{eq:univ-main-row}\\
|\widehat F_t|/N&\longrightarrow q_t,\label{eq:univ-main-free}\\
N^{-1}v_t^\top\widehat x_{A^Y}^{(t)}&\longrightarrow V_t.
\label{eq:univ-main-value}
\end{align}
Consequently, empirical averages of continuous functions of these trajectories with at most quadratic growth converge to the corresponding population averages.
\end{theorem}

\begin{proof}[Proof of Theorem~\ref{thm:intro-universality}]
Condition~\eqref{eq:stage1-feasibility} is a hypothesis of Theorem~\ref{thm:intro-state-evolution}, so it remains to verify the other two hypotheses of Theorem~\ref{thm:univ-exact}. Since the feasibility condition~\eqref{eq:stage1-feasibility} is strict, the first-round argument of Section~\ref{sec:finite-po} gives $\Delta_0>0$ such that the system $A^Gx\ge(\kappa_1+\Delta_0)\1$, $x\in[-1,1]^N$, is feasible with high probability. Starting from the empty history and applying Theorem~\ref{thm:explicit-population-recursion} in each round, we find that $\mathfrak h_t$ is a strict input for round $t+1$ for every $t<K$, so the Gaussian population histories through round $K$ are strict. Theorem~\ref{thm:univ-exact} therefore applies, with the same limits $\mu_t^{\rm col}$, $\mu_t^{\rm row}$, and $q_t$ as in Theorem~\ref{thm:intro-state-evolution}. Since $\widehat x_{A^Y}^{(t)}$ and $\widehat B_t$ are the iterates and frozen sets of \IterLP, its conclusions~\eqref{eq:univ-main-col} and~\eqref{eq:univ-main-row} are the first two limits of Theorem~\ref{thm:intro-universality}, and~\eqref{eq:univ-main-free} gives $|\widehat B_t|/N=1-|\widehat F_t|/N\to1-q_t$.
\end{proof}

We approximate the hard freezing rule by a smooth memory update. Choose even $C^3$ functions $\chi_{\delta_s}:[-1,1]\to[0,1]$ that vanish on $|a|\le1-2\delta_s$, equal one on $|a|\ge1-\delta_s$, and satisfy $\|\chi_{\delta_s}^{(r)}\|_\infty\le C_r\delta_s^{-r}$ for $1\le r\le3$. For a path $x$, we start from $m^{(0)}=c^{(0)}=0$ and update
\begin{equation*}
d^{(s)} =(1-m^{(s-1)})\odot\chi_{\delta_s}(x^{(s)}), \qquad (m^{(s)},c^{(s)}) =(m^{(s-1)}+d^{(s)},\ c^{(s-1)}+d^{(s)}\odot x^{(s)}).
\end{equation*}
Here $m$ records freezing weights and $c$ their weighted coordinate values. Write $H_s(x)=(m^{(s)},c^{(s)})$ for the memory and $H_s^+(H,x)$ for one update. Coordinatewise, $0\le m^{(s)}\le1$ and $|c^{(s)}|\le m^{(s)}$. Since $\delta_s$ affects only the memory after round $s$, $\delta_K$ is arbitrary.

At round $t$, let $0<\tau_{q,t},\eta_t\le1$ and $\beta_t>0$, and set
\begin{equation*}
F^0_{t,H}(x)=-v_t^\top x+\frac{\tau_{q,t}}2\|x\|^2
+\frac{\beta_t}2\|m\odot x-c\|^2,\qquad H=(m,c).
\end{equation*}
The ridge ensures uniqueness, and the memory penalty retains frozen values. Replacing row constraints by a quadratic penalty at scale $\eta_t$, define
\begin{align}
Q_{t,A,H}(x)&=F^0_{t,H}(x)
+\frac1{2\eta_t}\|(\kappa_t\1-Ax)_+\|^2,
\label{eq:univ-penalized-objective}\\
y_A^{(t)}&=\argmin_{x\in[-1,1]^N}Q_{t,A,H_{t-1}(y_A)}(x),
\label{eq:univ-stabilized-path}\\
\lambda_A^{(t)}&=\eta_t^{-1}(\kappa_t\1-Ay_A^{(t)})_+.
\label{eq:univ-soft-multiplier}
\end{align}
We call $y_A$ the surrogate path. Its stability under deletion of a row or zeroing of a column is quantified below. The approximation parameters will be chosen in the induction in Section~\ref{subsubsec:univ-parameter-choice}.

For comparison with the exact LP, define the linear row penalty with deterministic coefficient $\ell_t>0$ by
$F^\ell_{t,A,H}(x)=F^0_{t,H}(x) +\ell_t\|(\kappa_t\1-Ax)_+\|_1.$
For large finite $\ell_t$ and $\beta_t$, this objective retains the Gaussian LP localization gap; its bounded hinge slope permits entrywise replacement after smoothing. Thus $Q$ defines the stable path and $F^\ell$ compares it with the LP:
\begin{equation}
\widehat x_{A^G}
\xleftrightarrow[\text{LP--QP}]{F^\ell}
y_{A^G}
\xleftrightarrow[\text{comparison}]{\mathcal E}
y_{A^Y}
\xleftrightarrow[\text{QP--LP}]{F^\ell+b_t\mathcal E^{<t}}
\widehat x_{A^Y}.
\label{eq:univ-comparison-diagram}
\end{equation}
Here the residual energy $\mathcal E$ (Section~\ref{subsec:univ-residual}) controls distance to the surrogate path. In the last comparison, $b_t\mathcal E^{<t}$ with $b_t>0$ penalizes errors in the preceding candidate history.

\begin{lemma}[Stability of the memory]\label{lem:univ-state-stability}
For admissible states and points of the cube,
\[
\|H_s^+(H,x)-H_s^+(\widetilde H,\widetilde x)\|\le C\|H-\widetilde H\|+C(1+\delta_s^{-1})\|x-\widetilde x\|,
\]
where $C$ is independent of $\delta_s$. The minimizers of $Q_{t,A,H}$ and $Q_{t,A,\widetilde H}$ satisfy $\|y_H-y_{\widetilde H}\|\le C\beta_t\tau_{q,t}^{-1}\|H-\widetilde H\|$. For fixed $K$, the stability constants of the full recursion are polynomial in the parameters and their inverses.
\end{lemma}
\begin{proof}
The update rule gives the memory estimate. Adding the two minimizers' variational inequalities gives $\tau_{q,t}\|y_H-y_{\widetilde H}\|^2\le C\beta_t\|H-\widetilde H\|\|y_H-y_{\widetilde H}\|$, hence the second bound (also when the last norm is zero). Iterate over fixed $K$ for the last claim.
\end{proof}

\subsection{Auxiliary probabilistic estimates}
\label{subsec:univ-inputs}

Choose $C_A$ large enough that $\max\{\|A^G\|_{\op},\|A^Y\|_{\op}\}\le C_A$ with high probability. Fix $C_v>2$ and define
\begin{equation}\label{eq:objective-event}
 \mathcal O_{t,N}
 =\{\|v_t\|_2\le C_v\sqrt N,\ \|v_t\|_\infty\le C_v\sqrt{\log N}\},
 \qquad \mathcal O_N=\bigcap_{t\le K}\mathcal O_{t,N}.
\end{equation}
Gaussian tails give $\Pp(\mathcal O_N)\to1$. Independence of $v_t$ from its preceding history gives $\Pp_h(\mathcal O_{t,N}^c)=o(1)$ uniformly in $h$. Entrywise comparison conditions on all objectives in $\mathcal O_N$; a union bound permits intersection with the history events.

The low-$\ell_1$ estimates control exact-LP dual directions, and coordinate localization places the surrogate path in deterministic boxes. Restricted infima over these boxes can then be compared through the Lindeberg principle, using moment matching as in~\cite{HanShen2023}. The proofs are in Appendices~\ref{app:univ-low-l1}--\ref{app:repeated-design-comparison}.

\begin{lemma}[Low-$\ell_1$ estimates]
\label{lem:univ-low-l1}
The following hold with high probability.
\begin{enumerate}[label=(\roman*)]
\item There are $r_1,c_1>0$ such that
\begin{equation*}
\inf_{\substack{q\ge0,\ \|q\|_2=1\\ \|q\|_1\le r_1\sqrt N}}
\frac1N\sum_j|\langle Y_{:j},q\rangle|
\ge c_1.
\end{equation*}

\item For $2\le t\le K$, let $m_t=\E(1-|\bar X_{t-1}|)>0$. There are $r_t^{\rm rec},c_t^{\rm rec},a_t>0$, chosen before the approximation parameters, such that if
\[
\frac{\|\widehat x_{A^Y}^{(t-1)}-y_{A^Y}^{(t-1)}\|}{\sqrt N}\le a_t,
\qquad
\left|
\frac1N\sum_j(1-|\widehat x_{A^Y,j}^{(t-1)}|)-m_t
\right|
\le \frac{m_t}{8}
\]
hold with high probability, then so does
\begin{equation*}
\inf_{\substack{q\ge0,\ \|q\|_2=1\\ \|q\|_1\le r_t^{\rm rec}\sqrt N}}
\frac1N\sum_j
(1-|\widehat x_{A^Y,j}^{(t-1)}|)
|\langle Y_{:j},q\rangle|
\ge c_t^{\rm rec}.
\end{equation*}
This implication holds both for fixed and for polylogarithmically bounded approximation parameters.
\end{enumerate}
\end{lemma}

Write $(y_A^{(t,-i)},\lambda_A^{(t,-i)})$ for the path recomputed after deleting row $i$ (with multiplier coordinate zero), and $(y_A^{(t,[j])},\lambda_A^{(t,[j])})$ for the path recomputed after zeroing column $j$.

\begin{proposition}[Coordinate localization]
\label{prop:univ-loo}
Suppose that the parameters and their inverses are polylogarithmically bounded. Then there is $B_N=(\log N)^C$ such that, jointly for $A=A^G,A^Y$ and with high probability,
\begin{align}
\max_{i,t}\left\{
\|y_A^{(t)}-y_A^{(t,-i)}\|
+\sqrt{\eta_t}\|\lambda_A^{(t)}-\lambda_A^{(t,-i)}\|
\right\}
&\le B_N,
\label{eq:univ-loo-row}\\
\max_{j,t}\left\{
\|y_A^{(t)}-y_A^{(t,[j])}\|
+\sqrt{\eta_t}\|\lambda_A^{(t)}-\lambda_A^{(t,[j])}\|
\right\}
&\le B_N,
\label{eq:univ-loo-col}\\
\max_{t\le K}
\left\{
\|Ay_A^{(t)}\|_\infty
+\|\lambda_A^{(t)}\|_\infty
+\|A^\top\lambda_A^{(t)}\|_\infty
\right\}
&\le B_N.
\label{eq:univ-polylog}
\end{align}
The same bounds hold for the one-round problem with zero objective.
\end{proposition}

We apply entrywise comparison after restricting candidates to the deterministic boxes supplied by Proposition~\ref{prop:univ-loo}.

\begin{lemma}[Joint comparison of restricted infima]
\label{lem:repeated-design-comparison}
Fix $J$. Condition on objective vectors in $\mathcal O_N$; all bounds below are uniform over this event. Let $\mathcal D_{N,j}\subset[-P_N,P_N]^{d_{N,j}}$ be nonempty compact sets that do not depend on the design, with $d_{N,j}\le CN$ and $P_N$ polylogarithmic. Suppose that the costs $f_{N,j,Y}$ admit $C^3$ smoothings such that
\begin{equation*}
\sup_{Y,z}|f_{N,j,Y}^{\sigma}(z)-f_{N,j,Y}(z)|=o(1),
\qquad
\sup_{Y,z}|\partial_{Y_{ik}}^r f_{N,j,Y}^{\sigma}(z)|
\le N^{-1-r/2}(\log N)^{C_r},
\quad 1\le r\le3.
\end{equation*}
On the event $\max_{i,k}|Y_{ik}|\le C\sqrt{\log N}$, assume in addition a Lipschitz bound with constant $N^a(\log N)^C$ in the $\ell_\infty$ metric on candidates, uniformly over $\mathcal O_N$. If $V_N(Y) = \left( \inf_{\mathcal D_{N,j}}f_{N,j,Y} \right)_{j\le J},$ then, for every fixed $\Phi\in C_b^3(\R^J)$,
\begin{equation}
\label{eq:univ-joint-infimum-comparison}
\left|
\E\Phi(V_N(Y))-\E\Phi(V_N(G))
\right|
\longrightarrow0.
\end{equation}
In particular, the lemma applies to the energy $\mathcal E_{N,A}$ defined below, its prefix and row-lifted versions, and their sums with $N^{-1}F^\ell_{t,A,H}(x)$. In these applications the coefficients are nonnegative and polylogarithmic, and the states are generated by candidate histories.
\end{lemma}

\subsection{A residual for the surrogate path}
\label{subsec:univ-residual}

To apply comparison to the design-dependent QP optimizer, we introduce a nonnegative residual energy vanishing on its path. Strong monotonicity bounds path and memory errors by this energy. A candidate set separated from the Gaussian surrogate path therefore has positive residual infimum, which transfers to the subgaussian design.

Fix a memory state $H=(m,c)$. The identity $\frac{\|s_+\|^2}{2\eta_t} = \max_{\lambda\ge0} \left\{ \lambda^\top s-\frac{\eta_t}{2}\|\lambda\|^2 \right\}$ gives a saddle-point representation of the surrogate objective. Its saddle function and associated operator are
\begin{align}
\mathcal L_{t,A,H}(x,\lambda)
&=
-v_t^\top x+\frac{\tau_{q,t}}2\|x\|^2+\frac{\beta_t}2\|m\odot x-c\|^2+\lambda^\top(\kappa_t\1-Ax)-\frac{\eta_t}2\|\lambda\|^2,
\label{eq:univ-saddle}\\
T_{t,A,H}(x,\lambda)
&=
\begin{pmatrix}
-v_t-A^\top\lambda+\tau_{q,t}x+\beta_t m\odot(m\odot x-c)\\
Ax-\kappa_t\1+\eta_t\lambda
\end{pmatrix}.
\label{eq:univ-operator}
\end{align}
The matrix terms cancel. Thus, with $\mu_t=\min\{\tau_{q,t},\eta_t\}$,
$\langle T_{t,A,H}(z)-T_{t,A,H}(z'),z-z'\rangle \ge \mu_t\|z-z'\|^2.$
 On the event $\|A\|_{\op}\le C_A$, the operator is also Lipschitz in $z$ and in the state $H$, with constants controlled by the regularization parameters.

Let $B_N\ge1$ be as in Proposition~\ref{prop:univ-loo}, and set $\mathcal C_N=[-1,1]^N\times[0,2B_N]^M$. For a candidate path $z\in\mathcal C_N^K$ with components $z^{(t)}=(x^{(t)},\lambda^{(t)})$, define
\begin{equation}
R_{t,A}(z)
=
z^{(t)}
-
\Pi_{\mathcal C_N}
\bigl(
z^{(t)}-T_{t,A,H_{t-1}(x)}(z^{(t)})
\bigr).
\label{eq:univ-residual}
\end{equation}
The residual uses the candidate's own prefix memory, making comparison over a design-independent set possible. Write $z_A^{*,(t)}=(y_A^{(t)},\lambda_A^{(t)})$ for the surrogate path. On the localization event its multiplier cutoff is inactive, so $R_{t,A}(z_A^*)=0$ for every $t$.

\begin{proposition}[Path control by the residual]
\label{prop:univ-residual-bound}
There is a constant $C_{\rm path}$, polynomial in the regularization parameters, their reciprocals, and the Lipschitz constants of the memory update, such that the residual energy $\mathcal E_{N,A}(z) = \frac{C_{\rm path}}N \sum_{t=1}^K \|R_{t,A}(z)\|^2$ satisfies, on the operator-norm and localization events,
\begin{equation}
\frac1N
\sum_{t=1}^K
\left(
\|z^{(t)}-z_A^{*,(t)}\|^2
+
\|H_t(x)-H_t(y_A)\|^2
\right)
\le
\mathcal E_{N,A}(z).
\label{eq:univ-residual-bound}
\end{equation}
The same constant works for every prefix, and we write $\mathcal E_{N,A}^{<t}$ for the energy of the rounds up to $t-1$.
\end{proposition}

\begin{proof}
Let $z_*$ solve the one-round variational inequality with state $H_*$. Put $r=z-\Pi_{\mathcal C_N}(z-T_H(z))$ and $p=z-r$. Projection optimality gives $r-T_H(z)\in N_{\mathcal C_N}(p)$. If $L,L_H$ are the operator's Lipschitz constants in $z,H$, strong monotonicity of $T_{H_*}+N_{\mathcal C_N}$ yields
\begin{equation*}
\|z-z_*\|
\le
\left(1+\frac{L+1}{\mu}\right)\|r\|
+
\frac{L_H}{\mu}\|H-H_*\|.
\end{equation*}
Now write $e_t=\|z^{(t)}-z_A^{*,(t)}\|$, $h_t=\|H_t(x)-H_t(y_A)\|$, and $d_t=\|R_{t,A}(z)\|$. The one-round estimate and Lemma~\ref{lem:univ-state-stability} give $e_t\le a d_t+b h_{t-1}, $ $h_t\le L_{\rm st}(h_{t-1}+e_t), $ $h_0=0.$ Since $K$ is fixed, iterating these inequalities and applying the Cauchy--Schwarz inequality proves \eqref{eq:univ-residual-bound} with a constant $C_{\rm path}$ of the stated polynomial form.
\end{proof}

To control row-margin laws, add candidate row variables and penalize their discrepancy from the candidate matrix products. Let $U\in[-2B_N,2B_N]^{M\times K}$ be a candidate row path, and choose a $C^3$ truncation $\vartheta_N:\R\to[-2B_N,2B_N]$ that equals the identity on $[-B_N,B_N]$ and satisfies $\|\vartheta_N'\|_\infty\le C_\vartheta$ and $\|\vartheta_N^{(r)}\|_\infty\le CB_N^{1-r}$ for $r=2,3$. With $X(z)=[x^{(1)}|\cdots|x^{(K)}]$, define
\begin{equation*}
\mathcal E^{\rm row}_{N,A}(z,U)
=
(1+2C_\vartheta^2C_A^2)\mathcal E_{N,A}(z)
+
\frac2N
\|U-\vartheta_N(AX(z))\|_F^2.
\end{equation*}
On the localization event, $U_A^*=AX(z_A^*)$ lies in the region where $\vartheta_N$ is the identity. Moreover,
\begin{equation*}
\frac1N\|U-U_A^*\|_F^2
\le
\frac2N
\|U-\vartheta_N(AX(z))\|_F^2
+
2C_\vartheta^2C_A^2\mathcal E_{N,A}(z).
\end{equation*}
Hence $\mathcal E^{\rm row}_{N,A}$ vanishes at $(z_A^*,U_A^*)$ and controls the path, its memory, and the row margins simultaneously. The normalization by $N$ is equivalent to normalization by $M$ up to a bounded factor because $M/N\to\alpha>0$.

\subsection{Proof of universality}
\label{subsec:univ-proof}

\subsubsection{Comparison of the Gaussian LP and the Gaussian QP}
\label{subsec:univ-gaussian-gap}

The linear hinge and memory penalties soften the row constraints and frozen face. We show that the LP optimizer, QP optimizer, and their midpoint lie below the Gaussian localization gap; the midpoint identity then converts their law bounds into a bound on Euclidean distance with coordinates matched by index.

For cube-valued prefixes, define
\begin{equation*}
\mathcal B_{N,t}^{\rm col}(\eps)=\left\{x^{(1:t)}:\Wtwo\left(\frac1N\sum_{j=1}^N\delta_{(x_j^{(1)},\ldots,x_j^{(t)})},\mu_t^{\rm col}\right)\ge\eps\right\}.
\end{equation*}
Fix $t$, set $p=t-1$, abbreviate $(\tau_{q,t},\beta_t,\eta_t,\ell_t)$ by $(\tau_q,\beta,\eta,\ell)$, and let $H^*=(\1_B,h_B\1_B)$ be the exact hard memory on the Gaussian frozen set $B=\widehat B_p$, extended by zero outside $B$.

\begin{lemma}[Gap for the finite linear penalty]
\label{lem:univ-linear-gap}
For every $\eps>0$ there are $c>0$ and finite $\ell_0,\beta_0$ such that, for fixed $\ell\ge\ell_0$, $\beta\ge\beta_0$, and $\tau_q\ge0$,
\begin{equation*}
\inf_{\substack{x\in[-1,1]^N\\(\widehat x_{A^G}^{(1:p)},x)\in\mathcal B_{N,t}^{\rm col}(\eps)}}\frac1N F_{t,A^G,H^*}^{\ell}(x)\ge -V_t+8c
\end{equation*}
with high conditional probability, uniformly over $h\in\mathcal T_{p,N}(\mathfrak h_p)$ and over compact input families as specified in Section~\ref{sec:closing-gaussian-recursion}. \end{lemma}

\begin{proof}
We first keep the old frozen face exact and set $\tau_q=0$. On this face, an $\eps$-error in the law of the full trajectory forces an error of fixed positive size in the law of the free columns, because the frozen part has already converged. In the notation of Section~\ref{subsec:finite-ao-gaps}, set $F=B^c$ and $w=x_F-\widehat x_{A^G,F}^{(p)}$, and fix the gap cutoff $R=R_{\rm dual}$ of that section and a constant $R_m>C_A+1$. We retain the restrictions $\|u\|\le R_m\sqrt N,$ $N^{-1}L_p^\top u=\Gamma_p e_p+b_N(w).$ We replace the row obstacle by the penalty $\ell\|(\kappa_t\1-u)_+\|_1$. Since $u=Ax$ is admissible on the norm event, the lifted minimum is no larger than the original one. The normalized value of the corresponding auxiliary problem in Gordon's comparison is
\[
\inf_{x,u}\left\{-\frac{v_t^\top x}{N}+\frac\ell N\|(\kappa_t\1-u)_+\|_1
+R\left(\frac{\|Z_{N,h}(\omega_N(x))-u\|}{\sqrt N}-c_{g,N}(x)\right)_+\right\},
\]
with the same restrictions and with $c_{g,N}(x)=N^{-1}g^\top\Sigma_{p,F}^{1/2}w$. We use only the lower-tail comparison.

Suppose that no finite $\ell$ retains a positive fraction of the hard-constraint gap. On a bounded sublevel set, the nonnegative penalty terms and the objective-norm bound give $N^{-1}\|(\kappa_t\1-u)_+\|_1\le C/\ell$. We can therefore take $\ell_k,N_k\to\infty$ and bad near-minimizers with vanishing normalized row violation and uniformly bounded row second moments. Compactness of the marked column laws and weak compactness of the row laws yield a population candidate that remains in the bad set. The vanishing normalized row violation enforces the obstacle $\widetilde U\ge\kappa_t$.

For an old dual mark $\lambda$, Cauchy--Schwarz bounds the tail of a mixed moment by
\[
\left|\frac1N\sum_i\lambda_i u_i\1_{\{|\lambda_i|>L\}}\right|
\le R_m\left(\frac1N\sum_i\lambda_i^2\1_{\{|\lambda_i|>L\}}\right)^{1/2}.
\]
The right side vanishes as $N\to\infty$ and then $L\to\infty$ by the $W_2$ convergence of the old marks. Truncation therefore passes the moments involving the old dual variables to the limit. Lower semicontinuity bounds the limiting repair cost from below by the population repair cost. Replacing the limiting row variable by its conditional mean preserves feasibility and can only decrease this cost, contradicting the strict Gaussian value gap.

It remains to soften the frozen face. If $x^F$ is obtained from $x$ by replacing $x_B$ with $h_B$, then on the operator-norm and objective-norm events
\begin{equation*}
\begin{split}
|g_\ell(x)-g_\ell(x^F)|&\le L_\ell\sqrt N\|x_B-h_B\|,\\
g_\ell(x)+\frac\beta2\|x_B-h_B\|^2&\ge g_\ell(x^F)-\frac{L_\ell^2N}{2\beta},
\end{split}
\qquad g_\ell(x)=-v_t^\top x+\ell\|(\kappa_t\1-Ax)_+\|_1,
\end{equation*}
with $L_\ell=C_v+\ell\sqrt{M/N}\,C_A$. Thus, for large $\beta$, every point of the relevant low sublevel set lies arbitrarily close to the hard face, and projection onto that face would otherwise contradict the preceding gap. Finally, $N^{-1}|g_{\ell,A}(x)-g_{\ell,A'}(x)|\le\ell\sqrt{M/N}\|A-A'\|_{\op}$ uniformly on the cube, so the conditional Gaussian coupling transfers the gap from the Gaussian reference law to the selected Gaussian history.
\end{proof}

For $H=(m,c)$ and $e=N^{-1/2}(\|m-\1_B\|+\|c-h_B\1_B\|)$,
$\sup_{x\in[-1,1]^N}N^{-1}|F_{t,A,H}^{\ell}(x)-F_{t,A,H^*}^{\ell}(x)|\le2\beta e$.
Thus a small memory error preserves the gap. The midpoint will convert the resulting law bounds to Euclidean closeness.

\begin{lemma}[Low sublevel sets and the midpoint]
\label{lem:univ-midpoint}
Let $y$ minimize $Q_{t,A,H}$ and let $\widehat x$ satisfy the row constraints. Then
\begin{equation*}
F_{t,A,H}^{\ell}(y)\le F_{t,A,H}^{\ell}(\widehat x)+\frac M2\eta\ell^2.
\end{equation*}
The sublevel set at this level contains $\widehat x$, $y$, and $(\widehat x+y)/2$. If the current-round empirical laws of these three vectors are all within $\eps$ in $\Wtwo$ of a common law on $[-1,1]$, then
$N^{-1}\|\widehat x-y\|^2\le16\eps.$
\end{lemma}

\begin{proof}
Young's inequality gives $F^\ell\le Q+M\eta\ell^2/2$; convexity then puts the midpoint in the same sublevel set. Let $m_2$ be the second moment of the target law. Since all three empirical laws and the target are supported on $[-1,1]$, each $\Wtwo$ bound makes the corresponding normalized squared norm differ from $m_2$ by at most $2\eps$. Substituting these bounds into
\[
\frac{\|\widehat x-y\|^2}{N}=2\left(\frac{\|\widehat x\|^2}{N}-m_2\right)+2\left(\frac{\|y\|^2}{N}-m_2\right)-4\left(\frac{\|(\widehat x+y)/2\|^2}{N}-m_2\right)
\]
proves the claim.
\end{proof}

\begin{corollary}[Gaussian approximation at fixed accuracy]
\label{cor:univ-gaussian-proxy}
If Lemma~\ref{lem:univ-linear-gap} holds at radius $\eps$ with gap $8c$, the incoming memory error is at most $e$, and $2\beta e<c,$ $\frac{\tau_q}{2}+\frac\beta2e^2+\frac{M}{2N}\eta\ell^2<c,$ then, with high probability,
$N^{-1}\|y_{A^G}^{(t)}-\widehat x_{A^G}^{(t)}\|^2\le16\eps.$
Moreover,
\begin{equation*}
\frac{\|A^G(y_{A^G}^{(t)}-\widehat x_{A^G}^{(t)})\|^2}{N}\le16C_A^2\eps,\qquad \frac{|v_t^\top(y_{A^G}^{(t)}-\widehat x_{A^G}^{(t)})|}{N}\le4C_v\sqrt\eps.
\end{equation*}
\end{corollary}

\begin{proof}
The exact optimizer is feasible and has normalized value $V_t+o_{\Pp}(1)$. The memory perturbation, the ridge term, and the Young-inequality bound of the preceding lemma keep the exact optimizer, the QP optimizer, and their midpoint below the gap on the bad set. Their empirical laws are therefore $\eps$-close to the target, and Lemma~\ref{lem:univ-midpoint} gives the first claim. The last two bounds follow on the operator-norm and objective-norm events.
\end{proof}

\subsubsection{Universality of the surrogate path}
\label{subsubsec:univ-stabilized-proof}

A candidate whose law is separated from the Gaussian prediction is separated from the Gaussian surrogate path and has positive residual energy. Comparison of restricted infima transfers this gap to the subgaussian design, where the actual surrogate path has zero residual.

\begin{corollary}[Universality of the surrogate path]
\label{cor:univ-stabilized-transfer}
Under the assumptions of Theorem~\ref{thm:univ-exact}, and with the parameters chosen along the slow diagonal sequence constructed below, we have, jointly for $t\le K$,
\begin{align}
W_2\left(\frac1N\sum_j\delta_{(y_{A^Y,j}^{(1)},\ldots,y_{A^Y,j}^{(t)})},\mu_t^{\rm col}\right)&\longrightarrow0,\label{eq:univ-stabilized-col}\\
W_2\left(\frac1M\sum_i\delta_{((A^Yy_{A^Y}^{(1)})_i,\ldots,(A^Yy_{A^Y}^{(t)})_i)},\mu_t^{\rm row}\right)&\longrightarrow0,\label{eq:univ-stabilized-row}\\
N^{-1}v_t^\top y_{A^Y}^{(t)}&\longrightarrow V_t.\label{eq:univ-stabilized-value}
\end{align}
\end{corollary}

\begin{proof}
Fix $\varepsilon>0$ and choose fixed parameters so that the Gaussian QP column and row laws and normalized value have errors at most $\varepsilon/2$ with high probability. On the operator-norm and localization events, define
\[
 I_A^{\rm col}=\inf_{\substack{z\in\mathcal C_N^K\\
 x^{(1:t)}(z)\in\mathcal B_{N,t}^{\rm col}(\varepsilon)}}\mathcal E_{N,A}(z).
\]
Proposition~\ref{prop:univ-residual-bound} gives $I_{A^G}^{\rm col}\ge\varepsilon^2/4$.
For the row-law bad set, the same argument with $\mathcal E_{N,A}^{\rm row}$ gives a lower bound $\alpha_-\varepsilon^2/4$ when $M/N\ge\alpha_->0$.
For the value bad set
\[
 \mathcal B^{\rm val}_{N,t}(\varepsilon)
 =\{z\in\mathcal C_N^K:|N^{-1}v_t^\top x^{(t)}(z)-V_t|\ge\varepsilon\},
\]
the objective bound $\|v_t\|\le C_v\sqrt N$ forces normalized distance at least $\varepsilon/(2C_v)$ from the Gaussian QP optimizer. Its restricted residual infimum is therefore at least $\varepsilon^2/(4C_v^2)$.

Given the objectives, all three bad sets are closed and independent of the design; an empty set needs no comparison. Lemma~\ref{lem:repeated-design-comparison} transfers the positive lower bounds, with fixed buffers, to the subgaussian design. Its actual surrogate path has zero residual energy, including in the row lift, so it avoids all three bad sets. This proves the fixed-accuracy conclusions. A sufficiently slow diagonal choice of parameters makes them simultaneous over the fixed number of rounds.
\end{proof}

\subsubsection{Feasibility and dual localization for the exact LP}
\label{subsubsec:univ-dual-control}

The return to the exact LP requires $O(\sqrt N)$ control of its row multipliers. At the first round, a zero-objective QP transfers a strict feasibility reserve from the Gaussian design. At later rounds, the margin drop controls dual directions with large $\ell_1$ norm, and Lemma~\ref{lem:univ-low-l1} controls the remaining directions.

\begin{lemma}[First-round feasibility and dual localization]
\label{lem:univ-first-reserve}
Under the Gaussian reserve assumption of Theorem~\ref{thm:univ-exact}, there are $\Delta_*,c_*>0$ such that, with high probability,
\[
\max_{x\in[-1,1]^N}\min_i(A^Yx)_i\ge\kappa_1+\Delta_*,
\]
and
\begin{equation}
\inf_{q\ge0,\ \|q\|_2=1}\left\{\|(A^Y)^\top q\|_1-\kappa_1\|q\|_1\right\}\ge c_*\sqrt N.
\label{eq:univ-first-recession}
\end{equation}
Moreover, the exact LP of the first round has a row multiplier of norm $O_P(\sqrt N)$.
\end{lemma}

\begin{proof}
Let $\kappa_+=\kappa_1+\Delta_0$ and consider the zero-objective QP $y_A^0=\argmin_{x\in[-1,1]^N}\{\|x\|^2/2+\|(\kappa_+\1-Ax)_+\|^2/(2\eta_0)\}$. Under the Gaussian design, comparison with a feasible point bounds the violation by $\sqrt{\eta_0N}$. The one-round comparison of the row-lifted residual transfers the small-violation estimate to the subgaussian design: for any fixed $\varepsilon>0$, if $\eta_0$ is sufficiently small, then $\|e\|_2\le\varepsilon\sqrt N$ with high probability, where $e=(\kappa_+\1-Ay_A^0)_+$ and $A=A^Y$.

Let $r_1,c_1$ be the constants in Lemma~\ref{lem:univ-low-l1}(i). Choose $0<r\le r_1$ with $|\kappa_1|r\le c_1/2$, then $0<\varepsilon<\Delta_0r/2$, and finally $\eta_0$. For $q\ge0$ with $\|q\|_2=1$, testing against $y_A^0$ gives $\|A^\top q\|_1-\kappa_1\|q\|_1 \ge\Delta_0\|q\|_1-q^\top e \ge\Delta_0\|q\|_1-\varepsilon\sqrt N.$ This is at least $(\Delta_0r/2)\sqrt N$ when $\|q\|_1\ge r\sqrt N$. Otherwise, Lemma~\ref{lem:univ-low-l1}(i) gives the lower bound $(c_1-|\kappa_1|r)\sqrt N\ge(c_1/2)\sqrt N$. Thus \eqref{eq:univ-first-recession} holds with $c_*=\min\{c_1/2,\Delta_0r/2\}$. By homogeneity and the minimax theorem, the feasibility reserve is at least $c_*\sqrt{N/M}$, which is bounded away from zero. Comparing the exact dual objective at its minimizer with its value at zero then gives $\|\widehat\lambda_1\|_2\le2\|v_1\|_1/(c_*\sqrt N)=O_P(\sqrt N)$.
\end{proof}

For $t\ge2$, let $x^0=\widehat x_A^{(t-1)}$, $B=\widehat B_{t-1}$, $F=B^c$, and $h=\widehat x_{A,B}^{(t-1)}$. The reduced dual objective is
\[
D_t(\lambda)=\|v_{t,F}+A_{:,F}^\top\lambda\|_1+\langle h,v_{t,B}+A_{:,B}^\top\lambda\rangle-\kappa_t\1^\top\lambda,\qquad \lambda\ge0,
\]
and its recession function satisfies
\begin{equation*}
D_t^{\infty}(a)\ge\Delta_t\|a\|_1+\sum_j(1-|x_j^0|)|(A^\top a)_j|,\qquad a\ge0,
\end{equation*}
where $\Delta_t=\kappa_{t-1}-\kappa_t>0$. The first term controls directions with $\|a\|_1\ge r_t^{\rm rec}\sqrt N$. Lemma~\ref{lem:univ-low-l1}(ii) controls the complementary directions once the preceding LP and surrogate paths satisfy its two fixed-accuracy hypotheses. Consequently, with high probability, $D_t^{\infty}(a)\ge c_t^{\rm dual}\sqrt N\|a\|_2$ for all $a\ge0$, where the constant is fixed before the approximation parameters.

Let $\widehat\lambda_t$ minimize $D_t$, and let $\widehat\nu_t=(v_t+A^\top\widehat\lambda_t)_B$, extended by zero outside $B$, be the multiplier of the equality constraints. Comparison of the dual objective at its minimizer with its value at zero gives
\begin{equation}
\|\widehat\lambda_t\|_2+\|\widehat\nu_t\|_2\le C_t\sqrt N.
\label{eq:univ-exact-dual-localization}
\end{equation}

The multiplier bound controls the penalized cost of the exact optimizer. At $t=1$, we take $B=\varnothing$ and $e=0$.

\begin{lemma}[The exact optimizer as a feasible witness]
\label{lem:univ-exact-witness}
Let $\widehat x=\widehat x_A^{(t)}$, $y=y_A^{(t)}$, and $e=N^{-1/2}(\|m-\1_B\|+\|c-h\1_B\|)$. Under \eqref{eq:univ-exact-dual-localization},
\begin{equation*}
\frac1N F^\ell_{t,A,H}(\widehat x)\le-\frac1N v_t^\top y+\mathfrak a_t(e),\qquad \mathfrak a_t(e)=\tau_q+\beta e^2+2C_t e+2C_t^2(\eta+\beta^{-1}).
\end{equation*}
\end{lemma}

\begin{proof}
Put $r=(\kappa_t\1-Ay)_+$, $w=m\odot y-c$, and $d=N^{-1}v_t^\top(y-\widehat x)$. The KKT conditions give $d\le C_t(\|r\|+\|w\|)/\sqrt N+C_t e$. Write $S=\|r\|^2/(2\eta N)+\beta\|w\|^2/(2N)$ and $H_0=(\tau_q+\beta e^2)/2$. Optimality of the QP minimizer gives $S\le d+H_0$, while Young's inequality gives $d\le S/2+C_t e+C_t^2(\eta+\beta^{-1})$. Hence $F^\ell(\widehat x)/N+v_t^\top y/N\le d+H_0\le \mathfrak a_t(e)$.
\end{proof}

\subsubsection{Return to the exact subgaussian LP}
\label{subsubsec:univ-exact-return}

The round-$t$ memory depends on the preceding path and hence on the design, so it cannot be fixed while applying the restricted-infimum comparison. We include the preceding candidate history in the optimization and penalize its deviation from the surrogate path by the prefix residual energy. The candidate set is then independent of the design, while the penalty vanishes at the actual QP history. On the Gaussian side, a small prefix residual, together with the Gaussian LP--QP prefix and memory bounds, brings both the candidate history and its memory close enough to use the finite-penalty gap.

Let $\mathfrak Z_{N,t-1}$ be the product of the primal--dual boxes of the preceding rounds, and let $x^{<t}(z)$ denote the primal prefix of $z$. Define
\begin{equation}
\mathcal V^{\rm bad}_{t,A}(\varepsilon)
=
\inf_{\substack{z\in\mathfrak Z_{N,t-1},\,x\in[-1,1]^N\\
(x^{<t}(z),x)\in\mathcal B^{\rm col}_{N,t}(\varepsilon)}}
\left\{
b_t\mathcal E^{<t}_{N,A}(z)
+
\frac{F^\ell_{t,A,H_{t-1}(x(z))}(x)}{N}
\right\}.
\label{eq:univ-anchored-bad-value}
\end{equation}
At $t=1$, the prefix is empty and the residual term vanishes.

\begin{lemma}[Gap for candidate histories]
\label{lem:univ-history-anchor}
Fix $0<\varepsilon_0<\varepsilon$, and suppose that Lemma~\ref{lem:univ-linear-gap} gives a Gaussian gap $8c>0$ at radius $\varepsilon_0$. Write $H_{t-1}(y_{A^G})=(m_G,c_G)$ and define
\[
a_{G,t}=\left(N^{-1}\sum_{s<t}\|y_{A^G}^{(s)}-\widehat x_{A^G}^{(s)}\|_2^2\right)^{1/2},\qquad
e_{G,t}=N^{-1/2}\bigl(\|m_G-\1_B\|_2+\|c_G-h_B\1_B\|_2\bigr).
\]
Suppose that $a_{G,t}\le a$ and $e_{G,t}\le e$ with high probability, where $a,e\ge0$ are small enough that, for some $r_t>0$,
$a+\sqrt{r_t}<\varepsilon-\varepsilon_0,\allowbreak\quad 2\beta_t(e+C_H\sqrt{r_t})<c.$
Here $C_H=2$. Then one can choose finite $b_t>0$ such that
\begin{equation*}
\Pp\{\mathcal V^{\rm bad}_{t,A^G}(\varepsilon)<-V_t+3c\}\to0,
\qquad
\Pp\{\mathcal V^{\rm bad}_{t,A^Y}(\varepsilon)<-V_t+2c\}\to0.
\end{equation*}
\end{lemma}

\begin{proof}
Suppose first that $\mathcal E^{<t}_{N,A^G}(z)\le r_t$. Proposition~\ref{prop:univ-residual-bound} bounds the normalized distance from the candidate prefix to the Gaussian QP prefix by $\sqrt{r_t}$. It also bounds the concatenated state error by $\sqrt{r_t}$, so the sum of the two normalized state-component errors is at most $\sqrt{2r_t}\le C_H\sqrt{r_t}$. Thus the candidate prefix is within $a+\sqrt{r_t}$ of the exact prefix in normalized Euclidean distance, and every $\varepsilon$-bad candidate is $\varepsilon_0$-bad relative to the exact prefix. The memory perturbation bound gives
\[
\frac1N F^\ell_{t,A^G,H_{t-1}(x(z))}(x)
\ge \frac1N F^\ell_{t,A^G,H^*}(x)-2\beta_t(e+C_H\sqrt{r_t})
\ge -V_t+7c.
\]

If instead $\mathcal E^{<t}_{N,A^G}(z)>r_t$, the residual term contributes at least $b_tr_t$. On the norm event, $F^\ell(x)/N$ is bounded below by a fixed constant, so choosing $b_t$ sufficiently large gives the same lower bound. This proves a fixed gap for the restricted Gaussian infimum. Because its candidate set is independent of the design, Lemma~\ref{lem:repeated-design-comparison} transfers the gap to $A^Y$, with the buffer $c$ absorbing the comparison error.
\end{proof}

We evaluate the transferred bound at the actual subgaussian QP prefix $z_{A^Y}^*$, whose residual vanishes. Put $y=y_{A^Y}^{(t)}$, $\widehat x=\widehat x_{A^Y}^{(t)}$, $\varepsilon_v=|N^{-1}v_t^\top y-V_t|$, and $\alpha_N=M/N$. Lemma~\ref{lem:univ-exact-witness}, the Young-inequality bound, and convexity give
\[
\max_{x\in\{\widehat x,y,(\widehat x+y)/2\}}\frac{F^\ell_{t,A,H}(x)}N
\le -V_t+\varepsilon_v+\mathfrak a_t(e)+\frac{\alpha_N\eta\ell^2}{2}.
\]
The parameter choice in Section~\ref{subsubsec:univ-parameter-choice} makes the sum of the last three terms smaller than $2c$. Hence none of the three candidates lies in the bad set of \eqref{eq:univ-anchored-bad-value}, and Lemma~\ref{lem:univ-midpoint} gives $d_t^Y := \frac1N \|\widehat x_{A^Y}^{(t)}-y_{A^Y}^{(t)}\|_2^2 \le16\varepsilon.$

Thus the column law of the surrogate path transfers to the exact LP, with coordinates matched by index. The row law and the normalized objective follow from the same coupling. On the event $\|A^Y\|_{\op}\le C_A$, and provided $\alpha_N=M/N\ge\alpha_->0$,
\begin{equation}
\frac1M
\sum_{s\le t}
\|A^Y(\widehat x_{A^Y}^{(s)}-y_{A^Y}^{(s)})\|_2^2
\le
\frac{C_A^2}{\alpha_-}
\sum_{s\le t}d_s^Y,
\label{eq:univ-exact-row-coupling}
\end{equation}
while $\frac1N \left| v_t^\top (\widehat x_{A^Y}^{(t)}-y_{A^Y}^{(t)}) \right| \le C_v\sqrt{d_t^Y}.$

Boundary atoms prevent Wasserstein convergence alone from identifying exact freezing. Put $\widehat J_t=\{i:\widehat u_{A^Y,i}^{(t)}=\kappa_t\}$. Given the design and past, the independent continuous objective selects a unique vertex almost surely, so Lemma~\ref{lem:finite-active-count} gives $|\widehat F_t|\le|\widehat J_t|$. If the column and row marginal errors are $\omega_{\rm col},\omega_{\rm row}$, then, for every $a>0$,
\begin{equation*}
\frac{|\widehat F_t|}{N}
\le
\alpha_N
\left\{
\Pp(U_t\le\kappa_t+2a)
+
\frac{\omega_{\rm row}^2}{a^2}
\right\}.
\end{equation*}
The column coupling gives the matching lower bound. More quantitatively, if
$s_{t,N}(\delta) = \frac1N \#\left\{ j: 0<1-|\widehat x_{A^Y,j}^{(t)}|<2\delta \right\}$,
then
\begin{equation}
s_{t,N}(\delta)\le{} \Pp(1-3\delta<|\bar X_t|<1) + \alpha_N\Pp(\kappa_t<U_t\le\kappa_t+2a) + \frac{\omega_{\rm col}^2}{\delta^2} + \alpha_N\frac{\omega_{\rm row}^2}{a^2} + |\alpha_N-\alpha|. \label{eq:univ-finite-strip-bound}
\end{equation}
The identity $q_t=\Pp(|\bar X_t|<1)=\alpha\Pp(U_t=\kappa_t)$ from Lemma~\ref{lem:free-row-balance} matches the column lower bound and row upper bound. Choose $\delta,a$ first, then make the law errors small: $|\widehat F_t|/N\to q_t$, and the transition strip has vanishing density. The latter controls the next round's smooth-to-exact memory error.

\subsubsection{Choice of parameters and completion of the induction}
\label{subsubsec:univ-parameter-choice}

Given a tolerance $\varepsilon$ for the return step, let $8c_\varepsilon$ be the Gaussian gap at radius $\varepsilon/2$. Choose $\zeta<\varepsilon/16$ with $4C_v\sqrt\zeta<c_\varepsilon/16$, and let $8c_\zeta$ be the corresponding Gaussian gap. Take finite $\ell,\beta$ that are valid for both gaps. With $c_* =\min(c_\varepsilon,c_\zeta)$, $\bar\alpha>\sup_N\alpha_N$, and an allowance $E$ for the incoming memory error, we require
\[
2\beta E<c_*/4,\qquad
\frac{\tau_q+\beta E^2+\bar\alpha\eta\ell^2}{2}<c_*/4,\qquad
\mathfrak a_t(E)+\frac{\bar\alpha\eta\ell^2}{2}<c_\varepsilon/2.
\]
We first enlarge $\beta$ so that $2C_t^2/\beta<c_\varepsilon/4$ and then decrease $\tau_q,\eta,E$. Corollary~\ref{cor:univ-gaussian-proxy} bounds the error in the Gaussian value by $4C_v\sqrt\zeta+o(1)$, and the residual comparison at fixed accuracy gives $\varepsilon_v<c_\varepsilon/4$ with high probability. This closes the budget for the return step.

We add a round subscript to the tolerances above and choose them backwards: the requirements at a later round determine how accurately the preceding paths and memories must be matched. If $d_s=N^{-1}\|\widehat x_A^{(s)}-y_A^{(s)}\|^2,$ then the memory error entering round $t$ satisfies
\begin{equation}
e_t
\le
C_K
\left\{
\left(\sum_{s<t}s_{s,N}(\delta_s)\right)^{1/2}
+
\sum_{s<t}(1+\delta_s^{-1})\sqrt{d_s}
\right\}.
\label{eq:univ-memory-error-budget}
\end{equation}
In \eqref{eq:univ-memory-error-budget}, the first term controls the soft-to-hard discrepancy on the exact path through the transition strips, and the second controls the change in smooth memory between the exact and surrogate paths. The state Lipschitz constant in Lemma~\ref{lem:univ-state-stability} is independent of the strip widths. Once the later tolerances are fixed, choose the earlier strip widths and matching errors to satisfy this bound and Lemma~\ref{lem:univ-low-l1}(ii). Impose also $\sum_{s<t}d_s<\zeta_t^2/64$ and $C_A^2\sum_{s<t}d_s/\alpha_-<\omega_{{\rm row},t}^2/64$ for the prescribed row-law tolerance $\omega_{{\rm row},t}$. Since $a_{G,t}^2=\sum_{s<t}d_s^G$, these choices give $a_{G,t}<\zeta_t/8<\varepsilon_t-\varepsilon_t/2$ and $2\beta_t e_{G,t}<c_{*,t}/4$. Thus a sufficiently small $r_t>0$ satisfies both strict inequalities in Lemma~\ref{lem:univ-history-anchor}, with $\varepsilon_0=\varepsilon_t/2$. There are only finitely many such requirements because $K$ is fixed.

For each target accuracy $1/m$, fix the parameters and prescribed positive error bounds by this backward selection. With these choices fixed, the induction below proceeds forward: the preceding memory bound supplies the next Gaussian approximation, and the return estimate supplies the next memory bound. Only after this fixed-accuracy induction do we let the accuracy improve along a slow diagonal.

\begin{proof}[Proof of Theorem~\ref{thm:univ-exact}]
Proceed simultaneously for the Gaussian and subgaussian paths by induction on $t$, with all parameters fixed at the selected accuracy. The first-round reserve initializes the exact subgaussian path. Suppose the prescribed bounds hold through round $t-1$. The incoming Gaussian memory bound allows Corollary~\ref{cor:univ-gaussian-proxy} to control the Gaussian QP error at round $t$. The fixed-accuracy residual comparisons in the proof of Corollary~\ref{cor:univ-stabilized-transfer} then give the prescribed subgaussian QP law and value bounds. The margin drop and Lemma~\ref{lem:univ-low-l1} give feasibility and dual localization for the exact subgaussian LP. The feasible-witness bound and the candidate-history gap then put $N^{-1}\|\widehat x_{A^Y}^{(t)}-y_{A^Y}^{(t)}\|^2$ below its prescribed bound with high probability.

The row law and exact free fraction follow from \eqref{eq:univ-exact-row-coupling} and \eqref{eq:univ-finite-strip-bound}, while \eqref{eq:univ-memory-error-budget} supplies the accuracy needed for the next round. Iterating over the fixed number $K$ of rounds proves the fixed-accuracy bounds with high probability. Choose $N_m$ so that their combined failure probability is at most $1/m$ for $N\ge N_m$. A sufficiently slow diagonal $m=m(N)\to\infty$ also keeps all parameters, their inverses, and comparison coefficients polylogarithmic, as required by Proposition~\ref{prop:univ-loo} and Lemma~\ref{lem:repeated-design-comparison}. This proves all claims of Theorem~\ref{thm:univ-exact}.
\end{proof}

\subsection{Transfer of the rounding certificate}
\label{subsec:univ-stage2}

We transfer the discrepancy-based rounding certificate of~\cite{LSZ2025} using the terminal free fraction, primal norm, and row-margin law from universality. Uniform row-norm bounds over coordinate subsets permit application of the deterministic-matrix rounding results to the adaptively selected sets.

Set
$A=\frac{Y}{\sqrt N}, \allowbreak\quad \theta^{(0)}=\widehat x_A^{(K)}, \allowbreak\quad S_0=\widehat F_K, \allowbreak\quad q=q_K, \allowbreak\quad m_2=\E\bar X_K^2$.
Fix a target margin $\kappa<\kappa_K$. Theorem~\ref{thm:univ-exact} gives
\begin{equation*}
\frac{|S_0|}{N}\xrightarrow{\Pp}q,
\qquad
\frac{\|\theta^{(0)}\|^2}{N}\xrightarrow{\Pp}m_2,
\qquad
\frac1M\sum_{i=1}^M\delta_{(A\theta^{(0)})_i}
\xrightarrow{\Pp}\law(U_K).
\end{equation*}
Since $\theta_j^{(0)}\in\{\pm1\}$ on $S_0^c$, if $q>0$ we also have
$\frac{M}{|S_0|}\xrightarrow{\Pp}\frac{\alpha}{q}, \allowbreak\quad \frac{\|\theta^{(0)}_{S_0}\|^2}{|S_0|} \xrightarrow{\Pp} \frac{m_2-1+q}{q}$.
A \emph{strict finite certificate} specifies a decreasing sequence $R_0=q>\cdots>R_L>0$ of remaining free fractions, continuous budget functions $b_\ell\ge0$ computable to inverse-polynomial accuracy in polynomial time, and norm lower bounds $r_\ell\in[0,1)$. The initial bound satisfies $r_0=0$ or $0<r_0<(m_2-1+q)/q$. For a fixed reserve $\varepsilon_*>0$, the budgets must satisfy
\[
\sum_{\ell=1}^L b_\ell(u)\le u-\kappa-\varepsilon_*\qquad(u\ge\kappa_K).
\]
At each step, the certificate must satisfy one of the following conditions.
\begin{itemize}
\item An ordinary step satisfies \eqref{eq:univ-stage2-entropy}, $R_\ell/R_{\ell-1}\ge1/2$, and $r_\ell=0$.
\item For an improved step, evaluate the quantities in \cite[Definition~A.1]{LSZ2025} at
$(\alpha/R_{\ell-1},\allowbreak\law(b_\ell(U_K)/\sqrt{R_{\ell-1}}),\allowbreak r_{\ell-1})$.
They must have finite $T_1$ and satisfy the admissibility condition defining $T_2$. They must also satisfy $1-R_\ell/R_{\ell-1}<p_1$ and either $r_\ell=0$ or $0<r_\ell<p_0$.
\end{itemize}

For the uniform row-norm estimates, fix $K_\xi$ such that $\E e^{\xi^2/K_\xi^2}\le2$.

\begin{lemma}[Uniform bounds on restricted row norms]
\label{lem:univ-subset-row-norms}
For every fixed $r>0$, with high probability,
\begin{align}
\sup_{|S|\ge rN}
\frac1M\sum_{i=1}^M
\left(
\frac{\|Y_{i,S}\|^2}{|S|}-1
\right)^2
&\le
C_{\alpha,K_\xi}\frac{\log N}{rN},
\label{eq:univ-subset-row-average}\\
\|Y_{i,S}\|^2
&\le
CK_\xi^2|S|\log\frac{eN}{|S|}
\label{eq:univ-subset-row-uniform}
\end{align}
where the second bound holds simultaneously for every $i\le M$ and every nonempty $S\subset[N]$.
\end{lemma}

\begin{proof}
To prove \eqref{eq:univ-subset-row-average}, we truncate $Y_{ij}^2-1$ at a logarithmic level, recenter, and apply the matrix Bernstein inequality to the resulting $M\times N$ matrix. Its operator norm is $O(\sqrt{N\log N}+\log^2N)$ with high probability, and applying this matrix to $\1_S$ gives the bound uniformly over $|S|\ge rN$. For \eqref{eq:univ-subset-row-uniform}, the exponential-moment bound gives
\[
\Pp\left( \|Y_{i,S}\|^2 > CK_\xi^2|S|\log\frac{eN}{|S|} \right) \le e^{-c|S|\log(eN/|S|)}.
\]
A union bound over rows and subsets completes the proof.
\end{proof}

For $u_i=(A\theta^{(0)})_i$, set $c_{i\ell}=\sqrt N\,b_\ell(u_i)/\|Y_{i,S_{\ell-1}}\|$, taking $c_{i\ell}=0$ when the denominator is zero. If the remaining free fraction satisfies $|S_{\ell-1}|/N\to R_{\ell-1}>0$, Lemma~\ref{lem:univ-subset-row-norms} and continuity give
\[
\frac1M\sum_i\delta_{c_{i\ell}}\xrightarrow{\Pp}
\law\!\left(\frac{b_\ell(U_K)}{\sqrt{R_{\ell-1}}}\right).
\]
The strict condition for an ordinary step, from the one-sided partial-coloring lemma~\cite[Theorem~3.4]{LSZ2025}, is
\begin{equation}
\alpha\E\exp\!\left(-\frac{b_\ell(U_K)^2}{16R_{\ell-1}}\right)<\frac{R_{\ell-1}}8.
\label{eq:univ-stage2-entropy}
\end{equation}
The empirical entropy converges to the left-hand side, so the corresponding strict bound holds at finite size with high probability. Set $n_0=|S_0|$ and $n_\ell=\lfloor(R_\ell/R_{\ell-1})n_{\ell-1}\rfloor$, and remove $n_{\ell-1}-n_\ell$ nearly tight coordinates. The loss in each row is at most $b_\ell(u_i)$.

After the certified steps, we complete the remaining coordinates by ordinary partial coloring. When $n$ coordinates remain, a step uses $c(n)=4\sqrt{\log(e+8M/n)}$, which satisfies its entropy bound. By \eqref{eq:univ-subset-row-uniform}, with $r=n/N$, each such step loses at most
$CK_\xi\sqrt{r\log(e/r)\log(e+8\alpha_N/r)}$
per row. Summing over successive halvings starting from $n\le RN$ bounds the additional loss in every row by
\[
C_{\alpha}K_\xi\sqrt R
\log\frac{e^2(1+\alpha)}{R}
+
o(1)
\]
which tends to zero as $R\downarrow0$.

Throughout the rounding procedure, we use the tolerance $\delta_N=N^{-2}$, compatible step sizes, and inverse-polynomial precision, and we absorb numerical errors into the slack. We retry each step from the same point $O(\log N)$ times, checking the margin, freezing, and prescribed norm conditions. The final rounding to the nearest sign costs $O(\delta_N\sqrt N)=o(1)$. By the bounds of~\cite{LSZ2025}, the procedure succeeds with high probability in randomized polynomial time.

\begin{proposition}[Transfer of the rounding certificate]
\label{prop:univ-stage2-transfer}
Assume the hypotheses of Theorem~\ref{thm:univ-exact}, and suppose that the terminal Gaussian population state admits strict finite certificates as defined above, with a common reserve $\varepsilon_*>0$ and arbitrarily small $R_L>0$, given by a sequence of certificates with $R_L\downarrow0$ that can be generated and checked in time polynomial in $N$, uniformly in the index. Then, for every fixed centered subgaussian law with unit variance for the i.i.d.\ entries, rounding from $\widehat x_A^{(K)}$ produces $\sigma\in\{\pm1\}^N$ satisfying $A\sigma\ge\kappa\1$ with high probability, in randomized polynomial time.
\end{proposition}

\begin{proof}
The completion bound depends on the entry law through $K_\xi$, so we do not fix the certificate in advance. Let $\mathfrak C_1,\mathfrak C_2,\ldots$ be this sequence of strict finite certificates, with the common reserve $\varepsilon_*$ and terminal fractions $R_L(\mathfrak C_m)\downarrow0$. For $m=1,\ldots,\lceil\log N\rceil$, the procedure runs the certified steps of $\mathfrak C_m$ followed by ordinary partial coloring, with fresh randomness, and it outputs the first $\sigma\in\{\pm1\}^N$ with $A\sigma\ge\kappa\1$. Fix $m_0$ such that completion from $2R_L(\mathfrak C_{m_0})N$ remaining coordinates costs less than $\varepsilon_*/2$. For $N\ge e^{m_0}$, the run with $m=m_0$ takes place, and we analyze it. Condition on the matrix and all past rounding outputs before each step. The transferred quantities satisfy the hypotheses of the deterministic-matrix theorems of~\cite{LSZ2025}, which apply using fresh randomness at that step. Each attempt has a fixed positive success probability, so the checked $O(\log N)$ retries succeed with high probability. We first apply the finitely many certified steps.

For an improved step, choose $1-R_\ell/R_{\ell-1}<\pi'<p_1$. Then \cite[Theorem~A.1]{LSZ2025} provides $\pi'n$ nearly tight coordinates and a squared norm of at least $[\pi'+(1-\pi')p_0]n$. Removing the prescribed $d=n-n_\ell\le\pi'n$ coordinates leaves a squared norm of at least $[\pi'+(1-\pi')p_0]n-d\ge p_0(n-d),$ which propagates $r_\ell$, while ordinary steps reset it to zero.

Thus the finitely many certified steps leave $n_L/N\to R_L$ and margins at least $\kappa+\varepsilon_*-o(1)$. The completion bound and the checked retries show that this run outputs a solution with high probability, and the $\lceil\log N\rceil$ runs take randomized polynomial time in total.
\end{proof}

\section*{AI disclosure}
\addcontentsline{toc}{section}{AI disclosure}

We used large language models at several stages of this work, including GPT-5.6, GPT-6 Astra, and GPT-6.1 Terra from OpenAI and Claude Opus 5.5 from Anthropic. We first worked out how the single-round analysis of~\cite{LSZ2025} extends to subgaussian matrices, through an intermediate strongly convex quadratic program and the comparison argument of~\cite{HanShen2023}, and this strategy is unchanged in the present paper. For the iterated LP, we found the form of the conditional density of the matrix after the first round from numerical experiments and proved it by writing the KKT equalities in two coordinate systems. The models helped us refine this proof, extend it to $K$ rounds, and complete the coupling estimates of Section~\ref{sec:optimizer-coupling}. When we returned to the first round, the models pointed out the Gaussian-width interpretation of the function $\varphi_1(\rho)$ of~\cite{LSZ2025}, which reduces the first round to a scalar problem and, as we then found, carries over to later rounds. They also assisted with the resulting calculations and with organizing the later rounds in the same structure (Sections~\ref{sec:finite-po}--\ref{sec:closing-gaussian-recursion}). For universality, the models developed our quadratic approximation into the iterated surrogate with smooth freezing and constructed the auxiliary costs of Section~\ref{sec:universality}. The models also assisted with the technical proofs in the appendix. Claude and ChatGPT helped write the code for the numerical experiments, which is available at~\coderepo, and Claude helped polish the exposition. GPT-6.1 Terra also assisted with later revisions of the exposition in Sections~\ref{sec:finite-po}--\ref{sec:universality}. The authors take full responsibility for the content of this paper.

\bibliographystyle{alpha}
\bibliography{reference}

\begin{appendices}

  \section{Deferred technical proofs}\label{app:deferred-proofs}

  In this appendix, we prove the technical estimates used in Sections~\ref{sec:optimizer-coupling}, \ref{sec:finite-po}, and~\ref{sec:universality}. Throughout, the depth $K$ is fixed. Population inputs satisfy the assumptions of
Section~\ref{sec:finite-po}; compact input families and
uniform conditional convergence are defined in
Section~\ref{sec:closing-gaussian-recursion}.

  \subsection{Determinant moment estimate under the Gaussian reference law}
  \label{app:determinant-moment}

  \begin{proof}[Proof of Lemma~\ref{lem:determinant-fourth-moment}]
Set $C=A_{I_p,F_p}$ and $n=|I_p|=|F_p|$. By Definition~\ref{def:regular-history}, its determinant is nonzero at $A_0\in\mathcal A_{h,p}$. Full support of $Q_{h,p}$ on this affine space (Lemma~\ref{lem:history-gaussian-regression}) gives $\E_{Q_{h,p}}|\det C|^2>0$. The case $n=0$ is immediate, so assume $n\ge1$. Set $R_I=(P_{L_p}^\perp)_{I_p,I_p}$ and $R_F=(P_{X_p}^\perp)_{F_p,F_p}$. The regression representation gives, under $Q_{h,p}$,
    \begin{equation*}
      C
      \stackrel d=
      \bar C
      +
      \frac1{\sqrt N}R_I^{1/2}G_nR_F^{1/2},
    \end{equation*}
where $\bar C=(A_{h,p})_{I_p,F_p}$ and $G_n$ is i.i.d.\ standard Gaussian. Each covariance factor is the identity minus a rank-at-most-$p$ matrix, so $a=\dim\ker R_I\le p$ and $b=\dim\ker R_F\le p$. Choose orthogonal $U,V$ diagonalizing these factors, with positive eigenvalues first. Orthogonal invariance of the Gaussian law and $|\det C|$ gives
    \begin{equation*}
      U^\top C V=B+Z,
      \qquad
      Z_{[n-a],[n-b]}
      =\frac1{\sqrt N}D_I G_{n-a,n-b}D_F,
    \end{equation*}
All other entries of $Z$ vanish; $D_I,D_F$ are positive diagonal, $G_{n-a,n-b}$ is i.i.d.\ standard Gaussian, and $B=U^\top\bar C V$ has rank $r\le2p$. No singular covariance factor is inverted.

Expand $\det(B+Z)$ into products of complementary minors. Minors of $B$ above order $r$ vanish, and Gaussian minors lie in the positive block, giving
    \begin{equation}\label{eq:app-det-expansion}
      \det(B+Z)
      =
      \sum_{\alpha\in\mathcal I_N}
      c_\alpha\det G_{S_\alpha,T_\alpha},
      \qquad
      n-2p\le |S_\alpha|=|T_\alpha|\le n,
    \end{equation}
The coefficients $c_\alpha$ absorb signs, complementary $B$-minors, diagonal factors, and powers of $N^{-1/2}$. Their number satisfies
    \begin{equation}\label{eq:app-number-of-minors}
      |\mathcal I_N|
      \le
      \sum_{k=0}^{r}\binom nk^2
      \le N^{C_p}.
    \end{equation}
    In particular, the null directions remove at most $O(p)$ rows and columns from the Gaussian minors.

    After terms with identical index pairs are combined, the distinct minors in \eqref{eq:app-det-expansion} are pairwise orthogonal in $L^2$. Indeed, minors of different orders lie in different Wiener chaoses, and for two distinct minors of the same order, some row or column of the Gaussian matrix appears with odd degree in their product. Consequently,
    \begin{equation}\label{eq:app-det-second}
      \E|\det(B+Z)|^2
      =
      \sum_{\alpha\in\mathcal I_N}
      c_\alpha^2\,
      \E|\det G_{S_\alpha,T_\alpha}|^2.
    \end{equation}
    For an $m\times m$ matrix with i.i.d.\ standard Gaussian entries, $\E(\det G_m)^2=m!, $ $\E(\det G_m)^4=\frac12\,m!(m+2)!,$ and therefore
    \begin{equation}\label{eq:app-central-det-ratio}
      \frac{\E|\det G_m|^4}
      {\bigl(\E|\det G_m|^2\bigr)^2}
      =
      \frac{(m+1)(m+2)}2
      \le Cn^2 .
    \end{equation}
    Let $L=|\mathcal I_N|$.  By H\"older's inequality, $\left|\sum_{\alpha\in\mathcal I_N}x_\alpha\right|^4 \le L^3\sum_{\alpha\in\mathcal I_N}|x_\alpha|^4.$ Combining this inequality with \eqref{eq:app-number-of-minors}--\eqref{eq:app-central-det-ratio}, we obtain
    \begin{equation*}
\E|\det(B+Z)|^4 \le Cn^2L^3 \sum_{\alpha\in\mathcal I_N} c_\alpha^4 \bigl(\E|\det G_{S_\alpha,T_\alpha}|^2\bigr)^2 \le N^{C_p} \left( \sum_{\alpha\in\mathcal I_N} c_\alpha^2 \E|\det G_{S_\alpha,T_\alpha}|^2 \right)^2.
\end{equation*}
    Together with \eqref{eq:app-det-second} and the bound $p\le K$, this proves \eqref{eq:determinant-fourth-moment}.
  \end{proof}

  \subsection{Construction of a feasible point with positive population value}
  \label{app:positive-value-point}

  \begin{lemma}[A feasible population point with positive value]
    \label{lem:app-positive-value-point}
    Under the assumptions of Lemma~\ref{lem:positive-branch}, there is a column-feasible $\omega$ with $\cC_\omega\ne\varnothing$ and $\phi_t(\omega)>d_t(\omega)$; hence $V_t>0$.
  \end{lemma}

  \begin{proof}
    For $p=0$, the claim is \eqref{eq:stage1-feasibility} written
    using \eqref{eq:empty-history-functionals}. Assume henceforth $p\ge1$.

    Let $G\sim N(0,1)$ be independent of the old free-column mark, and set $L:=1-|X_p|$. For $0<\eps\le1$, define $Y_\eps := X_p+\eps L\tanh G .$ We show that the parameter $\omega_\eps$ induced by $Y_\eps$ has strictly positive population value for all sufficiently small $\eps$.

\paragraph{Column feasibility and a lower bound on the width.}
    The bound $|Y_\eps|\le|X_p|+\eps(1-|X_p|)\le1$ gives box feasibility.
    Since $W_\eps=\eps L\tanh G$, independence and $\E\tanh G=0$ give
    $q\E[XW_\eps]=q\E[RW_\eps]=0$, hence $a_\eps=b_\eps=0$ and
    \begin{align}
      \rho_{\omega_\eps}^2&=q\E W_\eps^2
       =q\eps^2\E L^2\,\E\tanh^2G,
       \label{eq:app-positive-value-rho}\\
      \phi_t(\omega_\eps)&\ge q\E[GY_\eps]
       =q\eps\,\E L\,\E(G\tanh G),
       \label{eq:app-positive-value-width}\\
      \phi_t(\omega_\eps)&\ge c_1\eps\qquad(c_1>0).
       \label{eq:app-positive-value-linear-gain}
    \end{align}
    The width bound uses the feasible test variable $Y_\eps$.
    Since $|X_p|<1$ almost surely by Definition~\ref{def:strict-history},
    $\E L>0$; also $\E(G\tanh G)>0$. Thus $\rho_{\omega_\eps}\asymp\eps$
    and the constant $c_1$ is positive.

    \paragraph{Row repair.}
    Since $a_\eps=b_\eps=0$, the row pre-projection margin \eqref{eq:population-row-baseline} reduces to $Z_\eps=U_p+\rho_{\omega_\eps}G_{\rm row}$, where $G_{\rm row}\sim N(0,1)$ is independent of the old row mark. Write $\rho:=\rho_{\omega_\eps}$. The margin constraint from the previous round gives $U_p \ge \kappa_{t-1} = \kappa_t+\Delta_t,$ and therefore $(\kappa_t-Z_\eps)_+ \le (-\Delta_t-\rho G_{\rm row})_+.$ Gaussian tail bounds imply that, for every fixed $m\ge1$,
    \begin{equation}
      \E(\kappa_t-Z_\eps)_+^m
      \le
      C_m\rho^m
      \exp\left\{
      -\frac{c_m\Delta_t^2}{\rho^2}
      \right\}.
      \label{eq:app-positive-value-obstacle-moment}
    \end{equation}

    The pointwise projection $\max\{\kappa_t,Z_\eps\}$ onto the obstacle need not satisfy the old-dual moment constraint exactly. To correct the moment, we introduce the map
    \begin{equation*}
      F_\eps(\eta)
      :=
      \alpha\E\!\left[
      \Lambda
      \max\{\kappa_t,Z_\eps-\eta^\top\Lambda\}
      \right]
      -
      \Gamma e_p,
      \qquad
      \eta\in\R^p.
    \end{equation*}
    Since the pre-projection margin already satisfies the old-dual moment constraint $\alpha\E(\Lambda Z_\eps)=\Gamma e_p$, we have $F_\eps(0) = \alpha\E[ \Lambda(\kappa_t-Z_\eps)_+ ].$ By the Cauchy--Schwarz inequality, the bound $\E\|\Lambda\|^2<\infty$, and \eqref{eq:app-positive-value-obstacle-moment},
    \begin{equation}
      \|F_\eps(0)\|
      \le
      C
      \exp\left\{
      -\frac{c\Delta_t^2}{\rho^2}
      \right\}.
      \label{eq:app-positive-value-defect}
    \end{equation}

    At every point of differentiability, $DF_\eps(\eta) = -\alpha\E\!\left[ \Lambda\Lambda^\top \1\{ Z_\eps-\eta^\top\Lambda>\kappa_t \} \right].$ For $\eps=0$ and $\eta=0$, we have $Z_0=U_p>\kappa_t $ $\text{a.s.},$ and hence $DF_0(0) = -\alpha\E[\Lambda\Lambda^\top] = -Q^\lambda.$ By Definition~\ref{def:strict-history}, $Q^\lambda\succ0$.  By dominated convergence, there exist constants $r_0,c_0>0$ such that, for all sufficiently small $\eps$, $-DF_\eps(\eta)\succeq c_0I, $ $\|\eta\|\le r_0.$ Thus $F_\eps$ is uniformly strongly monotone near the origin.  Combined with \eqref{eq:app-positive-value-defect}, a quantitative inverse function theorem yields a zero $\eta_\eps$ satisfying
    \begin{equation}
      F_\eps(\eta_\eps)=0,
      \qquad
      \|\eta_\eps\|
      \le
      C
      \exp\left\{
      -\frac{c\Delta_t^2}{\rho^2}
      \right\}.
      \label{eq:app-positive-value-eta}
    \end{equation}

    Define $U_\eps:=\max\{\kappa_t,Z_\eps-\eta_\eps^\top\Lambda\}$. By construction, $U_\eps\ge\kappa_t, $ $\alpha\E(\Lambda U_\eps)=\Gamma e_p,$ so $U_\eps\in\cC_{\omega_\eps}$. Moreover, $U_\eps-Z_\eps = (\kappa_t-Z_\eps+\eta_\eps^\top\Lambda)_+ - \eta_\eps^\top\Lambda.$ Using the $1$-Lipschitz property of $x\mapsto x_+$, together with \eqref{eq:app-positive-value-obstacle-moment} and \eqref{eq:app-positive-value-eta}, we obtain $\alpha\E(U_\eps-Z_\eps)^2 \le C \exp\left\{ -\frac{c\Delta_t^2}{\rho^2} \right\}.$ Since $d_t(\omega_\eps)$ is $\sqrt\alpha$ times the $L^2$ distance from $Z_\eps$ to $\cC_{\omega_\eps}$, $d_t(\omega_\eps) \le C \exp\left\{ -\frac{c\Delta_t^2} {\rho_{\omega_\eps}^2} \right\}.$

By \eqref{eq:app-positive-value-rho}, $\rho_{\omega_\eps}\asymp\eps$, so $d_t(\omega_\eps)=O(e^{-c/\eps^2})=o(\eps)$. Together with \eqref{eq:app-positive-value-linear-gain}, this gives $\phi_t(\omega_\eps)>d_t(\omega_\eps)$ for small $\eps$ and therefore $V_t\ge\Psi_t(\omega_\eps)>0$ by \eqref{eq:positive-recursion-value}.

On a compact family of strict inputs, $q$, $\Delta_t$, $\lambda_{\min}(Q^\lambda)$, and $\E(1-|X_p|)$ are bounded away from zero, and the squared row marks are uniformly integrable. Hence the preceding estimates hold uniformly over the family, which yields a common choice of $\eps$ and a uniform positive lower bound on $V_t$. This also proves Lemma~\ref{lem:positive-branch}.
  \end{proof}

  \subsection{Finite multipliers at the population optimizer}
  \label{app:population-multipliers}

  \begin{lemma}\label{lem:app-population-multipliers}
    Under the hypotheses of Lemma~\ref{lem:positive-branch}, the population maximum is attained. At every maximizer, the column and row dual optima are attained at finite multipliers, with $\tau_\omega>0$. Moreover, at every maximizer, the column width with $(a,b)$ held fixed is differentiable in the transverse standard deviation, with $\partial_\rho\phi_t=\tau_\omega\rho_\omega$. On compact families of such inputs, these multipliers can be chosen to be uniformly bounded, and $V_t$, $\rho_\omega$, $d_t(\omega)$, $\tau_\omega$, and $\gamma_\omega$ are bounded away from zero.
  \end{lemma}

  \begin{proof}
    We use the notation $Q,D,h,T$ of Section~\ref{subsec:population-uniqueness}. For $|Y|\le1$, let $a(Y),b(Y)$ denote the moments in \eqref{eq:population-column-moments}, and let $c(Y)=q\E[GY]$. By \eqref{eq:uniqueness-T}, $\|T(Y)\|^2=q\E Y^2-h(a(Y))$. Let $Z(Y,\rho)$ denote the right-hand side of \eqref{eq:population-row-baseline} with these moments, where $\rho$ is treated as an independent variable, and set $r=W-Z(Y,\rho)$ and $\mathcal H=\alpha\E[\Lambda W]-\Gamma e_p-b(Y)$. Consider the lifted convex problem
    \begin{equation*}
      V_L=\max v:\quad
      |Y|\le1,\quad W\ge\kappa_t,\quad \mathcal H=0,\quad
      \|T(Y)\|\le\rho,\quad
      \sqrt{v^2+\alpha\|r\|_2^2}\le c(Y).
    \end{equation*}
    Let $f_1,f_2$ denote the left-hand sides minus the right-hand sides of the last two inequalities. The construction in Appendix~\ref{app:positive-value-point} provides a feasible point with positive value.

    \emph{Existence and finite multipliers.}
    Let $C=q\E|G|$. Every feasible point satisfies $|v|,\sqrt\alpha\|r\|_2\le C$. The moments $a,b$ are bounded, so $z(Y)=Z(Y,\rho)-\rho G_{\rm row}$ is bounded in $L^2$. Since $W\ge\kappa_t$,
    \[
      \|r\|_2\ge\|(\kappa_t-z(Y)-\rho G_{\rm row})_+\|_2
      \ge\rho/\sqrt2-\|\kappa_t-z(Y)\|_2.
    \]
    Thus $\rho,W$ are bounded as well. The feasible set is therefore bounded and weakly closed in the product Hilbert space, hence weakly compact, so $V_L>0$ is attained.

Decrease $v$ slightly at a positive-value feasible point $z=(Y,W,\rho,v)$,
    keeping $v>0$, so $f_2(z)<0$. Use the anchor
    $z_0=(X_p,U_p,0,0)$ for $p\ge1$ and $z_0=(0,\kappa_1+1,0,0)$ for $p=0$.
    It satisfies the box, obstacle, and moment constraints, with $f_1(z_0)=0$
    and obstacle reserve $\delta_0=\Delta_t$ or $1$, respectively.
    For $z_\theta=(1-\theta)z+\theta z_0$, convexity and affinity give
    $f_1(z_\theta)\le0$, $\mathcal H(z_\theta)=0$, and
    $W_\theta\ge\kappa_t+\theta\delta_0$.
    Continuity in the product $L^2$ topology preserves $f_2<0$ for small
    fixed $\theta>0$. Increasing $\rho$ slightly then makes $f_1<0$
    while preserving $f_2<0$ and the unchanged box, moment, and obstacle constraints.

For $p\ge1$, choose $L<\infty$ with
    $K_L=\alpha\E[\Lambda\Lambda^\top\1_{\{\|\Lambda\|\le L\}}]\succ0$.
    The bounded correction
    \[
      B_h=\1_{\{\|\Lambda\|\le L\}}\Lambda^\top K_L^{-1}h,
      \qquad \alpha\E[\Lambda B_h]=h,\qquad \|B_h\|_\infty=O(\|h\|)
    \]
    changes $\mathcal H$ to $h$ when $W$ is replaced by $W+B_h$.
    For small $h$ it preserves the obstacle reserve and $f_2<0$, since
    $|f_2(Y,W+B_h,\rho,v)-f_2(Y,W,\rho,v)|\le\sqrt\alpha\|B_h\|_2$;
    $f_1$ is unchanged. At $p=0$ there is no moment constraint to correct.

    The convex perturbation function
    $p(u,h)=\inf\{-v:|Y|\le1,\ W\ge\kappa_t,\ f_i\le u_i, \ \mathcal H=h\}$
    is finite near zero in its finite-dimensional parameter space,
    with $p(u,h)\ge-C-u_2$. A subgradient at the origin and monotonicity
    in $u$ give $m,\ell\ge0$ and $\eta\in\R^p$ such that
    $\mathcal L=-v+mf_1+\ell f_2+\eta^\top\mathcal H\ge p(0).$
    Equality holds at every lifted optimum, so each minimizes the
    Lagrangian over the box and obstacle constraints and satisfies
    complementary slackness.

    \emph{Row and column optimality.}
    Fix a lifted optimum and write $v=V_L$, $c=c(Y)$, and $d=\sqrt\alpha\|r\|_2$.
    If $\rho=0$, then $T(Y)=0$. For $p\ge1$, the second component
    $D^{1/2}Q^{-1}a(Y)=0$ and $D\succ0$ imply $a(Y)=0$; the first
    component then gives $Y=X_p$. For $p=0$ the same formula gives
    $Y=0$. In either case the independence and centering of $G$
    imply $c(Y)=q\E[GY]=0$, contradicting $c\ge v>0$.
    Thus $\rho>0$. Moreover, $d>0$ because of the lower tail of the fresh Gaussian variable. The second norm constraint is active, so stationarity in $v$ gives $\ell=c/v$. The variational inequality for the row variable gives the finite multiplier $\eta_\lambda=v\eta,$ $W=\max\{\kappa_t,Z-\eta_\lambda^\top\Lambda\}.$ Together with $\mathcal H=0$, this shows that the row variable is the row projection and that the row dual is attained. Gaussian integration by parts, applied conditionally on the old marks, gives $\E[rG_{\rm row}]=-\rho\Pp(W=\kappa_t)$; here the conditional derivative is bounded by $\rho$, and the conditional law has no atom at the threshold. Stationarity in the unconstrained variable $\rho$, together with complementary slackness, now yields
    \begin{equation}\label{eq:app-multiplier-exactness}
      m=\frac{\alpha\rho}{v}\Pp(W=\kappa_t)>0,
      \qquad \rho=\|T(Y)\|.
    \end{equation}

    Let $\beta=Q^{-1}a$ and $B_x=Q^{-1}\{U-\Gamma^\top(Q^\lambda)^{-1}\Lambda\}$. The row moment identity gives $\alpha\E[\Lambda r]=0$, and $D_Y\|T(Y)\|[\dot Y] =\frac q\rho\E[(Y-X_p-\beta^\top X)\dot Y].$ Hence the variational inequality for the column variable gives
    \begin{equation*}
      \tau=\frac m{\ell\rho}>0,\qquad
      \eta_x=\tau(e_p+\beta)+\frac\alpha c\E[rB_x],\qquad
      \eta_r=\frac{\eta_\lambda}{c},
    \end{equation*}
    and $Y=\clip((G+\eta_x^\top X+\eta_r^\top R)/\tau,-1,1)$. All coefficients are finite by the $L^2$ bounds above. Maximizing the quadratic pointwise and then cancelling the fixed moments shows that $c=\phi_t(\omega)$ and that the infimum in \eqref{eq:population-column-dual} is attained. By \eqref{eq:app-multiplier-exactness}, this optimum is feasible for the original maximization over parameters, with $v=\Psi_t(\omega)$. Conversely, for any parameter with positive value in the original problem, a maximizing sequence for the column width, combined with the row projection, gives values of the lifted problem converging to $\Psi_t(\omega)$. Hence $V_L=\max_\omega\Psi_t(\omega)$.

    Now fix any maximizer $\omega^\star$ of the original problem. Realize a maximizing sequence $Y_n$ for its column width using a common auxiliary seed, and set $W=P_{\mathcal C_{\omega^\star}}Z_{\omega^\star}$. A weak $L^2$ subsequential limit $Y$ still satisfies the box constraint, has the same moments $a,b$ and the same value $c=\phi_t(\omega^\star)$, and satisfies $\|T(Y)\|\le\rho_{\omega^\star}$. Thus $(Y,W,\rho_{\omega^\star},V_L)$ is optimal for the lifted problem. By \eqref{eq:app-multiplier-exactness}, equality holds in the transverse-variance constraint, so $q\E Y^2=s^\star$. Consequently, every maximizer of the original problem admits the finite multipliers above. The two Gaussian integration-by-parts identities give
    $q\Pp(|Y|<1)=\tau c=\alpha\Pp(W=\kappa_t)>0.$

Finally, let $A=\{|Y|\le1/2\}$, $M=\E[Y\1_A\mid X,R]/\Pp(A\mid X,R)$, and $H=\1_A(Y-M)$. The clipping formula gives $\E[H\mid X,R]=0$ and $e:=\E H^2=\E[YH]>0$. For all sufficiently small $u$ of either sign, $Y+uH$ remains in the box and preserves $a,b$, while $s(u)=s+qe(2u+u^2)$ and $c(u)=c+q\tau e u$. Hence the feasible second moments contain a two-sided neighborhood of $s$. Evaluating the column dual at the multipliers for $s$ bounds $\phi_t(s',a,b)$ from above by $c+\tau(s'-s)/2$, and the perturbation gives the same bound from below up to an error $O((s'-s)^2)$. Consequently, $\partial_s\phi_t=\tau/2$ and $\partial_\rho\phi_t=\tau\rho$, as claimed.

    \emph{Uniformity on compact families.}
    The input assumptions and Appendix~\ref{app:positive-value-point} give
    uniform bounds on the inputs and $V_t\ge v_*>0$.
    The strictly feasible point and, when needed, the bounded correction $B_h$ above can be chosen with a common perturbation radius, because uniform integrability permits a common truncation level with $K_L\succeq cI$. The convex perturbation functions are uniformly bounded on this neighborhood, so their subgradients, and hence $m,\ell,\eta$, are uniformly bounded. The coercivity estimates above also bound $\rho$ and $\|W\|_2$ uniformly.
    The uniformly positive residual column covariance gives
    \[
      \rho^2\ge c_0q\E(Y-X_p)^2,\qquad
      c(Y)=q\E[G(Y-X_p)]\le\sqrt q\,\sqrt{q\E(Y-X_p)^2}.
    \]
    Since $c(Y)\ge V_t\ge v_*$, this bounds $\rho$ away from zero. The row means and the multiplier $\eta_\lambda=v\eta$ are uniformly bounded, in $L^2$ and in norm, respectively. Conditioning on the row means and using the fresh Gaussian variable, we therefore bound both $\Pp(W=\kappa_t)$ and $d$ away from zero. The identities $\tau=\frac{\alpha\Pp(W=\kappa_t)}{c(Y)}, $ $\gamma=\frac d{V_t}$ give the remaining lower bounds, and the formulas above then show that the column and row multipliers are uniformly bounded.
  \end{proof}

  \begin{lemma}[Continuity in the input]\label{lem:app-input-continuity}
  Fix the dual cutoff $R_{\rm dual}=C/v_*$ of Section~\ref{subsec:population-geometry}, which is uniform over the family, and let $\mathfrak h_n\to\mathfrak h$ in a compact family of inputs satisfying the assumptions of Section~\ref{sec:finite-po}. Then $V_{t,n}\to V_t$, $\omega_{t,n}^\star\to\omega_t^\star$, and the laws $\law(X,R,V,Y_t)$ and $\law(U,\Lambda,Z_t,U_t)$ converge in $\Wtwo$. Moreover, for every $\eps>0$ there is $c(\eps)>0$, uniform over the family, such that $\Psi_{t,R_{\rm dual}}(\omega)\le V_t-c(\eps)$ whenever $\|\omega-\omega_t^\star\|\ge\eps$.
  \end{lemma}

  \begin{proof}
  The subscript $n$ indicates the input. We couple the marks in $L^2$ using common fresh Gaussian variables. The bounds of Lemma~\ref{lem:app-population-multipliers} yield convergent subsequences of optimal parameters and multipliers, with $\tau_n$ bounded away from zero. The clipping and projection formulas then give strong $L^2$ convergence, so the lifted KKT conditions pass to the limit and the laws $\law(X,R,V,Y_t)$ and $\law(U,\Lambda,Z_t,U_t)$ converge in $\Wtwo$. By Proposition~\ref{prop:population-optimizer-unique}, $V_n\to V$ and $\omega_n^\star\to\omega^\star$.

  For near-optimal parameters, the boundedness of the column variables and of the row costs yields subsequential limits $\omega_n\to\omega$ and subsequential limits of the joint laws; the moments of the latter converge by the $\Wtwo$ convergence of the old marks. The column correlations converge and the row cost is lower semicontinuous, so $\limsup_n\Psi_{n,R_{\rm dual}}(\omega_n)\le\Psi_{R_{\rm dual}}(\omega)$. Thus $\omega=\omega^\star$. If the gap were not uniform, compactness would yield such a sequence with $\|\omega_n-\omega_n^\star\|\ge\eps$, a contradiction.
  \end{proof}

  \subsection{Finite-to-population convergence and recovery}
  \label{app:finite-population}

  We prove Proposition~\ref{prop:finite-ao-gaps}. Set $p=t-1$ and suppress $R_m$ from $d_{N,h}^{(R_m)}$. The fixed cutoffs satisfy $R_{\rm dual}>\gamma_t$ and $R_m^2>\alpha\E U_t^2$ with uniform positive slack over the prescribed compact input family.

  \begin{proposition}[Finite AO approximation and restricted gaps]
    \label{prop:app-finite-ao-gaps}
    The value convergence and the parameter, column-law, and row-law gaps in Proposition~\ref{prop:finite-ao-gaps} hold uniformly over its prescribed family of finite histories.
  \end{proposition}

  The row estimate relies on the stability of the finite row projection for order parameters near the optimal one. We write $L_p$ for the old multiplier matrix, with rows $\Lambda_{p,i}^{\top}$.

  \begin{lemma}[Stability of the finite row projection]
    \label{lem:app-row-projection-stability}
    Let $\omega_N=(s_N,a_N,b_N)$ be admissible finite order parameters, possibly depending on the fresh Gaussian variables, with $\omega_N\xrightarrow{\Pp_h}\omega_t^\star$, and set $Z_N=Z_{N,h}(\omega_N)$. Then the row set $\mathcal C_{N,h}^{(R_m)}(b_N)$ is nonempty with high conditional probability, and the Euclidean projection $\widehat m_N=P_{\mathcal C_{N,h}^{(R_m)}(b_N)}Z_N$ onto this set satisfies, in conditional probability,
    \begin{equation}\label{eq:app-row-projection-stability}
      \frac1M\sum_{i=1}^M
      \delta_{(U_{p,i},\Lambda_{p,i},Z_{N,i},\widehat m_{N,i})}
      \xrightarrow{\Wtwo}\law(U,\Lambda,Z_t,U_t),
      \qquad
      d_{N,h}(\omega_N)\longrightarrow d_t^\star.
    \end{equation}
  \end{lemma}

  \begin{proof}
    The coefficients of $z_{N,h}(a_N,b_N)$ and $\rho_{N,h}(\omega_N)$ converge to their population values. We couple the fresh Gaussian row vector as $Q_{L,p}G_{\rm row} =P_{L,p}^{\perp}\widetilde G_{\rm row}, $ $\widetilde G_{\rm row}\sim N(0,I_M).$ Since $P_{L,p}$ has rank $p$, $N^{-1}\|P_{L,p}\widetilde G_{\rm row}\|^2=o_{\Pp_h}(1)$. The convergence of the old row laws and the conditional law of large numbers give
    \begin{equation}\label{eq:app-recovery-row-law}
      \frac1M\sum_{i=1}^M
      \delta_{(U_{p,i},\Lambda_{p,i},Z_{N,i})}
      \xrightarrow{\Wtwo}\law(U,\Lambda,Z_t).
    \end{equation}
    Perturbations of the coefficients have vanishing normalized $L^2$ norm, so this conclusion holds for every convergent sequence $\omega_N$, even when $\omega_N$ depends on the fresh Gaussian variables.

At $p=0$ both multipliers are empty and no moment equation is needed.
    For $p\ge1$, define the finite moment map and its limit by
    \begin{align*}
      F_N(\eta)&=\frac1N\sum_i\Lambda_{p,i}
       \max\{\kappa_t,Z_{N,i}-\eta^\top\Lambda_{p,i}\}-(\Gamma_pe_p+b_N),\\
      F(\eta)&=\alpha\E[\Lambda\max\{\kappa_t,Z_t-\eta^\top\Lambda\}]
                     -(\Gamma e_p+b_t^\star),
      \qquad F(\eta_{\lambda,t})=0.
    \end{align*}

    The integrand has quadratic growth and is Lipschitz in $\eta$ with constant $\|\Lambda\|^2$. Hence \eqref{eq:app-recovery-row-law}, combined with a finite-net argument on each fixed ball in $\eta$, implies
    \begin{equation}\label{eq:app-recovery-F-uniform}
      \sup_{\|\eta-\eta_{\lambda,t}\|\le r}
      \|F_N(\eta)-F(\eta)\|\xrightarrow{\Pp_h}0.
    \end{equation}
    Since the fresh Gaussian component has variance $\rho_t^2>0$, the argument of the maximum has no conditional atom at the threshold, and
    \[
      DF(\eta_{\lambda,t})
      =-\alpha\E\!\left[\Lambda\Lambda^{\top}
      \1\{Z_t-\eta_{\lambda,t}^{\top}\Lambda>\kappa_t\}\right]
      \prec0.
    \]
    Indeed, the conditional probability of the event in the indicator is positive for every old row mark, and $Q^\lambda\succ0$. By dominated convergence, $DF$ is continuous near $\eta_{\lambda,t}$. Hence there are $c,r>0$ such that
    \begin{equation}\label{eq:app-recovery-strong-monotonicity}
      (\eta-\eta_{\lambda,t})^{\top}F(\eta)
      \le-c\|\eta-\eta_{\lambda,t}\|^2,
      \qquad \|\eta-\eta_{\lambda,t}\|\le r.
    \end{equation}
    Consequently, with high conditional probability, $F_N$ points strictly inward on the boundary of this ball. Brouwer's fixed-point theorem then yields a zero $\eta_{\lambda,N}$ in the ball. Indeed, if $F_N$ had no zero, the map $\eta\mapsto\eta_{\lambda,t}+rF_N(\eta)/\|F_N(\eta)\|$ would have a fixed point on the boundary, contradicting the inward inequality. At this zero, strong monotonicity and Cauchy--Schwarz give
    \[
      c\|\eta_{\lambda,N}-\eta_{\lambda,t}\|
      \le\sup_{\|\eta-\eta_{\lambda,t}\|\le r}\|F_N(\eta)-F(\eta)\|
      \xrightarrow{\Pp_h}0.
    \]
    This completes the multiplier construction for $p\ge1$.

For either depth set $\widehat m_{N,i}=\max\{\kappa_t,Z_{N,i}-\eta_{\lambda,N}^\top\Lambda_{p,i}\}$.
    Its old-dual moments are exact by $F_N(\eta_{\lambda,N})=0$ at positive depth;
    at depth zero it is simply $\max\{\kappa_1,Z_{N,i}\}$.
    For any obstacle-feasible $m$ with the same moments,
    \[
     \langle\widehat m_N-Z_N+L_p\eta_{\lambda,N},m-\widehat m_N\rangle\ge0,
     \qquad L_p^\top(m-\widehat m_N)=0.
    \]
    The second identity cancels the multiplier term. Expanding the square gives
    $\|m-Z_N\|^2\ge\|\widehat m_N-Z_N\|^2+\|m-\widehat m_N\|^2$,
    identifying the unique projection before the row cutoff is imposed.

    The map $(\Lambda,Z)\mapsto\max\{\kappa_t,Z-\eta^{\top}\Lambda\}$ is Lipschitz for each fixed $\eta$. Under $\eta_{\lambda,N}\to\eta_{\lambda,t}$, its normalized squared perturbation is at most $\|\eta_{\lambda,N}-\eta_{\lambda,t}\|^2\|L_p\|_F^2/M=o_{\Pp_h}(1)$. Consequently, \eqref{eq:app-recovery-row-law} yields the $\Wtwo$ convergence of the marked empirical measures in \eqref{eq:app-row-projection-stability}, together with
    $\frac1N\|\widehat m_N-Z_N\|^2\longrightarrow(d_t^\star)^2, \allowbreak\quad \frac1N\|\widehat m_N\|^2\longrightarrow\alpha\E U_t^2<R_m^2$.
    The projection therefore lies within the row cutoff, and hence it coincides with the projection onto $\mathcal C_{N,h}^{(R_m)}(b_N)$.
  \end{proof}

  \begin{lemma}[Finite recovery of the optimizer]
    \label{lem:app-finite-recovery}
    There exist $y_N^\star\in[-1,1]^{F_p}$ and $m_N^\star\in\mathcal C_{N,h}^{(R_m)}(b_N^\star)$ such that, with $\omega_N^\star$ the order parameter induced by $y_N^\star$,
    \[
      \omega_N^\star\longrightarrow\omega_t^\star,
      \qquad c_{v,N}(y_N^\star)\longrightarrow V_t,
      \qquad c_{g,N}(y_N^\star)\longrightarrow d_t^\star,
      \qquad
      \frac{\|m_N^\star-Z_{N,h}(\omega_N^\star)\|}{\sqrt N}
      \longrightarrow d_t^\star.
    \]
    All convergences hold in conditional probability, uniformly over $h\in\mathcal T_{p,N}(\mathfrak h_p)$.
  \end{lemma}

  \begin{proof}
    Let $\widetilde g=\Sigma_{p,F_p}^{1/2}g$ and $S_{N,j}=(v_{t,j}+\gamma_t\widetilde g_j)/\sqrt{1+\gamma_t^2}$, and define
    \begin{equation}\label{eq:app-recovery-column}
      y_{N,j}^\star=\clip\left(
      \frac{S_{N,j}+\eta_{x,t}^{\top}X_{p,j}
      +\eta_{r,t}^{\top}R_{p,j}}{\tau_t},-1,1\right),
      \qquad j\in F_p.
    \end{equation}
    Since $\Sigma_{p,F_p} =I-X_{p,F_p}(X_p^{\top}X_p)^{-1}X_{p,F_p}^{\top},$ the matrix $\Sigma_{p,F_p}^{1/2}-I$ has rank at most $p$ and bounded operator norm. Hence $N^{-1}\|\widetilde g-g\|^2=o_{\Pp_h}(1)$. Replacing $\widetilde g$ by $g$ in \eqref{eq:app-recovery-column} changes $y_N^\star$ by $o_{\Pp_h}(\sqrt N)$ in Euclidean norm, because clipping is Lipschitz and $\tau_t>0$. The conditional law of large numbers for $(v_t,g)$ and the convergence of the old marked laws therefore give
    \[
      \frac1{|F_p|}\sum_{j\in F_p}
      \delta_{(X_{p,j},R_{p,j},v_{t,j},g_j,y_{N,j}^\star)}
      \xrightarrow{\Wtwo}\law(X,R,V,G,Y_t).
    \]
    In particular, $\omega_N^\star\longrightarrow\omega_t^\star.$ With $w_N^\star=y_N^\star-x_{F_p}^{(p)}$, the same convergence gives
    $c_{v,N}(y_N^\star)\longrightarrow q\E[VY_t], \allowbreak\quad c_{g,N}(y_N^\star) =\frac1N\widetilde g^{\top}w_N^\star \longrightarrow q\E[GY_t]$.
    The envelope formula for the support function and the stationarity condition in the scalar variable imply
    \[
      q\E[GY_t]
      =\frac{\gamma_t}{\sqrt{1+\gamma_t^2}}\phi_t^\star=d_t^\star,
      \qquad
      q\E[VY_t]
      =\sqrt{1+\gamma_t^2}\phi_t^\star-\gamma_td_t^\star=V_t.
    \]
    Thus $c_{v,N}(y_N^\star)\longrightarrow V_t, $ $c_{g,N}(y_N^\star)\longrightarrow d_t^\star.$ Finally, we apply Lemma~\ref{lem:app-row-projection-stability} to $\omega_N^\star$ and take $m_N^\star=\widehat m_N$.
  \end{proof}

  \begin{proof}[Proof of Proposition~\ref{prop:app-finite-ao-gaps}]
    We use the subsequence criterion for convergence in probability. Passing to a further subsequence, we may assume that the marked Gaussian environments converge in $\Wtwo$ and that the normalized finite-rank errors vanish almost surely. The compactness arguments below are carried out pathwise for these environments.

    \paragraph{Value convergence.}
    Take a near-maximizing sequence for \eqref{eq:finite-hinge}. The finite order parameters range over a compact set. Along a subsequence, the order parameters and the column correlations satisfy $\omega_N\to\omega, $ $c_{v,N}\to q\E[VY], $ $c_{g,N}\to q\E[GY]$ for some population-feasible $Y$. The marked column laws converge in $\Wtwo$, because $Y$ is bounded and the old marks and fresh Gaussians have uniformly integrable squares.

Bounded row costs give a weakly convergent subsequence of marked row laws, and the obstacle passes to its limit. To pass the old-dual moments, truncate the old multiplier marks: bounded marks and the row $L^2$ bound make the products uniformly integrable, while Cauchy--Schwarz and uniform integrability of the squared old marks control the tails. Replacing the limiting row variable by its conditional mean given the old marks and Gaussian pre-projection margin preserves the obstacle and moments and decreases squared cost by Jensen's inequality. Thus the limit is population-feasible and $\liminf_N d_{N,h}(\omega_N)\ge d_t(\omega)$. Hence
    \[
      \limsup_N\left\{
      c_{v,N}-R_{\rm dual}(d_{N,h}(\omega_N)-c_{g,N})_+
      \right\}
      \le\mathcal J_t(Y,\omega)\le V_t.
    \]
    For the reverse inequality, Lemma~\ref{lem:app-finite-recovery} provides a feasible pair with $c_{v,N}\to V_t$ and $d_{N,h}(\omega_N^\star)-c_{g,N}\to0$, whose value therefore converges to $V_t$.

    \paragraph{Parameter and column restrictions.}
    Consider finite feasible pairs whose normalized values tend to $V_t$. The preceding upper bound and the parameter gap in Proposition~\ref{prop:population-restricted-gaps}(i) force $\omega_N\to\omega_t^\star$. For a column-law bad sequence, the $\Wtwo$ limit of the marked laws remains at distance at least $\eps'$ from the selected law. Setting the scalar in the inner minimization equal to $\gamma_t$, we bound the limiting value of the sequence by
    \[
      q\E[(V+\gamma_tG)Y]-\gamma_td_t^\star
      \le V_t-\frac{q\tau_t\sqrt{1+\gamma_t^2}}2(\eps')^2,
    \]
    where the inequality follows from Proposition~\ref{prop:population-restricted-gaps}(ii). Hence neither bad set contains near-maximizers.

    \paragraph{Row restrictions.}
    Let $(y_N,m_N)$ be a near-maximizing row-law bad sequence. By the preceding step, $\omega_N\to\omega_t^\star$. Let $\widehat m_N$ be the projection from Lemma~\ref{lem:app-row-projection-stability}. Its marked law converges in $\Wtwo$ to $\law(U,\Lambda,U_t)$, whereas the marked law of $m_N$ stays at distance at least $\eps$ from this limit. The triangle inequality, together with the coupling that pairs coordinates with the same index, gives eventually $\frac1M\|m_N-\widehat m_N\|^2\ge(\eps')^2.$ The variational inequality for the finite projection gives
    \begin{equation*}
      \frac1N\|m_N-Z_N\|^2
      \ge d_{N,h}(\omega_N)^2
      +\frac1N\|m_N-\widehat m_N\|^2.
    \end{equation*}
    Since $M/N\to\alpha$, this yields $\liminf_N\frac1N\|m_N-Z_N\|^2 \ge(d_t^\star)^2+\alpha(\eps')^2.$ For $D_N=\|m_N-Z_N\|/\sqrt N$ and $0\le\gamma_t\le R_{\rm dual}$, we have $c_{v,N}-R_{\rm dual}(D_N-c_{g,N})_+ \le c_{v,N}+\gamma_tc_{g,N}-\gamma_tD_N.$ The limiting column coordinate has the selected moments, so the support-function bound gives $\limsup_N(c_{v,N}+\gamma_tc_{g,N}) \le\sqrt{1+\gamma_t^2}\phi_t^\star.$ Therefore the limit superior of the restricted values is at most
    $V_t-\gamma_t\left( \sqrt{(d_t^\star)^2+\alpha(\eps')^2}-d_t^\star\right)<V_t$,
    which contradicts near-maximality.

    If no positive constant as in \eqref{eq:app-finite-restricted-gap} existed, a diagonal argument would produce a sequence of bad near-maximizers contradicting one of these three conclusions. Applying the same argument to convergent sequences whose population inputs lie in the prescribed compact family, and using Lemma~\ref{lem:app-input-continuity}, yields the stated uniform bound, and hence Proposition~\ref{prop:finite-ao-gaps}.
  \end{proof}

  \subsection{Proof of the low-\texorpdfstring{$\ell_1$}{l1} recession estimates}
  \label{app:univ-low-l1}

  In this subsection we prove Lemma~\ref{lem:univ-low-l1}. For a matrix $Z$ with i.i.d.\ centered subgaussian entries of unit variance, set $T_r=\{q:\|q\|_2\le1,\ \|q\|_1\le r\sqrt N\}$ and, for a Lipschitz function $\psi:\mathbb R\to[0,1]$, let
  \[
    \mathcal E_{Z,r}(\psi)=\sup_{q\in T_r}\left|
    N^{-1}\sum_j\{\psi(\langle Z_{:j},q\rangle)
    -\mathbb E\psi(\langle Z_{:1},q\rangle)\}\right|.
  \]
  Its expectation is at most $C\|\psi\|_{\rm Lip}N^{-1/2}\mathbb E\sup_{q\in T_r}|\langle W,q\rangle|$ by symmetrization and the contraction principle, where $W=N^{-1/2}\sum_j\varepsilon_jZ_{:j}$ and the $\varepsilon_j$ are independent Rademacher signs. Conditionally on the signs, $W$ has independent coordinates with uniformly bounded subgaussian norms. With $k=\lceil r^2N\rceil$, we have
  \[
    \sup_{q\in T_r}|\langle W,q\rangle|
    \le \Bigl(\sum_{i\le k}|W|_{(i)}^2\Bigr)^{1/2}
    +r\sqrt N\,|W|_{(k+1)},
  \]
  and the right-hand side has expectation at most $Cr\sqrt{N\log(C/r^2)}$. Since changing one column changes $\mathcal E_{Z,r}$ by at most $N^{-1}$, we obtain
  \begin{equation}
    \mathbb E\mathcal E_{Z,r}(\psi)
    \le C\|\psi\|_{\rm Lip}r\sqrt{\log(C/r^2)}+O(N^{-1/2}),
    \qquad
    \mathbb P(\mathcal E_{Z,r}>\mathbb E\mathcal E_{Z,r}+u)
    \le e^{-cNu^2}.
    \label{eq:a4-empirical-process}
  \end{equation}
  The constants depend only on the entry law and on the range of the aspect ratio. Here $r>0$ is fixed and small enough that $r^2<\alpha_-/2$, where $\alpha_-\le M/N$, so that $k\le M$ for all sufficiently large $N$. All later choices of $r$ may be decreased further so that this bound holds.

  For part~(i), the variable $Z_q=\langle Y_{:1},q\rangle$ has unit second moment and fourth moment at most $C_4$, uniformly over unit vectors $q$. The Paley--Zygmund inequality gives $\mathbb P(|Z_q|\ge1/2)\ge p_0=9/(16C_4)$. Take a $4$-Lipschitz cutoff function $\phi\in[0,1]$ that vanishes on $[-1/4,1/4]$ and equals one outside $(-1/2,1/2)$. Then $\phi(s)/4\le|s|$ and $\mathbb E\phi(Z_q)\ge p_0$. Applying \eqref{eq:a4-empirical-process} to $1-\phi$, which is also $4$-Lipschitz with values in $[0,1]$, shows that the empirical average of $\phi$ is at least $p_0/2$ for sufficiently small fixed $r_1$; this proves (i) with constant $p_0/8$.

  For part~(ii), fix $t\ge2$ and write $\mathfrak m=\mathbb E(1-|\bar X_{t-1}|)>0$, $w_j=1-|\widehat x_{A^Y,j}^{(t-1)}|$, and $\widetilde w_j=1-|y_{A^Y,j}^{(t-1)}|$. Assume that the following bounds from the preceding round
  \begin{equation*}
    \left|N^{-1}\sum_jw_j-\mathfrak m\right|\le\mathfrak m/8,
    \qquad N^{-1/2}\|\widehat x_{A^Y}^{(t-1)}-y_{A^Y}^{(t-1)}\|_2
    \le d_{\rm sb}
  \end{equation*}
  hold with high probability; we choose $d_{\rm sb}$ below, independently of the surrogate parameters. Taking $d_{\rm sb}\le\mathfrak m/8$, we obtain $N^{-1}\sum_j\widetilde w_j\ge\mathfrak m/2$ and
  \begin{equation}
    N^{-1}\#\{j:\widetilde w_j\ge\zeta\}\ge\rho,
    \qquad \zeta=\mathfrak m/4,\quad
    \rho=(\mathfrak m/2-\zeta)/(1-\zeta)>0.
    \label{eq:a4-positive-density}
  \end{equation}
  Let $\widetilde w^{(-I)}$ denote the weights obtained by deleting the rows in $I$ from every surrogate round. By Appendix~\ref{app:univ-loo}, for each fixed $k$,
  \begin{equation}
    \max_{|I|\le k}N^{-1/2}\|\widetilde w-\widetilde w^{(-I)}\|_2=o_P(1).
    \label{eq:a4-deleted-weight-close}
  \end{equation}
  Hence, with high probability, the events $E_I=\{N^{-1}\sum_j\widetilde w_j^{(-I)}\ge\mathfrak m/3\}$ hold simultaneously for all $|I|\le k$. Each $E_I$ is measurable with respect to $\mathcal G_I=\sigma(Y_{I^c,:},v_1,\ldots,v_K)$.

  Let $c_\xi=\inf_{d,\|u\|_2=1}\mathbb E|\sum_{i\le d}u_i\xi_i| \ge C_4^{-1/2}$, where the lower bound follows by moment interpolation. Choose $C_A$ such that $\mathbb P(\|Y\|_{\rm op}\le C_A\sqrt N)\to1$, and set $c_{\rm fr}=c_\xi\mathfrak m/12$; this constant does not depend on $k$. We first choose $\sigma\in(0,1/2)$ and then choose $\theta,a>0$ small enough that
  \begin{equation}
    \frac{c_{\rm fr}}2\sqrt{1-\sigma^2}-C_A\sigma\ge c_{\rm fr}/4,
    \qquad C_\xi(a+\theta)/\sigma\le\rho/16,
    \label{eq:a4-diffuse-scales}
  \end{equation}
  where $C_\xi$ is a Berry--Esseen constant. We then fix $k=\lceil\theta^{-2}\rceil$. All of these choices are made before the surrogate parameters.

  For $1\le|I|\le k$ and a unit vector $u\in\mathbb R^I$, set $F_I(u)=N^{-1}\sum_j\widetilde w_j^{(-I)}|\langle Y_{I,j},u\rangle|$. Conditionally on $\mathcal G_I$, its summands are independent with uniformly bounded subexponential norms, and on $E_I$ its conditional mean is at least $c_\xi\mathfrak m/3$. Bernstein's inequality applied conditionally, followed by integration against the $\mathcal G_I$-measurable indicator of $E_I$, gives
  $\mathbb P(E_I\cap\{F_I(u)<c_\xi\mathfrak m/6\})\le e^{-cN}.$
  On the operator-norm event, $F_I$ is $C_A$-Lipschitz. Nets of mesh $c_\xi\mathfrak m/(12C_A)$ on all of these spheres have at most $C^kN^k$ points in total. A union bound therefore yields $\inf_{I,u}F_I(u)\ge c_{\rm fr}$ with high probability. By \eqref{eq:a4-deleted-weight-close}, replacing the deleted weights by $\widetilde w$ changes these averages by $o_P(1)$ uniformly, so
  \begin{equation}
    \inf_{\substack{1\le|I|\le k\\\|u\|_2=1}}
    N^{-1}\sum_j\widetilde w_j|\langle Y_{I,j},u\rangle|
    \ge c_{\rm fr}/2.
    \label{eq:a4-heavy-true-weight}
  \end{equation}

  For a unit vector $q$, let $I(q)=\{i:|q_i|>\theta\}$ and $R(q)=I(q)^c$. If $\|q_{R(q)}\|_2<\sigma$, then $|I(q)|\le k$, and the triangle inequality together with \eqref{eq:a4-heavy-true-weight} and \eqref{eq:a4-diffuse-scales} bounds the weighted average below by $c_{\rm fr}/4$. Otherwise, $s=\|q_{R(q)}\|_2\ge\sigma$. Conditionally on $Y_{I(q),j}$, the remaining sum has variance $s^2$, and its sum of third absolute moments is at most $C\theta s^2$. The Berry--Esseen theorem therefore gives, uniformly in the shift contributed by $I(q)$,
  \[
    \mathbb P(|\langle Y_{:j},q\rangle|\le2a\mid Y_{I(q),j})
    \le C_\xi(a+\theta)/s\le\rho/16.
  \]
  Choose an $a^{-1}$-Lipschitz cutoff function satisfying $\1_{\{|z|\le a\}}\le\psi_a(z)\le\1_{\{|z|\le2a\}}$, and apply \eqref{eq:a4-empirical-process} to $\psi_a$. For sufficiently small fixed $a_{\rm sb}>0$, with high probability, at most $\rho N/8$ columns satisfy $|\langle Y_{:j},q\rangle|\le a$, uniformly over all such directions with $\|q\|_1\le a_{\rm sb}\sqrt N$. Combined with \eqref{eq:a4-positive-density}, this bounds the weighted average below by $7\rho\zeta a/8$.

  Set $c_*=\min\{c_{\rm fr}/4,7\rho\zeta a/8\}$ and $d_{\rm sb}=\min\{\mathfrak m/8,c_*/(2C_A)\}$. In both cases, the average weighted by the surrogate weights is at least $c_*$, while replacing $\widetilde w$ by $w$ changes it by at most $N^{-1}\|w-\widetilde w\|_2\|Y^\top q\|_2\le C_Ad_{\rm sb}$. Thus (ii) holds with $r_t^{\rm rec}=a_{\rm sb}$, $c_t^{\rm rec}=c_*/2$, and $a_t=d_{\rm sb}$, all of which are chosen before the surrogate parameters. Restricting to $q\ge0$ gives the stated estimates.

  \subsection{Fixed-depth leave-one-out localization}
  \label{app:univ-loo}

  In this subsection we prove Proposition~\ref{prop:univ-loo}. We write $y^{(t)}=y_A^{(t)}$ and $\lambda^{(t)}=\lambda_A^{(t)}$, and assume
  $\max_{t\le K}\{\tau_{q,t}^{-1},\eta_t^{-1},\beta_t,\delta_t^{-1}\} \le(\log N)^C, \allowbreak\quad 0<\tau_{q,t},\eta_t\le1$.
  We write $P_N$ for a polylogarithmic bound whose exponent may increase finitely many times. The superscript $(-I)$ denotes deletion of the rows in $I$ from every quadratic penalty, with zero multipliers on $I$, and the superscript $[j]$ denotes replacement of column $j$ by zero throughout the recursion. In both cases, the objectives and parameters are unchanged, and the entire memory path is recomputed.

  We work on the event where $\|A\|_{\rm op}\le C_A$ and $\max_t\|v_t\|_2\le C_v\sqrt N$, which holds with high probability for either design. Comparing the optimal value with the value at zero, and using the nonnegativity of the penalties and $|v_t^\top y^{(t)}|\le C_vN$, we obtain
  \begin{equation}
    \|\lambda^{(t)}\|_2\le\sqrt N P_N.
    \label{eq:app-loo-coarse-dual}
  \end{equation}
  The same bound holds for every deleted problem, because deleting rows or zeroing a column cannot increase the operator norm, and $|y_j|,|c_j|\le1$.

  For states, we use the norm $\|H-\widetilde H\|=\|m-\widetilde m\|_2+ \|c-\widetilde c\|_2$. The memory gradients differ by at most $C\beta_t\|H-\widetilde H\|$, uniformly on the cube. For two $\tau_{q,t}$-strongly convex objectives, the variational inequalities at their minimizers bound the distance between the minimizers by $\tau_{q,t}^{-1}$ times the difference of the two gradients at either minimizer. Let $e_t^x,e_t^H$ denote the primal and state errors. Lemma~\ref{lem:univ-state-stability} then yields, with $D_t$ a bound on the change of the design-dependent gradient at the deleted optimizer,
  \begin{equation}
    e_t^x\le\tau_{q,t}^{-1}(C\beta_t e_{t-1}^H+D_t),
    \qquad
    e_t^H\le C(1+\delta_t^{-1})(e_{t-1}^H+e_t^x),\quad e_0^H=0.
    \label{eq:app-loo-state-update}
  \end{equation}
  Thus, if $D_t$ is polylogarithmic, then so are the primal and state errors up to the fixed depth $K$.

  \paragraph{Rows.}
  Conditionally on the objectives and all other rows, the deleted path $y^{(t,-i)}$ is independent of row $i$. Since $\|y^{(t,-i)}\|_2\le\sqrt N$, the leave-one-out score is uniformly subgaussian. A union bound gives $\max_{i,t}|A_{i:}y^{(t,-i)}|\le C\sqrt{\log N}$ with high probability. Restoring row $i$ adds a gradient term of norm $\eta_t^{-1}(\kappa_t-A_{i:}y^{(t,-i)})_+\|A_{i:}\|_2\le P_N$, so \eqref{eq:app-loo-state-update} yields $\max_{i,t}\|y^{(t)}-y^{(t,-i)}\|_2\le P_N$. Hence $|A_{i:}y^{(t)}|\le|A_{i:}y^{(t,-i)}| +C_A\|y^{(t)}-y^{(t,-i)}\|_2\le P_N$, and the formula \mbox{$\lambda^{(t)}=\eta_t^{-1}(\kappa_t\1-Ay^{(t)})_+$} also bounds $\|\lambda^{(t)}\|_\infty$ by $P_N$. On the remaining rows, the Lipschitz property of the positive part gives $\eta_t\|\lambda_{-i}^{(t)}-\lambda_{-i}^{(t,-i)}\|_2 \le C_A\|y^{(t)}-y^{(t,-i)}\|_2$, and the omitted coordinate has already been bounded. Consequently,
  $\max_{i,t}\bigl(\|y^{(t)}-y^{(t,-i)}\|_2 +\sqrt{\eta_t}\|\lambda^{(t)}-\lambda^{(t,-i)}\|_2\bigr)\le P_N$.
  For fixed $k$, we repeat the argument with the rows in $T\cup\{i\}$ deleted, where $|T|<k$ and $i\notin T$. Row $i$ is still independent of the deleted path, and a union bound over the $O(N^k)$ pairs bounds all scores by $C_k\sqrt{\log N}$. Restoring one row at a time and telescoping, we obtain $\max_{|I|\le k}\max_{t\le K} \|y^{(t)}-y^{(t,-I)}\|_2\le(\log N)^{C_k}.$ This proves the estimate for finitely many deleted rows used in Appendix~\ref{app:univ-low-l1}; here $k$ is fixed and does not depend on $N$.

  \paragraph{Columns.}
  Let $\widetilde A$ denote the design with column $j$ replaced by zero. The multiplier of the deleted problem is independent of $A_{:j}$, so conditional subgaussian tail bounds and \eqref{eq:app-loo-coarse-dual} give $\max_{j,t}|A_{:j}^\top\lambda^{(t,[j])}|\le P_N$ with high probability. Here we first restrict to the event on which the coarse bound holds for the deleted problem; this event does not depend on column $j$. The union of the complements of these events is contained in the exceptional event for the common bounds on the operator norm and the objective norms.

  Set $\widetilde y=y^{(t,[j])}$, and define $\lambda_B(x)=\eta_t^{-1}(\kappa_t\1-Bx)_+$ for $B=A,\widetilde A$. Since $(A-\widetilde A)\widetilde y=A_{:j}\widetilde y_j$, we have $\|\lambda_A(\widetilde y)-\lambda_{\widetilde A}(\widetilde y)\|_2 \le\eta_t^{-1}C_A$. With $P_{t,B}(x)=\|(\kappa_t\1-Bx)_+\|_2^2/(2\eta_t)$, we obtain
  \[
    \|\nabla P_{t,A}(\widetilde y)-\nabla P_{t,\widetilde A}(\widetilde y)\|_2
    \le C_A\|\lambda_A(\widetilde y)-\lambda_{\widetilde A}(\widetilde y)\|_2
    +|A_{:j}^\top\lambda^{(t,[j])}|\le P_N.
  \]
  The recursion \eqref{eq:app-loo-state-update} bounds the primal error by $P_N$. Moreover, $\eta_t\|\lambda^{(t)}-\lambda^{(t,[j])}\|_2 \le C_A(\|y^{(t)}-y^{(t,[j])}\|_2+1)$, and therefore
  $\max_{j,t}\bigl(\|y^{(t)}-y^{(t,[j])}\|_2 +\sqrt{\eta_t}\|\lambda^{(t)}-\lambda^{(t,[j])}\|_2\bigr)\le P_N$.
Finally, $|(A^\top\lambda^{(t)})_j|\le|A_{:j}^\top\lambda^{(t,[j])}|+C_A\|\lambda^{(t)}-\lambda^{(t,[j])}\|_2$. Together with the row estimates, this gives \eqref{eq:univ-polylog} for both designs, with $B_N=(\log N)^{C'}$, completing Proposition~\ref{prop:univ-loo}.

  \subsection{Fixed-depth comparison of restricted infima}
  \label{app:repeated-design-comparison}

We prove Lemma~\ref{lem:repeated-design-comparison} following the general-cost argument of~\cite{HanShen2023}, then verify its hypotheses for our costs. Condition on objectives in $\mathcal O_N$, so candidate sets are design-independent. As before, $P_N$ is a polylogarithmic bound whose exponent may increase finitely many times. Derivatives are with respect to standardized entries $Y_{ik}$, where $A=Y/\sqrt N$.

  \begin{proof}[Proof of Lemma~\ref{lem:repeated-design-comparison}]
    For each $\mathcal D_{N,j}$, choose an $h_N$-net $\mathcal N_{N,j}\subset\mathcal D_{N,j}$ in the $\ell_\infty$ norm, where $h_N=N^{-a-3}$. A maximal separated net satisfies $\log|\mathcal N_{N,j}|\le CN\log N$. Write
    $\varepsilon_{N,\mathrm{sm}}=\max_{j\le J}\sup_{Y,z}|f_{N,j,Y}^{\sigma}(z)-f_{N,j,Y}(z)|=o(1)$. On the bounded-entry event $\max_{i,k}|Y_{ik}|\le C\sqrt{\log N}$, the Lipschitz bound for the original costs and uniform approximation give
    \begin{equation*}
      0\le\min_{\mathcal N_{N,j}}f_{N,j,Y}^{\sigma}
      -\inf_{\mathcal D_{N,j}}f_{N,j,Y}^{\sigma}\le N^{-3}P_N+2\varepsilon_{N,\mathrm{sm}}=o(1).
    \end{equation*}
    This event has high probability under either design. For $\theta_N=N(\log N)^D$ with $D>1$, define the soft minimum
    \[
      S_{N,j}(Y)=-\theta_N^{-1}\log
      \sum_{z\in\mathcal N_{N,j}}e^{-\theta_Nf_{N,j,Y}^{\sigma}(z)}.
    \]
    The error of the soft minimum is at most $\theta_N^{-1}\log|\mathcal N_{N,j}|\le C(\log N)^{1-D}$. For a fixed entry, let $b_r=N^{-1-r/2}P_N$ be a bound on the $r$th derivative of the cost with respect to that entry. Differentiating the Gibbs weights gives
    \begin{equation*}
      |S_{N,j}'|\le b_1,\quad
      |S_{N,j}''|\le b_2+2\theta_Nb_1^2,\quad
      |S_{N,j}'''|\le b_3+6\theta_Nb_1b_2+8\theta_N^2b_1^3.
    \end{equation*}
The third derivative is the Gibbs mean of the third cost derivative, minus $3\theta_N$ times the covariance of its first two derivatives, plus $\theta_N^2$ times the third central moment of its first derivative. Thus the chain rule gives, for fixed $J$ and $\Phi\in C_b^3(\mathbb R^J)$,
    \begin{equation}
      \sup_Y|\partial_{Y_{ik}}^3\Phi(S_N(Y))|
      \le C_\Phi P_N(N^{-5/2}+\theta_NN^{-7/2}+\theta_N^2N^{-9/2}).
      \label{eq:app-hs-composite-derivative}
    \end{equation}
    This bound holds uniformly over all matrices. We replace the $MN$ entries one at a time, expanding to second order in the replaced entry. Since the two entry laws have matching means and variances, the first- and second-order Taylor terms cancel; the uniformly bounded third moments and \eqref{eq:app-hs-composite-derivative} then bound the total error by
    \begin{equation}
      C_\Phi N^{-1/2}P_N(1+\theta_N/N+\theta_N^2/N^2)
      \le N^{-1/2}(\log N)^{C'}=o(1).
      \label{eq:app-hs-replacement-error}
    \end{equation}
    Finally, we remove the soft minimum, the net, and the scalar smoothing, using their uniform error bounds and the boundedness of $\nabla\Phi$. The complements of the bounded-entry and objective events cost $o(1)$ because $\Phi$ is bounded. This proves \eqref{eq:univ-joint-infimum-comparison}.
  \end{proof}

  To verify the hypotheses for our costs, we convolve the positive part with a nonnegative, compactly supported mollifier at scale $\sigma$. The resulting function $h_\sigma$ satisfies
  \begin{equation}
    \|h_\sigma-(\cdot)_+\|_\infty\le C\sigma,
    \quad \|h_\sigma'\|_\infty\le1,
    \quad \|h_\sigma''\|_\infty\le C\sigma^{-1},
    \quad \|h_\sigma'''\|_\infty\le C\sigma^{-2}.
    \label{eq:app-hs-hinge-smoothing}
  \end{equation}
  Applying the same smoothing to each scalar projection preserves its range and its unit Lipschitz constant, at the cost of an error $C\sigma$. We use these smoothed projections in \eqref{eq:univ-residual}; the row truncation $\vartheta_N$ is already of class $C^3$.

  For a fixed candidate, the memory states do not depend on the matrix. A single entry $Y_{ik}$ appears in only $O_K(1)$ scalar summands, since it affects only the $i$th coordinate of $Ax^{(s)}$ and the $k$th coordinate of $A^\top\lambda^{(s)}$, with derivatives bounded by $2B_N/\sqrt N$. The projected residual coordinates are bounded by $4B_N$ for every matrix, and the first three derivatives of their squared losses with respect to the fields are bounded by $P_N(1+\sigma^{-2})$. The row lift satisfies the same bounds, because both its candidate variable and its truncated field lie in $[-2B_N,2B_N]$. The linear hinge term has first derivative bounded by $\ell_t$, and its higher derivatives are controlled by \eqref{eq:app-hs-hinge-smoothing}; the ridge and memory terms do not depend on the matrix at a fixed candidate. Accounting for the normalization and all polylogarithmic coefficients, we obtain
  \begin{equation*}
    \begin{aligned}
      \sup_{Y,z}|\partial_{Y_{ik}}^rf_{N,j,Y}^{\sigma}(z)|
      &\le N^{-1-r/2}P_N(1+\sigma^{-2}),\qquad 1\le r\le3,\\
      \sup_{Y,z}|f_{N,j,Y}^{\sigma}-f_{N,j,Y}|&\le P_N\sigma.
    \end{aligned}
  \end{equation*}
  Choosing $\sigma_N=(\log N)^{-S}$ with a sufficiently large fixed $S$ makes the smoothing error vanish, while the derivative bounds remain polylogarithmic; these factors are absorbed by the factor $N^{-1/2}$ in \eqref{eq:app-hs-replacement-error}.

  Finally, on the bounded-entry event, $\|A\|_{\infty\to\infty}+\|A^\top\|_{\infty\to\infty} \le C\sqrt{N\log N}$. Hence two candidates at $\ell_\infty$ distance $h$ have field differences at most $\sqrt N P_Nh$ and state differences at most $P_Nh$, since the state updates act columnwise at fixed depth. Because the residuals are bounded and the projections are unit-Lipschitz, the normalized squared losses differ by at most $\sqrt N P_Nh$. The hinge term satisfies the same bound, and the ridge, memory, and objective terms contribute at most $P_Nh$. Thus the Lipschitz bound in the candidate variable required by the lemma holds with $a=1$.

\subsection{Proof of Theorem~\ref{thm:intro-algorithm}}
\label{app:cold-rounding-transfer}

In this appendix, we prove Theorem~\ref{thm:intro-algorithm}. The partial-coloring step of Section~\ref{subsec:num-publication-finishing} makes at least three quarters of the remaining coordinates nearly tight. It is not one of the two step types of the strict finite certificates in Section~\ref{subsec:univ-stage2}, so we verify its Gaussian-tail condition directly and then apply the small-set completion bound of that section. Since this bound depends on the entry law, the rounding stage tries every number of partial-coloring steps up to $\lceil\log N\rceil$ and outputs the first solution it finds, so that the algorithm does not use the entry law.

\begin{corollary}[Rounding guarantee]
\label{cor:cold-rounding-transfer}
Assume the hypotheses of Theorem~\ref{thm:univ-exact}, and let $q_K>0$
be the terminal population free fraction. If
\begin{equation}
 \kappa_K-\kappa>\mathcal D_\alpha(q_K),
 \label{eq:app-cold-strict}
\end{equation}
where $\mathcal D_\alpha$ is the rounding cost~\eqref{eq:rounding-cost}, then for every fixed
centered, unit-variance subgaussian law of the i.i.d.\ entries, rounding from
$\widehat x_A^{(K)}$ produces, in randomized polynomial time and with high
probability, a vector $\sigma\in\{\pm1\}^N$ such that
$A\sigma\ge\kappa\1$.
\end{corollary}

As in the proof of Theorem~\ref{thm:intro-universality} in Section~\ref{subsec:univ-paths}, the hypotheses of Theorem~\ref{thm:intro-algorithm} imply those of Theorem~\ref{thm:univ-exact}, so Corollary~\ref{cor:cold-rounding-transfer} proves Theorem~\ref{thm:intro-algorithm}.

\begin{proof}
Write $q=q_K$ and $T=21/5$. Since $\mathcal D_{a,T}(q)$ is continuous in $a$, the strict inequality~\eqref{eq:app-cold-strict} persists if we replace $a=\max\{1,\alpha\}$ by a slightly larger rational number $a$. Write $A=Y/\sqrt N$ and
$S_0=\widehat F_K$. Theorem~\ref{thm:univ-exact} gives
$|S_0|/N\xrightarrow{\Pp}q$, while feasibility of the exact LP gives
$A\widehat x_A^{(K)}\ge\kappa_K\1$ with high probability.
For a fixed number $L$ of partial-coloring steps, we prescribe
$n_0=|S_0|,\allowbreak\quad n_{j+1}=\lfloor n_j/4\rfloor,\allowbreak\quad r_{j,N}=\frac{n_j}{N},\allowbreak\quad r_j=q4^{-j}, \allowbreak\quad 0\le j<L$.
At step $j$, the remaining set $S_j$ has size $n_j$, and we allocate to every
row the same loss $b_{j,N}=\sqrt{2Tr_{j,N}\log(200a/r_{j,N})}.$ These budgets are computable from the observed set sizes and converge in
probability to $b_j=\sqrt{2Tr_j\log(200a/r_j)}$.

For each row with a nonzero restriction, set
$c_{ij}=\sqrt N\,b_{j,N}/\|Y_{i,S_j}\|$; rows with zero restriction impose no
constraint and are omitted. Lemma~\ref{lem:univ-subset-row-norms} holds
uniformly over the adaptively chosen sets $S_j$. Since $r_j>0$ for fixed
$j$, its averaged estimate, together with the boundedness and continuity of the
standard Gaussian distribution function $\Phi$, gives
\begin{equation*}
\frac2N\sum_{i:\,\|Y_{i,S_j}\|>0} \Phi\!\left(-\frac{c_{ij}}{\sqrt T}\right) \xrightarrow{\Pp} 2\alpha\Phi\!\left(-\frac{b_j}{\sqrt{Tr_j}}\right) \le\alpha\exp\!\left(-\frac{b_j^2}{2Tr_j}\right) =\frac{\alpha r_j}{200a} \le\frac{r_j}{200}<\frac{r_j}{100}.
\end{equation*}
Consequently, the finite-dimensional tail condition of~\cite[Theorem~3.11]{LSZ2025} holds with high probability for $K_1=T$, $K_2=100$, $K_3=4/3$. Its other numerical condition holds because $1/T+1/100+3/4=524/525<0.999$. From any point in the current cube, without a norm lower bound, a successful step makes at least $3n_j/4$ coordinates nearly tight with row loss at most $b_{j,N}$. Freeze exactly $n_j-\lfloor n_j/4\rfloor$ of them. A fixed positive success probability and $O(\log N)$ checked retries from the same point give high-probability success. With $\delta_N=N^{-2}$, fixing these coordinates at the nearby endpoints costs $O(\delta_N\sqrt N)=o(1)$ per row by \eqref{eq:univ-subset-row-uniform}. Thus a fixed $L$-step prefix leaves $n_L/N\to r_L$ in probability and loses at most $\sum_{j<L}b_j+o(1)$ per row.

Let $\ell=\log(200a/q)$. The elementary inequality
$\sqrt{\ell+j\log4}\le\sqrt\ell+j\log4/(2\sqrt\ell)$ yields
\begin{equation*}
\sum_{j=0}^{\infty}b_j =\sqrt{2Tq}\sum_{j=0}^{\infty}2^{-j} \sqrt{\ell+j\log4} \le\sqrt{2Tq}\left(2\sqrt\ell+ \frac{\log4}{\sqrt\ell}\right) =\mathcal D_{a,T}(q).
\end{equation*}
Let $\varepsilon=\kappa_K-\kappa-\mathcal D_{a,T}(q)>0$.
The uniform small-set completion estimate of
Section~\ref{subsec:univ-stage2} bounds the additional row loss incurred in
rounding at most $RN$ remaining coordinates by
\[
 B(R)+o(1),\qquad
 B(R)=C_\alpha K_\xi\sqrt R
       \log\frac{e^2(1+\alpha)}{R}\longrightarrow0
       \quad(R\downarrow0).
\]
Since $B$ depends on the entry law through $K_\xi$, so does the number of
partial-coloring steps that suffices. The rounding stage therefore runs, for
each $L=1,\dots,\lceil\log N\rceil$ and with fresh randomness, the $L$-step
prefix followed by ordinary partial coloring. A run ends its prefix early once no free coordinates remain, and it aborts as soon as a step
fails all its checked retries, and it outputs the first
$\sigma\in\{\pm1\}^N$ with $A\sigma\ge\kappa\1$. Fix $L_0$ such that
$2r_{L_0}<1$ and $B(2r_{L_0})<\varepsilon/2$, and let $N\ge e^{L_0}$, so that
the run with $L=L_0$ takes place. With high probability, its prefix leaves at
most $2r_{L_0}N$ free coordinates, and every row margin is at least
$\kappa+\varepsilon-o(1)$. Ordinary partial coloring then rounds these
coordinates within the remaining reserve, so this run, and hence the rounding
stage, outputs a solution with high probability. The checked retries and the
inverse-polynomial precision of Section~\ref{subsec:univ-stage2} bound the
running time of each run by a polynomial, so the $\lceil\log N\rceil$ runs take
randomized polynomial time in total; the resulting rounding errors are $o(1)$
and are absorbed by the strict reserve.
\end{proof}

\end{appendices}

\end{document}